\documentclass{aa}

\usepackage{graphicx}
\usepackage{txfonts}
\usepackage[]{hyperref}
\usepackage{natbib}
\usepackage{xspace}

\newcommand{\Msol}{\ensuremath{\,\mathrm{M_\odot}}\xspace}

\newcommand{\Lsol}{\ensuremath{\,\mathrm{L_\odot}}\xspace}

\newcommand{\kk}{\,kK\xspace}

\newcommand{\Myr}{\,Myr\xspace}

\newcommand{\kms}{\,km\,s$^{-1}$\xspace}

\newcommand{\GG}[1]{}

 \def\simle{\mathrel{\hbox{\rlap{\hbox{\lower4pt\hbox{$\sim$}}}\hbox{$<$}}}}
 \def\simgr{\mathrel{\hbox{\rlap{\hbox{\lower4pt\hbox{$\sim$}}}\hbox{$>$}}}}

\usepackage{amstext}

\begin{document}

%\title{The Small Magellanic Cloud has no clear deficiency of very massive stars, but also no excess}
   \title{A dearth of young and bright massive stars in the Small Magellanic Cloud}
%\title{No excess of very massive stars in The Small Magellanic Cloud}
\author{A. Schootemeijer\inst{1} \and N. Langer\inst{1,2} \and D. Lennon\inst{3,4} \and C. J. Evans\inst{5} \and P. A. Crowther\inst{6} \and S. Geen\inst{7} \and I. Howarth\inst{8} \and A. de Koter\inst{7} \and K. M. Menten\inst{2} \and J. S. Vink\inst{9}} %\and (?? non-final; exact order tbd)}
\institute{Argelander-Institut f\"{u}r Astronomie, Universit\"{a}t Bonn, Auf dem H\"{u}gel 71, 53121 Bonn, Germany\\  \email{aschoot@astro.uni-bonn.de} \and
Max-Planck-Institut für Radioastronomie, Auf dem Hügel 69, 53121 Bonn, Germany \and
Instituto de Astrofísica de Canarias, E-38200 La Laguna, Tenerife, Spain \and
Departamento de Astrof\'{i}sica, Universidad de La Laguna, E-38205 La Laguna, Tenerife, Spain \and
UK Astronomy Technology Centre, Royal Observatory, Blackford Hill, Edinburgh EH9 3HJ, UK \and
Department of Physics and Astronomy, University of Sheffield, Sheffield, S3 7RH, UK \and
Anton Pannekoek Institute for Astronomy, University of Amsterdam, Science Park 904, 1098 XH Amsterdam, The Netherlands \and
University College London, Gower Street, London WC1E 6BT, UK \and
Armagh Observatory, College Hill, Armagh BT61 9DG, UK
}
%\authorrunning{Schootemeijer et al.}
%\titlerunning{A dearth of young and bright massive stars in the SMC}
   \date{Received September -- ; accepted --}

   \abstract{ 
%Stars more massive than several tens of Solar masses in low-metallicity environments are the progenitors of many exciting transient phenomena. With their ionizing radiation and mechanical feedback, they are also thought to have played a dominant role in shaping galaxies in the early universe -- especially because such stars are expected to be more common at low metallicity. However, no proof for an overabundance of such massive stars has been found in the Small Magellanic Cloud (SMC; the only metal-poor, actively star-forming environment in the Galactic neighborhood) in earlier studies.
Massive star evolution at low metallicity is closely connected to many fields in high-redshift astrophysics, but is poorly understood so far.
Because of its metallicity of $\sim$0.2\,$Z_\odot$, its proximity, and because it is currently forming stars, the Small Magellanic Cloud (SMC) is a unique laboratory in which to study metal-poor massive stars.
}
{We seek to improve the understanding of this topic using available SMC data and a comparison to stellar evolution predictions.} %We find that massive stars in the SMC are not as rare as previously thought.
{We used a recent catalog of spectral types in combination with GAIA magnitudes to calculate temperatures and luminosities of bright SMC stars. By comparing these with literature studies, we tested the validity of our method, and using GAIA data, we estimated the completeness of stars in the catalog as a function of luminosity.
%Although this method is coarse\textbf{r} than a detailed atmosphere analysis, it
This allowed us to obtain a nearly complete view of the most luminous stars in the SMC. %As a result, w
We also calculated the extinction distribution, the ionizing photon production rate, and the star formation rate.
}
{%\textbf{M}assive stars with a luminosity of $\log (L/L_\odot) > 5.5$ ($M \gtrsim 40$\Msol) \textbf{appear to be relatively rare} in the SMC. We also find a \textbf{strikingly low number} of young massive stars. This confirms that this peculiarity, which has already been reported in the Milky way and Large Magellanic Cloud, extends to SMC metallicity. We find a lack of progenitors of helium-burning stars, especially at high luminosities. From the massive star population, we derive a star formation rate that is slightly lower than, but consistent with, literature values. The derived H\,I ionizing photon production rate is also in agreement with the SMC's H$\alpha$ luminosity.
Our results imply that the SMS hosts only $\sim$30 very luminous main-sequence stars ($M \geq 40$\Msol; $L \gtrsim 3 \cdot 10^5$\Lsol), which are far fewer than expected from %earlier work. 
the number of stars in the luminosity range $ 3 \cdot 10^4 < L/L_\odot < 3 \cdot 10^5$ and from the typically quoted star formation rate in the SMC.
Even more striking, we find that for %luminosities
masses above $M \gtrsim 20$\Msol, stars in the first half of their hydrogen-burning phase are almost absent. This mirrors a qualitatively similar peculiarity that is known for the Milky Way and Large Magellanic Cloud. This amounts to a lack of hydrogen-burning counterparts of helium-burning stars, which is more pronounced for higher luminosities. We derived the H I ionizing photon production rate of the current massive star population. It agrees with the H\,$\alpha$ luminosity of the SMC.
}
{%We argue that the first explanation for these two main results that should be considered, and also the most likely one, is a bias against the detection of bright young stars. More specifically, the young bright stars could still be buried in their birth clouds.
%Alternative explanations to invoke could be the star formation history (although deemed unlikely) or a steeper initial mass function. To claim the latter, given its enormous astrophysical implications, very strong evidence would be required.
%\textbf{The dearth of young and bright massive O stars in the SMC provides an interesting puzzle. %None of the explanations we discuss seems satisfactory at the current time. %Perhaps the most plausible is a prolonged embedding phase in birth clouds.}
%\textbf{A steeper initial mass function (IMF) cannot explain the missing young stars. There are various arguments against a halt in star formation. }
%We argue that the dearth of young and bright massive stars in the SMC is likely due to an observational bias and that many young bright stars could still be buried in their birth clouds. We find that a declining star formation rate or a steep initial mass function are less likely explanations. 
We argue that a declining star formation rate or a steep initial mass function are unlikely to be the sole explanations for the dearth of young bright stars. 
%Perhaps more likely is that many of these stars are embedded in their birth clouds, although observational evidence for this is lacking.
Instead, many of these stars might be embedded in their birth clouds, although %\textbf{clear} 
observational evidence for this is weak.
%A lack of objects manifesting themselves as hot bright stars at low metallicity would have 
We discuss
implications for the role that massive stars played in cosmic reionization, and for 
%initial mass function measurements based on star counting.
the top end of the initial mass function.
}
%To get the IMF more reliably, also ages would be necessary.

   \keywords{Stars: massive -- Stars: early-type -- Stars: evolution -- Galaxies: star formation -- Galaxies: stellar content}% -- Ultraviolet: general}

\maketitle

\section{Introduction \label{sec:introduction}}
Massive stars in low-metallicity environments are linked to spectacular astrophysical phenomena, such as mergers of two 
black holes \citep{Abbott16}, long-duration gamma-ray bursts \citep{Graham17}, and superluminous supernovae \citep{Chen17}. Moreover, massive stars are thought to have played a crucial role in providing ionizing radiation and mechanical feedback in galaxies in the early universe \citep{Hopkins14}. %stars that contain less heavy elements, t

Massive star evolution at low metallicity therefore is of great importance, but it is also poorly understood.
For example, some theoretical predictions suggest that very massive stars are more more likely to form in low-metallicity environments \citep[e.g.,][]{Larson71, Abel02}, although the arguments for this are debated \citep{Keto06, Krumholz09, Bate09, Kuiper18}.
This in turn would have important consequences for the extent to which these astrophysical phenomena take place.
While studies of individual stars in the early universe are currently not feasible, a recent analysis of the %supermassive star cluster 
starburst region 30 Doradus in the nearby Large Magellanic Cloud (half the solar metallicity) did indeed find an overabundance of very massive stars \citep{Schneider18}. This overabundance would be in line with the predictions mentioned above, but it is currently unknown if metallicity effects and/or environmental effects cause this excess of massive stars. Neither do we know whether this trend continues toward lower metallicity.
Because it has only about one-fifth of the solar metallicity \citep{Hill95, Korn00, Davies15}, a better understanding of the Small Magellanic Cloud (SMC) is a crucial stepping stone.
Earlier studies have been unable to find indications of an overabundance of massive stars in the SMC, however \citep{Blaha89, Massey95}. 
%For field stars in the SMC, the initial mass function (IMF) with $dN \propto M^\Gamma d\log M$ even appears to be significantly steeper than the canonical Salpeter exponent, which has a value of $\Gamma = -1.35$. \cite{Lamb13} find $\Gamma = -2.3$ for these stars.
For field stars in the SMC, \cite{Lamb13} derived an exponent of $\Gamma = -2.3$ for the initial mass function (IMF), which is much steeper than the canonical Salpeter exponent of $\Gamma = -1.35$ (with $dN \propto M^\Gamma d\log M$, where $N$ is the number of stars that are born, and $M$ is the stellar mass). %even appears to be significantly , which has a value . 

%In addition to the IMF, a more complete picture of the massive star content of the SMC is instructive. 
In addition to the IMF, there is much to gain from a more complete picture of the massive star content of the SMC. 
A prevalence of blue supergiants can constrain internal mixing \citep{Schootemeijer19, Higgins20} and binary interaction \citep{Justham14}. Moreover, models of gravitational wave progenitors \citep[e.g.,][]{deMink16, Marchant16, Belczynski16, Kruckow18}, of which the physical assumptions are extrapolated toward low metallicity, can be put to the test.
%Furthermore, one can investigate if the aforementioned ionizing radiation sources actually become more common towards lower metallicity.
%\textbf{Furthermore, one can test the prediction that massive stars at low metallicity emit more ionizing radiation.}
Furthermore, we can investigate the amount of ionizing radiation that is emitted by massive stars at low metallicity.

A main tool that is used to test stellar evolution predictions is the Hertzsprung-Russell diagram (HRD).
This has been applied for the SMC massive star population in dedicated studies published about two to three decades ago. They used %as a whole has been subjected to dedicated studies. 
two different approaches: photometry, and spectroscopy. %(to obtain spectral types). 
\cite{Massey02} and \cite{Zaritsky02} took the first approach and mapped $UBV$ magnitudes of SMC sources, which are indications for the location of these sources in the HRD. While these studies can be expected to be highly complete, they lack accuracy in predicting the effective temperatures and luminosities of hot stars \citep[e.g.,][]{Massey03}. 
Effective temperatures can be predicted more accurately from spectral types, which then also reduces the uncertainty in luminosity.
\cite{Blaha89} and \cite{Massey95} have used this method to compile HRDs of bright SMC stars. Inevitably, the completeness of their input catalogs is lower than for the photometry catalogs. The main focus of these studies was the IMF%(\textit{extinction?})
, which they found to be consistent with the IMF in the Milky Way (see the discussion above).

Since then, various developments in the field of observational astronomy have taken place that allow a major leap forward.
First, spectral types of many more SMC stars have become available, which have been compiled in the catalog of \cite{Bonanos10}, hereafter referred to as B10. Second, the second data release (DR2) of the all-sky GAIA survey \citep{Gaia18} provides a spatially complete catalog of SMC sources with information on their magnitudes and motions, thereby yielding information about which sources are in the foreground. Third, detailed atmosphere analyses on subsets of massive SMC stars \citep{Trundle04,Trundle05,Mokiem06,Hunter08b,Bouret13,Dufton19,Ramachandran19} can improve our understanding of how effective temperature correlates with spectral type. They also
enable a systematic verification of the reliability  of the HRD positions based on spectral types.
In summary, new observations allow a large improvement in terms of sample size, accuracy, completeness assessment, and identification of foreground sources. 
Our goal is to fully exploit these new observations. We use them to provide an extinction distribution, ionizing photon emission rates, and the relations of spectral type and temperature for massive SMC stars, and most importantly, also the improved HRD. 
%Our main, and unexpected, result is that this HRD contains a low number of young massive stars and a low number of bright massive stars.  For comparison: based on photometry, \cite{Massey10} reports 330 O-type stars in the SMC with a mass $M > 50$\Msol. Our results imply that this number is of the order of ten. Alternatively, we will show that if one would assume that the SMC currently forms stars at the typical literature rate of 0.05\Msol\,yr$^{-1}$ and has a Salpeter IMF, one would expect $\sim$300 stars with $M > 40$\Msol. We only infer the presence of a total of $\sim$40 of these stars (of which $\sim$30 are on on the main sequence). We will spend a significant part of the paper on considering different explanations. 
This paper is organized in the following way. In Sect.\,\ref{sec:methods} we describe our methods and the catalogs we used. Then, in Sect.\,\ref{sec:results}, we present the general properties of the stars in our sample. In Sect.\,\ref{sec:hrd_features} we perform an in-depth analysis of features of the massive star population in the SMC. We find that it contains only a few bright stars and young stars. This is the main result of this paper. We further discuss our main result in Sect.\,\ref{sec:numbers}, where we compare numbers. In Sect.\,\ref{sec:discussion} we consider possible explanations for our main result: a steeper IMF, model uncertainties, star formation history, observational biases, unresolved binaries, and embedding in birth clouds. Finally, we present our conclusions in Sect.\,\ref{sec:conclusions}.

\section{Methods \label{sec:methods}}
To achieve our goal of providing a more complete picture of luminous SMC stars, we employ three data sets in this study. We describe them in detail in Appendix\,\ref{sec:data_sets}. The first and most essential data set is retrieved from the B10 spectral type catalog (their table\,1). We then cross-correlate it with the GAIA DR2 catalog. Out of the 5324 B10 sources, we find a match in 5304 cases. All of these have listed $G$ magnitudes in GAIA DR2. We note that only $\sim$3000 sources have a $V$ magnitude listed in the B10 catalog. As a result, the use of $G$ magnitudes improves the number of sources for which we can calculate a luminosity. Out of the 5304 matched sources, 5269 pass our foreground test (Appendix\,\ref{sec:data_sets}). We discuss potential biases in the B10 catalog in Sect.\,\ref{sec:obs_biases}. 

The second data set is %the part of the GAIA catalog that is relevant for our purposes 
retrieved from GAIA DR2 alone
(again, see Appendix\,\ref{sec:data_sets} for details). The completeness of GAIA DR2 \citep[see][]{Gaia18, Arenou18} is essentially 100\% in the magnitude range of our sources of interest, which extends to $G \approx 16$. Therefore we can use this data set to estimate the completeness of the B10 catalog.

The third data set is a compilation of literature data in which atmosphere analyses have been performed on a sample of bright SMC stars. This data set is referred to as the various spectroscopic studies (VSS) sample, and is extracted from the studies of \cite{Trundle04}, \cite{Trundle05}, \cite{Mokiem06}, \cite{Hunter08b}, \cite{Bouret13}, \cite{Dufton19}, and \cite{Ramachandran19}. The VSS data set contains temperatures and luminosities, while the other two do not. It consists of 545 sources. Of these, 160 fall in the luminosity range of our main HRD ($\log (L/L_\odot) > 4.5$; Fig.\,\ref{fig:hrd_corr}).

Our aim is to also calculate temperatures and luminosities for stars in the B10 data set, which contains many more stars than the VSS sample.
%We use the spectral types from the B10 catalog and the relation between spectral types and effective temperatures of the stars in the VSS sample to calculate temperatures for all the B10 sources.
We use relations of spectral type and temperature to calculate effective temperatures of the stars in the B10 data set. For OB-type stars, we derive our own relations based on data from the VSS sample. For A-types and later, we use existing relations.
Then, we use their GAIA $G$ magnitudes to obtain a luminosity and place the B10 sources in an HRD. 
%In addition, we use GAIA photometry to estimate the completeness of the B10 catalog. Below, we describe these procedures in more detail.
We describe these procedures in more detail in Sect.\,\ref{sec:meth_lt}, and we explain our estimation of the B10 completeness with the GAIA data set in Sect.\,\ref{sec:f_complete}. 

\subsection{Deriving effective temperature and luminosity \label{sec:meth_lt}}
We used the existing studies from the VSS sample of SMC stars (last paragraph of Appendix\,\ref{sec:data_sets}), which provide spectral types as well as temperatures, %, with in addition ten A-type stars from \cite{Venn99},
to derive empirical relations of spectral types and effective temperatures ($T_\mathrm{eff}$). For each spectral type, we took the average derived temperature. We did this separately for stars with luminosity class (LC) V+IV, LC III+II, and LC I. When only one star of a certain spectral type was available, the temperature that we adopted was the average of its derived temperature and the temperatures we found for the two neighboring types. As an example: for spectral type B0, a neighboring type is O9.7, and a neighbor of a neighbor would be O9.5. 
%To make the relations smoother, we use the temperatures of neighboring types and types that are neighbors of neighbors in all possible cases (by possible cases, we mean types that have two neighboring types on both sides). While doing that, we give these a statistical weight of 0.5 and 0.25, respectively. For example, the smoothed temperature for B0-type stars is calculated as: $ T_\mathrm{eff, B0, smooth} = (0.25T_\mathrm{eff, O9.5} + 0.5T_\mathrm{eff, O9.7} + T_\mathrm{eff, B0} + 0.5T_\mathrm{eff, B0.2} + 0.25T_\mathrm{eff, B0.5}) / 2.5$.
%We only include spectral types that are mentioned in the studies described above. 
Next, for all spectral types that have neighbors and neighbors-of-neighbors at either side, we smoothed the relations of spectral type and temperature. Illustrated for type B0, the applied formula is as follows: $ T_\mathrm{eff, B0, smooth} = (0.25T_\mathrm{eff, O9.5} + 0.5T_\mathrm{eff, O9.7} + T_\mathrm{eff, B0} + 0.5T_\mathrm{eff, B0.2} + 0.25T_\mathrm{eff, B0.5}) / 2.5$.
%The values \textbf{that} we obtain \textbf{from this procedure} are displayed in Table.\,\ref{tab:spt_teffs}. 

%______________________________________________

   \begin{figure}
   \centering
   \includegraphics[width = \linewidth]{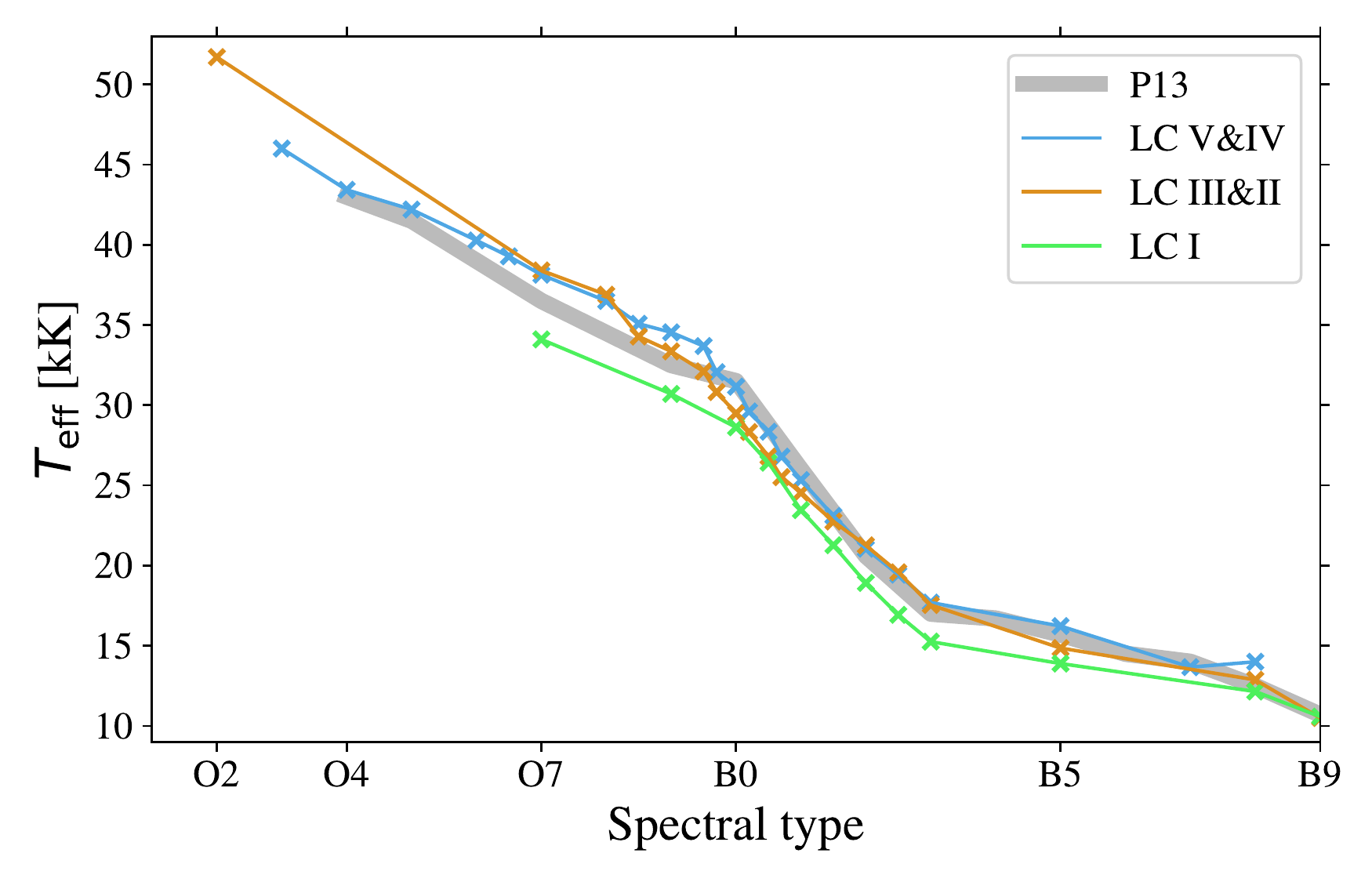}
   \caption{Derived relations of spectral types and effective temperatures (crosses) for different luminosity classes (LCs). The solid gray line shows the relation for Galactic dwarf stars \citep[P13;][]{Pecaut13}. }
             \label{fig:spt_teff}%
    \end{figure}
%______________________

The resulting spectral type - $T_\mathrm{eff}$ relations are shown in Fig.\,\ref{fig:spt_teff} and Table\,\ref{tab:spt_teffs}. For stars of A-type and later, we used the SMC spectral type - temperature relations of \cite{Evans03} and \cite{Tabernero18}. We compared our results with relations for Galactic dwarf (LC V) stars \citep[from][their extended table\,5]{Pecaut13}%\footnote{{\tt http://www.pas.rochester.edu/$\sim$emamajek/\newline EEM\_dwarf\_UBVIJHK\_colors\_Teff.txt}}.
\footnote{\url{http://www.pas.rochester.edu/~emamajek/EEM\_dwarf\_UBVIJHK\_colors\_Teff.txt}}.
The evolved LC I stars are cooler than dwarf stars that have the same spectral type. %\citep[in line with the results of][]{Mokiem07},
Moreover, early-type dwarfs at low metallicity are hotter than their Galactic counterparts with the same spectral type. These two trends are in line with the trends shown in \cite{Trundle07}. 
The trend for types earlier than O7 is not in line with these trends; the line for LC\,III and II stars lies above the values for LC\,V and IV stars. %The reason is probably that there, the LC\,III and II line is based on only one data point - a consequence of how rare such stars are.
However, the LC\,III and II line there has only one data point as a consequence of how rare such stars are, so that some scatter can be expected.
The typical differences between the \cite{Pecaut13} spectral type - $T_\mathrm{eff}$ relations and ours do not exceed $1-2$\kk.

We were unable to convert the spectral type of 114 sources in the B10 data set into a temperature (e.g., `Be? + XRB'). We applied the spectral type - $T_\mathrm{eff}$ relations to infer effective temperatures for the remaining 5155 sources. 
%{From the 304 sources that we do not include, 192 are omitted because they have spectral type of F or later. We focus on  A-type and earlier stars ($T_\mathrm{eff} \gtrsim 10$\kk), because a rather complete picture of the cooder bright stars already exists in the SMC \citep[e.g.,][]{Neugent10, Davies18, Yang19}. 
%The other 114 
When the temperature was known, the bolometric correction (BC) in the $G$ band was taken from the MIST \citep{Dotter16, Choi16} website\footnote{\url{http://.waps.cfa.harvard.edu/MIST/model_grids.html}, where we take those with the label `DR2Rev'}.
Because the surface gravity of most of the sources in the B10 data set is unknown, we used the BCs for $\log g = 3$. At higher and lower $\log g$, these BCs match temperature values that are typically well within 1\kk. %http://waps.cfa.harvard.edu/MIST/model_grids.html
In the B10 data set, 116 sources have the label `binary'. We used the spectral type that is mentioned first for them and further treated them as single stars. We highlight them in Fig.\,\ref{fig:hrd_asc10}.
%We can expect about half the sources to be a binary system \citep{Sana13}. For an equal-mass binary, we will derive a luminosity that is 0.3\,dex higher than that of its individual components. Given that luminosity scales strongly with mass, the error will be significantly smaller in most cases.
%We use the blackbody approximation to calculate the GAIA $G_\mathrm{BP} - G_\mathrm{RP}$ colors, $G$ band fluxes, and bolometric corrections (BCs) over the relevant temperature range. 
%With the temperatures given from the spectral types, we proceed as follows to estimate the luminosities. 
%\footnote{{\tt http://waps.cfa.harvard.edu/MIST}}.

%We compare the effective temperatures of stars in the VSS sample to those predicted by their $G_\mathrm{BP} - G_\mathrm{RP}$ colors. %, againusing the MIST BCs. 
%\textbf{The predicted colors are obtained by calculating the differences between the $G_\mathrm{BP}$ and $ G_\mathrm{RP}$ BCs from MIST.}
For stars in the VSS sample, we compared the GAIA colors that are predicted for their effective temperatures to their observed GAIA colors. The predicted colors are calculated as $\mathrm{BC}(G_\mathrm{BP}) - \mathrm{BC}(G_\mathrm{RP}$) as given by MIST.
%To be more precise, we obtain the predicted color by subtracting the $G_\mathrm{RP}$ BC from the $G_\mathrm{BP}$ BC for a
%We find that the temperatures and colors match best for a reddening of $E(G_\mathrm{BP} - G_\mathrm{RP}) \approx 0.14$. This translates to $A_G = 0.28$ and $A_V = 0.35$ \citep[table\,3 from][see also \cite{Gordon03}]{Wang19}.
%We find that the temperatures and colors match best for an extinction of $A_V = 0.35$ which implies a reddening of $E(G_\mathrm{BP} - G_\mathrm{RP}) = 0.14$, and $A_G = 0.28$ \citep[table\,3 from][see \cite{Gordon03}, also]{Wang19}. % reddening of $E(G_\mathrm{BP} - G_\mathrm{RP}) \approx 0.14$. This translates to $A_G = 0.28$ and $A_V = 0.35$ \citep[table\,3 from][see also \cite{Gordon03}]{Wang19}.
We find that the predicted and observed colors match best for a reddening of $E(G_\mathrm{BP} - G_\mathrm{RP}) = 0.14$. This value corresponds to an extinction of $A_G=0.28$ in the $G$ band and $A_V=0.35$ in the $V$ band \citep[table\,3 from][see also \citealt{Gordon03}]{Wang19}. We use these extinction values throughout the paper.

%______________________________________________

   \begin{figure}
   \centering
   \includegraphics[width = \linewidth]{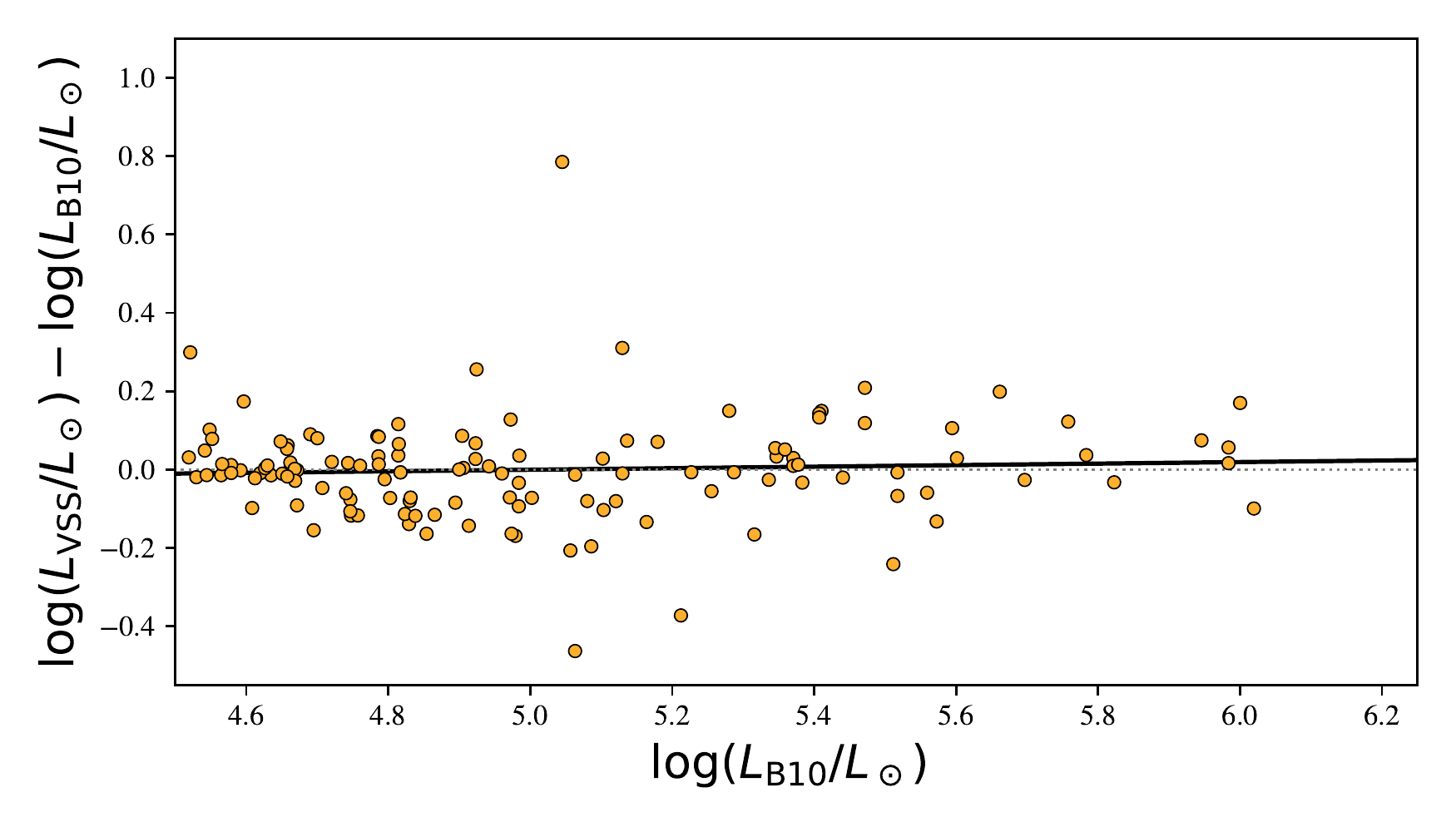}
   \caption{Difference between the logarithm of the luminosities reported in the various spectroscopic studies (VSS) sample and those derived by our method using spectral types of B10 as a function of $\log (L_\mathrm{B10}/L_\odot)$. Each dot represents an individual source. The dotted line shows where $ L_\mathrm{B10} = L_\mathrm{VSS}$; the solid line is a linear fit to the scatter points.
   }
             \label{fig:logl_dlogl}%
    \end{figure}
%______________________

%n extinction that would be of magnitude $A_V = 0.35$ in the $V$ band. 
Furthermore, we adopted a distance modulus (DM) of the SMC of 18.91 \citep{Hilditch05}.
%Then, using the DM and the BC, we calculate the luminosity using the observed $G$ magnitude.
Then we calculated the absolute bolometric magnitude of a source using
\begin{equation}
    M_\mathrm{bol} = m_\mathrm{G} + \mathrm{BC} - \mathrm{DM} - A_G,
\end{equation}
which is translated into luminosity using $\log (L/L_\odot) = -0.4(M_\mathrm{bol} - M_\mathrm{bol, \, \odot})$. Here, we adopt a solar value of $M_\mathrm{bol, \, \odot} = 4.74$.

To test this method, we show the luminosities of the sources brighter than $\log (L/L_\odot) = 4.5$ in Fig.\,\ref{fig:logl_dlogl} for the sources that are included in the VSS sample and in the B10 data set. On the y-axis we show the difference between the luminosity reported in the VSS sample and the luminosity we obtained with our method based on the B10 data set. The black line, showing a linear fit, indicates that there is no systematic offset between the luminosities derived by our B10 method and the VSS literature values. The values of $\log (L_\mathrm{VSS}/L_\odot) - \log (L_\mathrm{B10}/L_\odot)$ have a standard deviation of $\sigma = 0.13$\,dex. % with respect to the fitted line.

\subsection{Investigating the completeness with GAIA photometry \label{sec:f_complete}}
Although the B10 catalog contains more stars than the VSS sample, it still does not contain all of the brightest stars in the SMC. We can expect the GAIA data to be much more complete. This is supported by the fact that we found a GAIA counterpart for 99.6\% of the B10 sources; see also a discussion of this in Appendix\,\ref{sec:fcompl_tests}. Unfortunately, however, GAIA colors are less accurate in determining effective temperatures (and therefore luminosities) than spectral types, and the results strongly depend on extinction. This is especially true for the hotter stars.

Using GAIA colors and magnitudes, we therefore cannot individually derive reliable luminosities for our sources. However, %we argue that 
the colors and magnitudes can be used to estimate the completeness of an ensemble of stars, in this case, the B10 catalog.
%To clarify: w
With `completeness' we mean the completeness fraction of hot, bright stars that can be identified as such by the relevant observational approaches, that is, spectroscopy, and/or photometry in the optical.

%______________________________________________

   \begin{figure}
   \centering
   \includegraphics[width = \linewidth]{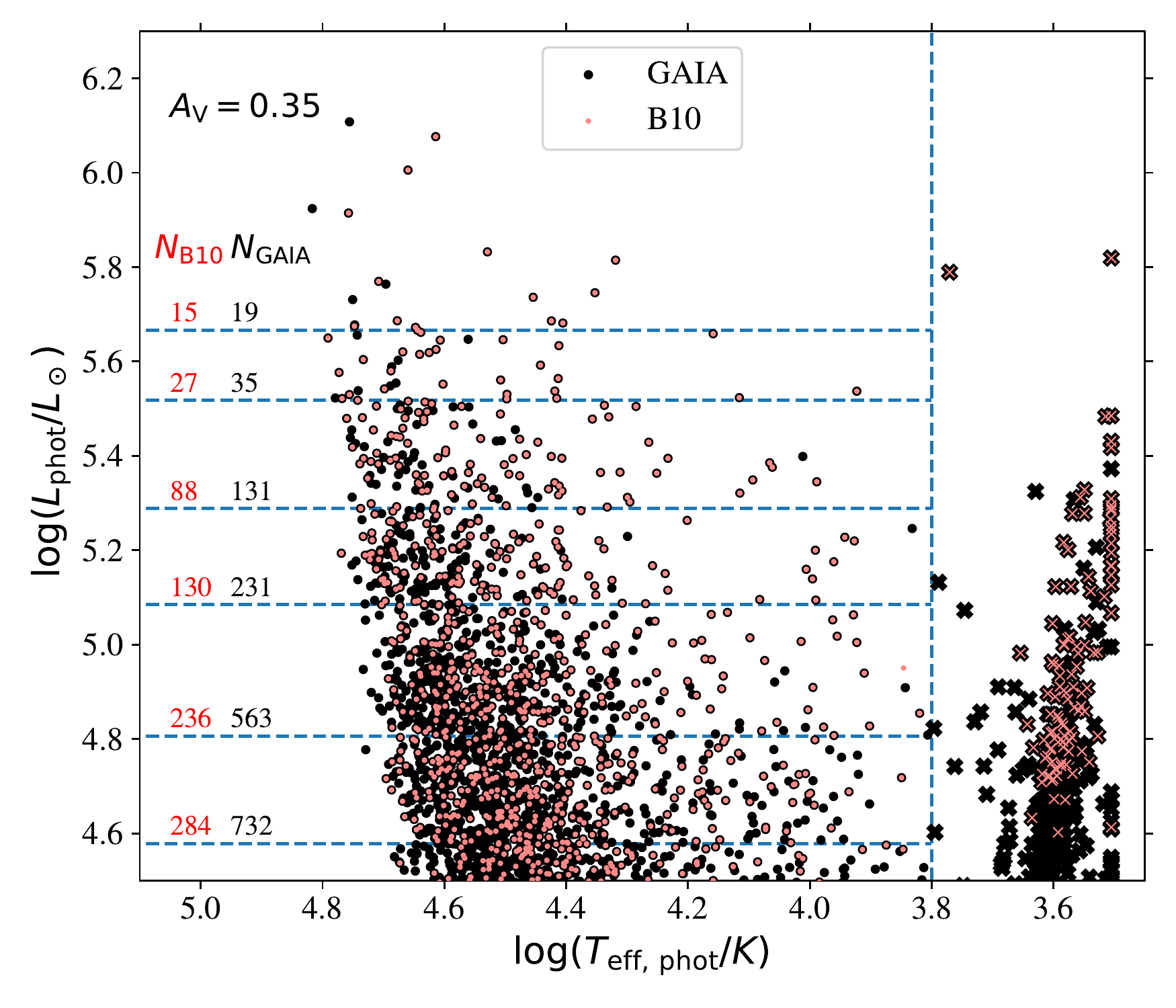}
   \caption{Hertzsprung-Russell diagram of bright SMC sources constructed with GAIA photometry. Black markers indicate sources from the GAIA sample. The smaller red markers are sources in the B10 sample. Therefore, red points with black edges (all but one of the red points, see Appendix\,\ref{sec:data_sets}) are sources that are listed in both samples.
   We count the number of sources that are hotter than $10^{3.8}$\,K for both samples in different luminosity intervals. The numbers are displayed on the left side of the plot. The luminosity intervals are indicated by dashed blue lines. 
   For completeness, we also show the cool sources with crosses. %The colors have the same meaning.}
   }
             \label{fig:gaia_hrd}%
    \end{figure}
%______________________

\begin{figure*}
   \centering
   \includegraphics[width = \linewidth]{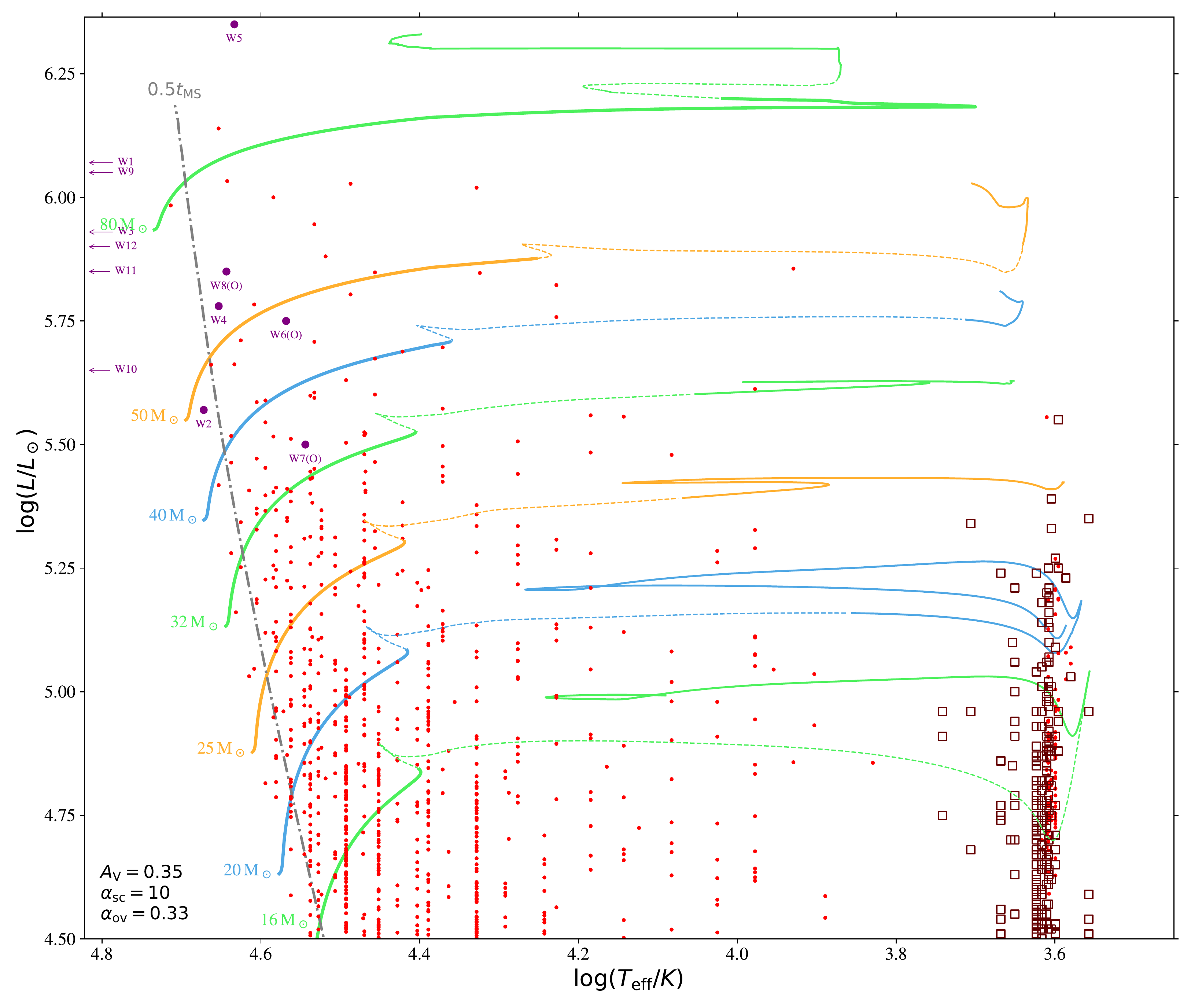}
   \caption{
   Hertzsprung-Russell diagram of luminous stars in the SMC. 
   %Red dots are sources with observables derived using `B10' \citep{Bonanos10} and GAIA data.
   Red dots represent sources from the B10 data set.
   Open dark red squares are red supergiants from \cite{Davies18}, where the $T_\mathrm{eff}$ is obtained using relations from \cite{Tabernero18}. We show the Wolf-Rayet stars \citep{Hainich15, Shenar16, Shenar17} labeled with a `W' and their identifying number. Whe  they have an O-star companion that is brighter in the $V$-band, we show this instead (indicated by `(O)'). We also show evolutionary tracks of \cite{Schootemeijer19} with a semiconvection parameter of $\alpha_\mathrm{sc} = 10$, and an overshooting parameter of $\alpha_\mathrm{ov} = 0.33$. Solid lines indicate hydrogen- and helium-core burning phases; dashed lines indicate the in between phase. The dash-dotted gray line shows the location of these models halfway through their main-sequence (MS) lifetime, $t_\mathrm{MS}$.
   %Hertzsprung-Russell diagram of massive stars in the Small Magallanic Cloud. Red dots are sources with observables derived using \cite{Bonanos10} and GAIA data. If in the VSS sample an atmospheric analysis has been performed on a source, we point an arrow to the temperature and luminosity derived there. The studies that we use for this are referred to as H08 \citep{Hunter08}, R19 \citep{Ramachandran19}, M06 \citep[][only the AzV stars, since the others are already in the H08 catalog]{Mokiem06}, B13 \citep{Bouret13}, T04/05 \citep{Trundle04, Trundle05}. In addition, we show the Wolf-Rayet stars \citep{Hainich15, Shenar16, Shenar17} labeled with a `W' and their identifying number. If they have an O-star companion that is brighter in the $V$-band, we show it instead (indicated by `(O)'). Also shown are evolutionary tracks of \cite{Schootemeijer19} with a semiconvection parameter of $\alpha_\mathrm{sc} = 1$, and for overshooting $\alpha_\mathrm{ov} = 0.33$.
   }
             \label{fig:hrd_corr}%
    \end{figure*}

We used GAIA photometry, and again MIST BCs, to calculate the photometric temperature and luminosity of the B10 and GAIA sources\footnote{The observed GAIA color is used to infer a temperature. Apart from this, the procedure is the same as described in Sect.\,\ref{sec:meth_lt}.}: $T_\mathrm{eff, \, phot}$ and $L_\mathrm{phot}$. We then plotted them in an HRD that we refer to as the `photometric HRD' (Fig.\,\ref{fig:gaia_hrd}). %In this HRD, 
%There, we show both the GAIA sources and the B10 sources (that are also in the GAIA catalog).
Next, we used this HRD to estimate the completeness of the B10 catalog as a function of luminosity.
For this estimate we also used an HRD for which we used B10 spectral types instead of GAIA colors (Fig.\,\ref{fig:hrd_corr}), as we explain below.
We only considered temperatures above $T_\mathrm{eff} = 10^{3.8}$\,K 
%We do not believe that with our method we can do better in terms of completeness and accuracy than the existing work of 
because for cooler temperatures we rely on the results of \cite{Davies18} on cool, bright SMC stars. The way we operate is described below. 

%__________________________________________________ One column table
\begin{table}
\centering    
    \caption[]{Number of stars with $T_\mathrm{eff} \gtrsim 10^{3.8}$\,K counted in different luminosity intervals. %See text for explanation.
    }
\begin{tabular}{l | r r r r }
\hline
\hline
            \noalign{\smallskip}
            $\log (L_\mathrm{B10}/L_\odot)$ & $N_\mathrm{B10}$ & $N_\mathrm{GAIA,\, phot}$ & $N_\mathrm{B10} / N_\mathrm{GAIA,\, phot}$\\
            \noalign{\smallskip}
            \hline
            \noalign{\smallskip}
            5.75+ & 15 & 19 & 0.79 \\
            5.50-5.75 & 27 & 35 & 0.77 \\
            5.25-5.50 & 88 & 131 & 0.67\\
            5.00-5.25 & 130 & 231 & 0.56 \\
            4.75-5.00 & 236 & 563 & 0.42 \\
            4.50-4.75 & 284 & 732 & 0.39\\ %            %\noalign{\smallskip}
            \noalign{\smallskip}
            \hline
            \noalign{\smallskip}
            Total & 780 & 1640 & 0.47%              
            %\hline
            \label{tab:f_complete}

\end{tabular}
\end{table}
%______________________

%______________________________________________

   \begin{figure}[t]
   \centering
   \includegraphics[width = \linewidth]{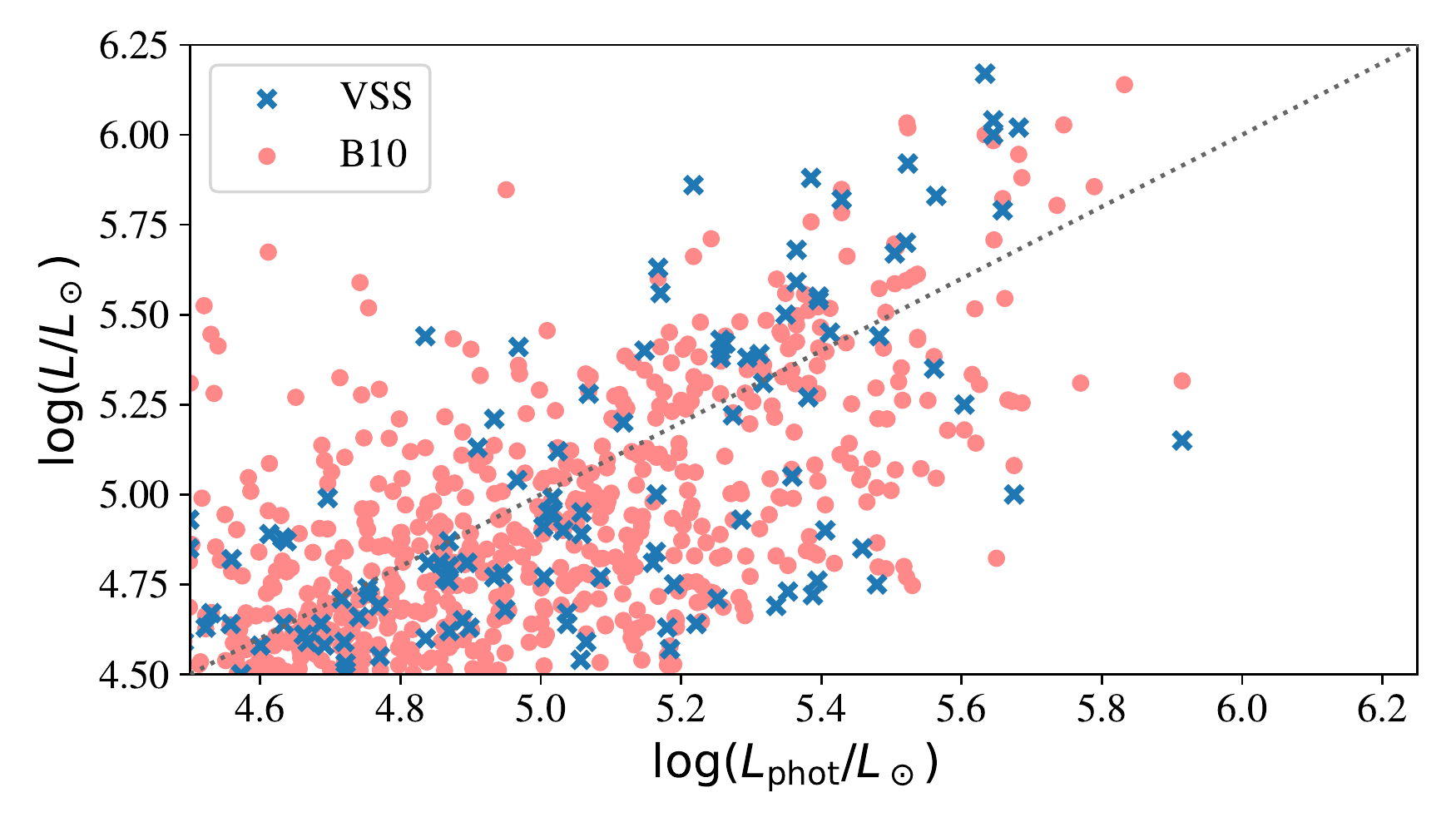}
   \caption{Correlation between the luminosity of sources as calculated using their GAIA color and magnitude ($L_\mathrm{phot}$)
   and the luminosity of the same source derived in various spectroscopic studies (VSS) or its luminosity based on the temperature derived from the spectral type listed in B10.
   %obtained by other means. These other means are looking up the luminosity reported in the various spectroscopic studies (VSS), or the calculating the luminosity based on a temperature derived from the spectral type listed in \citealt{Bonanos10} (B10).
   %and their luminosity derived in our sample using the \cite{Bonanos10} catalog ($L_\mathrm{B10}$). The dashed line shows where $L_\mathrm{phot} = L_\mathrm{B10}$.)
   }
             \label{fig:lvsl}%
    \end{figure}
%______________________

\begin{enumerate}
    \item We counted the number of sources in Fig.\,\ref{fig:hrd_corr} %(that we will introduce in Fig.\,\ref{fig:hrd_corr}) 
    in different luminosity intervals. The numbers are listed under $N_\mathrm{B10}$ in Table\,\ref{tab:f_complete}. For example, in Fig.\,\ref{fig:hrd_corr} we count 15 stars above $\log (L_\mathrm{B10}/L_\odot) = 5.75$.
    \item Then we examined the photometric HRD (Fig.\,\ref{fig:gaia_hrd}), from high to low luminosity. In this example we considered the 15 brightest stars in B10, which represent the $\log (L_\mathrm{B10}/L_\odot) > 5.75$ bin.
    \item We counted the number of GAIA sources until we found 15 sources that were also in the B10 catalog.
    We took this approach to have the same number of B10 sources in each luminosity bin.
    %encounter the 16$^\mathrm{th}$ brightest star from the B10 catalog.
    %Therefore, the counting in Fig.\,\ref{fig:gaia_hrd} does not per definition take place in the same luminosity intervals as those listed in the left column of Table\,\ref{tab:f_complete}.
    This counting in Fig.\,\ref{fig:gaia_hrd} took place in the intervals that are separated by dashed blue lines. They do not by definition coincide exactly with the luminosity intervals listed in the left column of Table\,\ref{tab:f_complete}. 
    In this example at the bright end, %, where the B10 catalog should be almost complete, 
    we count $N_\mathrm{GAIA,\, phot} = 19$. 
    For the $\log (L_\mathrm{B10}/L_\odot) > 5.75$ bin, we thus estimate that it has a completeness fraction of $ 15 / 19 = 0.79$.
    %Thus, in the (GAIA-derived) luminosity range of the 12 brightest sources in the B10 catalog method, there are four sources that are in the GAIA catalog but not in the B10 catalog.
    \item We repeated this for each luminosity interval (i.e., also for $5.50 < \log (L_\mathrm{B10}/L_\odot) < 5.75$ with the 16th$^\mathrm{}$ to 42nd$^\mathrm{}$ brightest star from the B10 catalog method, etc.) listed in Table\,\ref{tab:f_complete} to calculate the completeness fraction of the B10 catalog: $N_\mathrm{B10}/N_\mathrm{GAIA,\, phot}$.
\end{enumerate}

%______________________________________________

%______________________

Fig.\,\ref{fig:gaia_hrd} and Table\,\ref{tab:f_complete} show that the completeness level of the B10 catalog is 70\% to 80\% for the brightest sources. The completeness drops below 40\% around $\log (L/L_\odot) = 4.5$. %Later on in Sect.\,\ref{sec:ldist}, we will consider this when we discuss the luminosity distribution of bright stars in the SMC.

%The question is: how reliable is this method? %A requirement is that the order of the luminosities is not mixed up too much by taking the photometrically derived value. 
For this estimate to be reliable, the values of $L_\mathrm{B10}$ and $L_\mathrm{phot}$ need to be similar for most of the stars.
We note that if $L_\mathrm{phot}$ had no predictive power, we would expect to measure the same completeness in all luminosity bins. %A test in Appendix\,\ref{sec:fcompl_tests} shows (Fig.\,\ref{fig:order_of_logl}) that 
We investigate this further in Fig.\,\ref{fig:lvsl}, where we compare $L_\mathrm{phot}$ of the sources to their VSS luminosity and the luminosity calculated using their B10 spectral type. This figure shows that while the scatter is large, the luminosities obtained with GAIA photometry on average give a good indication in which luminosity segment most sources belong.
%the order of $L_\mathrm{phot}$ correlates reasonably well with the order of the luminosities as they are given in literature for our VSS sample (see Fig.\,\ref{fig:order_of_logl}). $L_\mathrm{phot}$ correlates reasonably well with $L_\mathrm{B10}$, although the scatter is sizable. %\textbf{(i.e., the luminosities that we deem most reliable). This is shown in Fig.\,\ref{fig:order_of_logl}.} \textit{idea: scatter L instead?}
%\textit{Furthermore, earlier on we have found that there are 780 sources in the B10 catalog above the luminosity limit of $\log (L/L_\odot) = 4.5$. Making the same assumptions for extinction and no further tuning, we find a similar value for this luminosity limit (i.e., above which there are 780 B10 sources) of $\log (L_\mathrm{phot}/L_\odot) = 4.586$. This strengthens our assumption that $L_\mathrm{phot}$ is applicable for this completeness estimate, where we consider a population rather than individual sources. \textbf{(Norbert: I'm inclined to kill the italic part, ok?)}}

In Appendix\,\ref{sec:fcompl_tests} we provide another (simpler) test in which we count blue sources in the B10 and GAIA DR2 catalogs. The trends in this second test are very similar to those described in this section. %In Appendix\,\ref{sec:fcompl_tests}, we also briefly discuss the completeness of the GAIA catalog.}

Both of our completeness tests imply a higher completeness than the completeness quoted for O stars in B10 itself ($\sim$4\%). However, their number is based on an estimate of 2800 O stars of $M > 20$\Msol from the conference proceedings of \cite{Massey10}. This number is in turn based on $UBV$ photometry from \cite{Massey02}. In their table\,8b, all\textit{} 70 blue stars that are known to have O types have a photometry-derived temperature in the O-star regime. Because of the uncertainties in determining the temperatures of hot stars with photometry \citep[e.g.,][]{Massey03}, this means that flagging many B-type stars as O-type stars seems unavoidable with this method. This has also been argued by \cite{Smith19}, and we refer to Appendix\,\ref{sec:fcompl_tests} for a quantitative discussion that supports this view. 
It therefore seems likely that the O-star completeness fraction quoted in B10 is severely underestimated.
%Thus, there are good reasons to believe that the O-star completeness fraction quoted in B10 is severely underestimated.

In the Simbad catalog\footnote{\url{http://simbad.u-strasbg.fr}}, we inspected the four most luminous sources in Fig.\,\ref{fig:gaia_hrd} that are not in the B10 catalog. From high to low luminosity in Fig.\,\ref{fig:gaia_hrd}, these are i) Sk\,177 \citep{Sanduleak69} with the unusable spectral type information of `OB'; ii) the O5.5\,V star of $\log (L/L_\odot) \approx 5.3$ in \cite{Lamb13}; iii) Sk\,183, which is an O3\,V star with $\log (L/L_\odot) =5.66$ \citep{Evans12}; and iv) a source without information on spectral type. %Thus, where information is available, these sources are bright (as their color and magnitude indicate) but not exceptionally bright.

We return to the luminosity-dependent completeness fraction presented in Table\,\ref{tab:f_complete} in Sect.\,\ref{sec:ldist}. There we discuss the luminosity distribution of stars in the B10 sample and compare it with theoretical predictions.

%We notice a subtle trend in $\Delta\log L$ in Fig.\,\ref{fig:logl_dlogl}, where our method using the B10 catalog more strongly under-predicts the luminosity of the brightest sources. %Below, we give two possible reasons for this.
%The first is that stars of the same temperature that are larger (and thus have a higher $\log L$ and in practice a lower $\log g$) tend to have weaker line blanketing (and thus, higher flux) in the UV than their smaller, dimmer counterparts \citep[see, e.g., the synthetic spectra of][]{Hainich19}\footnote{Available at {\tt www.astro.physik.uni-potsdam.de/$\sim$wrh/PoWR}}. As a result, larger, brighter stars need a larger BC. %The BC that we use, however, is not dependent on $\log g$ because we use the blackbody approach. Therefore, our method would under-predict the luminosity of the brightest stars, compared to less bright stars.

%A possible reason for the brightest stars to appear relatively more dim is that they might be more likely to reside in regions with a relatively high dust production.This would mean they are more prone to have relatively strong extinction.

%______________________________________________

   \begin{figure}[t]
   \centering
   \includegraphics[width = \linewidth]{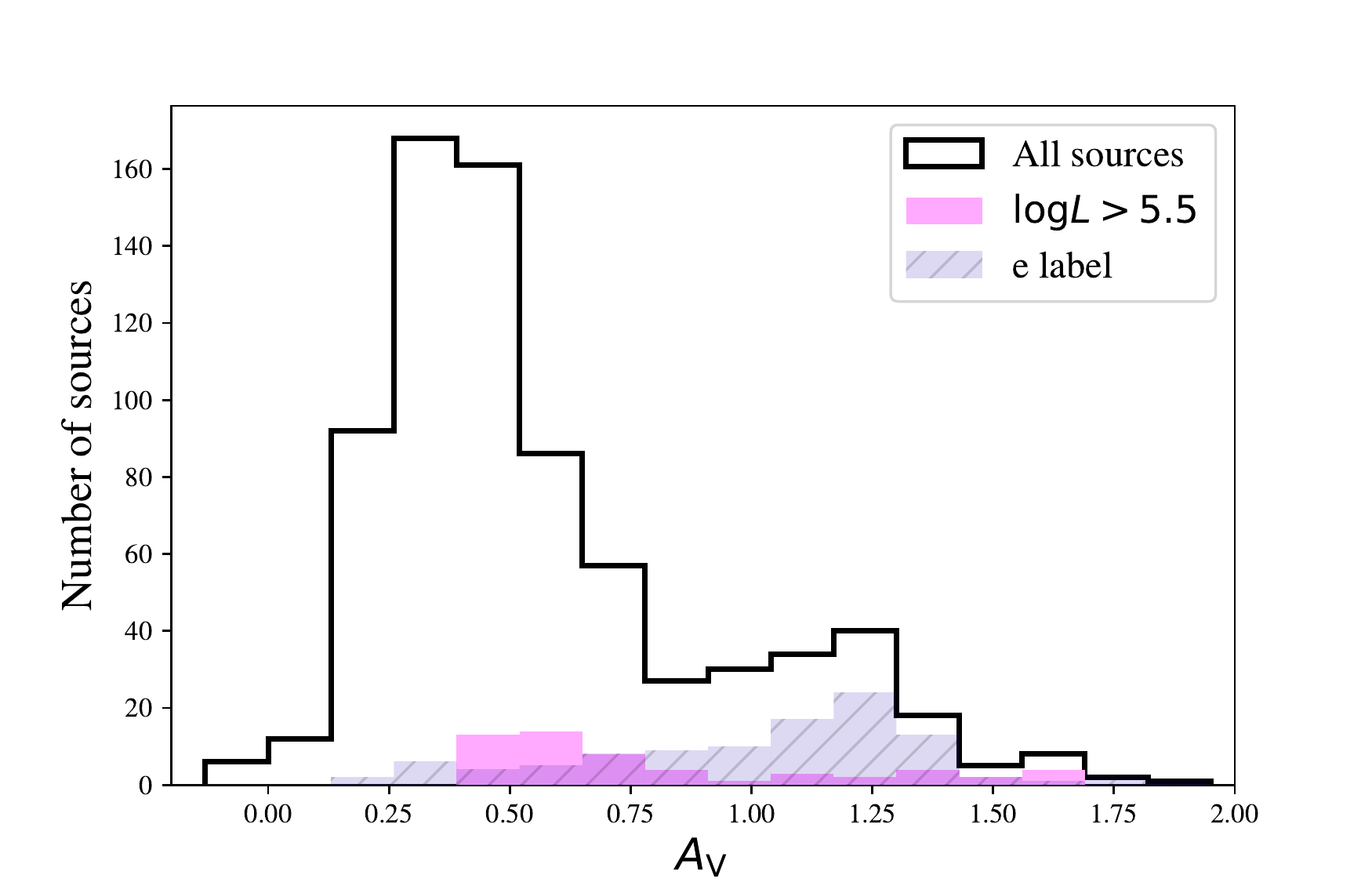}
   \caption{Distribution of the visual extinction $A_V$ in the B10 sample when measured for each source individually. We also show the $A_V$ distribution of the subset for which we derive a luminosity higher than $\log (L/L_\odot) = 5.5$, and the 
   subset of stars with an `e' label in the B10 data set.}
             \label{fig:AVdist}%
    \end{figure}
%______________________

%\section{Results \label{sec:results}}
\section{General population properties \label{sec:results}}

%\subsection{Presentation in the Hertzsprung-Russell diagram}

%In this section we present the results of our analysis in the form of an HRD. %We also use the sources in our sample to calculate the distribution of the visual extinction $A_V$, the ionizing photon production rate, and to make an estimate for the star formation rate.

%\textbf{Afterwards, in Sect.\,\ref{sec:hrd_features}, we will discuss the main features of the stellar distribution in the HRD. These include a high number of apparently post-MS blue supergiants, an apparent lack of young sources, and an apparent lack of bright sources.}

%In the remainder of this section, we \textbf{will present general properties of the stellar population in our HRD (Fig.\,\ref{fig:hrd_corr}): the} visual extinction \textbf{distribution}, ionizing photon production rate and star formation rate (SFR). 

Fig.\,\ref{fig:hrd_corr} shows the distribution of the B10 sources in the HRD. 
It contains 780 stars with $T_\mathrm{eff} > 10^{3.8}$\,K that are more luminous than 10$^{4.5}$\Lsol.
Below $T_\mathrm{eff} = 10^{3.8}$\,K, we rely on the red supergiant (RSG) sample of \cite{Davies18}, which we overplot on the HRD. To avoid duplicates, the B10 sources in this low-temperature regime are not considered in the rest of the paper.
In Sect.\,\ref{sec:hrd_features} we discuss the main features of the stellar population shown in Fig.\,\ref{fig:hrd_corr}. We first describe a number of checks that we performed.

%In Appendix \ref{sec:app_a}, we show and discuss a number of test HRDs; %In those, we do, for example, a more detailed comparison with observations, and we test how our results depend on assumptions that we made in the `Methods' section (Sect.\,\ref{sec:methods}).} %We discuss these in Sect.\,\ref{sec:test_hrds}}.
%here, we summarize our findings from these tests. 
In Fig.\,\ref{fig:hrd_arr} we compare the HRD positions of stars in Fig.\,\ref{fig:hrd_corr} to their HRD positions according to the VSS sample (if they are in the VSS sample). We note that %the HRD positions of our stars in Fig.\,\ref{fig:hrd_corr} match reasonably well to those in the VSS sample 
they match reasonably well,
as is the case for the derived luminosities (Fig.\,\ref{fig:logl_dlogl}). 

When we calculate the extinction for each star individually (Fig.\,\ref{fig:hrd_varAV}, and see Sect.\,\ref{sec:ex_etc}) instead of assuming a constant $A_V = 0.35$, the shape of the population does not significantly change compared to the population shown in Fig.\,\ref{fig:hrd_corr}. A difference is that with individually calculated extinction, we find more stars above the 40\Msol track. However, compared to the VSS sample, the luminosities of the brightest stars are then systematically overestimated (Fig.\,\ref{fig:logl_dlogl_varAV}). On the other hand, a constant $A_V = 0.35$ does not result in a systematic offset for the brightest stars (Fig.\,\ref{fig:logl_dlogl}). 
Moreover, with variable extinction, the standard deviation in $\log (L_\mathrm{VSS}/L_\odot) - \log (L_\mathrm{B10}/L_\odot)$ is the same (0.13\,dex) as with the constant $A_V = 0.35$. For these reasons, we use $A_V = 0.35$ in our main HRD (Fig.\,\ref{fig:hrd_corr}).

The shape of the population also remains almost intact when different spectral type - temperature relations are used, as we show in Fig.\,\ref{fig:hrd_p13}. The same is true when we take a different input catalog (Fig.\,\ref{fig:hrd_sim}, where we use Simbad instead of B10). This demonstrates that the results we present later on are robust against the choice of assumptions described in Sect.\,\ref{sec:methods}.
For the sources shown in Fig.\,\ref{fig:hrd_corr}, we also calculated the distribution of the visual extinction $A_V$ (Sect.\,\ref{sec:ex_etc}), the ionizing photon production rate (Sect.\,\ref{sec:ionizing_radiation}), and estimate the implied star formation rate (SFR; Sect.\,\ref{sec:sfr}).

%We discuss the method to calculate the general properties of the massive star population in the SMC --visual extinction, ionizing photon production rate and star formation rate (SFR) -- in more detail in Appendices\,\ref{sec:app_a} and \ref{sec:app_b}. 

\subsection{Extinction \label{sec:ex_etc}}

%\textbf{Using the stars in our HRD (Fig.\,\ref{fig:hrd_corr}), we can make estimates for their extinction, the ionizing radiation emitted by them, and the star formation rate in the SMC. We discuss the method in more detail in Appendix\,\ref{sec:app_a} and \ref{sec:app_b}.}

%The distribution of the individually calculated extinction $A_V$ of the sources B10 sample (that are shown in Fig.\,\ref{fig:hrd_varAV}) is displayed in Fig.\,\ref{fig:AVdist}. 
%We calculated the visual extinction $A_V$ of the individual B10 sources by comparing their expected intrinsic GAIA color to the observed GAIA color (Appendix\,\ref{sec:xtinxion}).

We calculated the extinction sources in the B10 data set with a luminosity of $\log ( L / L_\odot ) > 4.5$. In order to do so, %we first calculate the reddening $E(G_\mathrm{BP} - G_\mathrm{RP})$. In this procedure 
we employed the effective temperature and resulting expected intrinsic color $(G_\mathrm{BP} - G_\mathrm{RP})_\mathrm{int}$, which we obtained as described in Sect.\,\ref{sec:meth_lt}. %The expected intrinsic color $(G_\mathrm{BP} - G_\mathrm{RP})_\mathrm{int}$ is calculated as also described in Sect.\,\ref{sec:meth_lt}. 
We then obtained the reddening as $E(G_\mathrm{BP} - G_\mathrm{RP}) = (G_\mathrm{BP} - G_\mathrm{RP})_\mathrm{obs} - (G_\mathrm{BP} - G_\mathrm{RP})_\mathrm{int}$, where $(G_\mathrm{BP} - G_\mathrm{RP})_\mathrm{obs}$ is the observed GAIA color. Then we calculated the extinction in the $V$ band as $A_\mathrm{V} = 2.42 \cdot A(G_\mathrm{BP} - G_\mathrm{RP})$. This relation is obtained from values in table\,3 of \cite{Wang19}. More details are given in Appendix\,\ref{sec:xtinxion}.

Fig.\,\ref{fig:AVdist} displays the resulting $A_V$ distribution. %of  the sources shown in the resulting HRD (Fig.\,\ref{fig:hrd_varAV}).
It shows that most stars have an extinction of $A_V = 0.3$ to 0.5. The distribution has a second peak slightly above $A_V = 1$ that is mainly caused by Be stars; these tend to be redder than what would be expected for their spectral type. This results in a higher inferred value for $A_V$. The brightest stars also tend to have a slightly higher extinction, although only modestly so. For $A_V / E(B-V) = 2.74$ in the SMC \citep[][]{Gordon03}, the peak in the $A_V$ distribution that we find agrees reasonably well with the canonical SMC reddening value of $E(B-V) \approx 0.1$ \citep[e.g.,][]{Massey95, Gorski20}.

%______________________________________________

   \begin{figure}
   \centering
   \includegraphics[width = \linewidth]{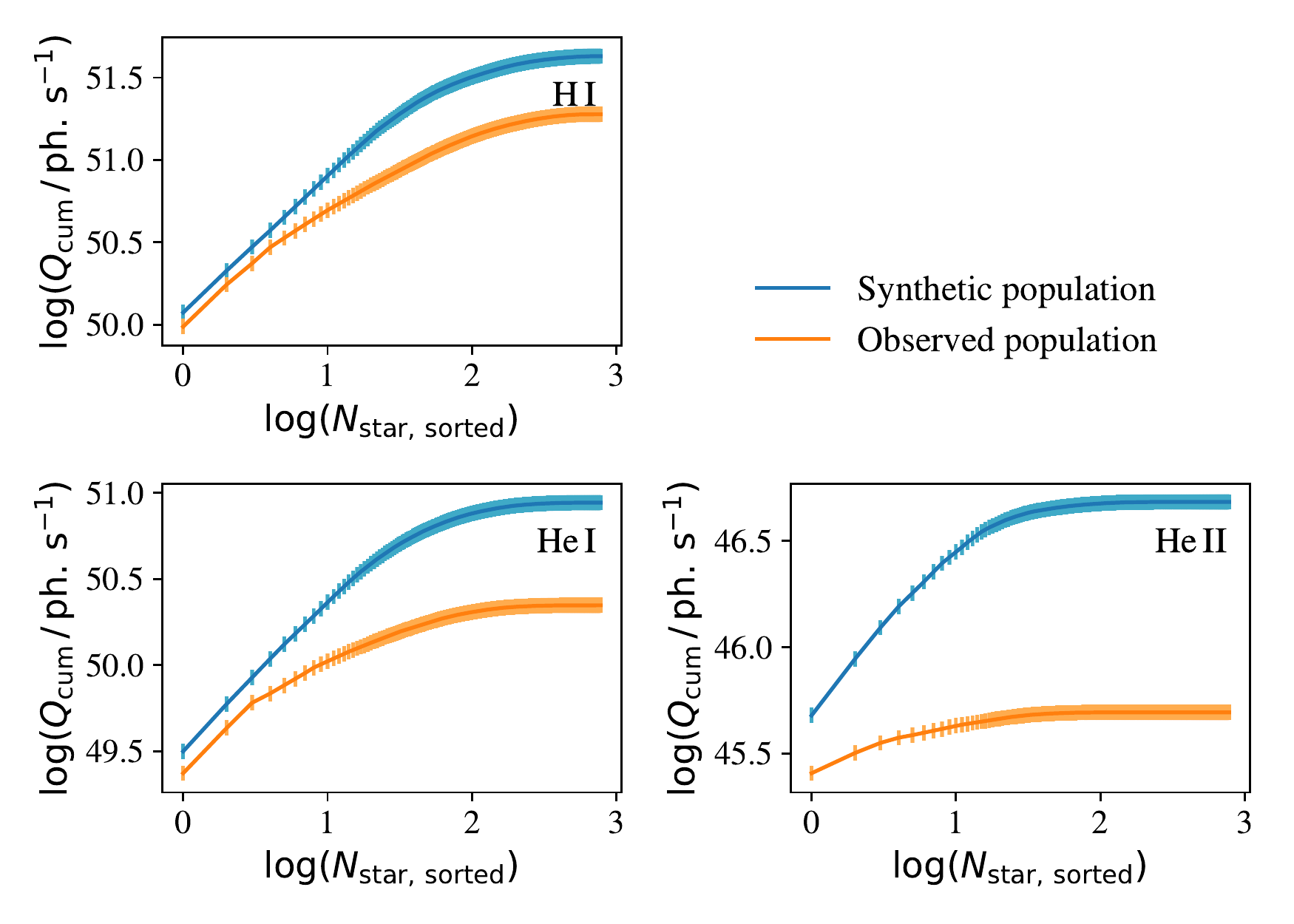}
   \caption{Diagram of the cumulative distribution of the ionizing photon production rate $Q$. The stars in this cumulative distribution are sorted from a high to low $Q$.
   }
             \label{fig:iotons}%
    \end{figure}
%______________________

\subsection{Ionizing radiation and its escape fraction \label{sec:ionizing_radiation}}

%POWR models \citep{Hainich19} 
Potsdam Wolf-Rayet (POWR) stellar atmosphere models \citep{Hamann06, Todt15, Hainich19} 
provide predictions for the ionizing photon production rate ($Q$) of bright SMC stars. We used them to calculate the H\,I, He\,I, and He\,II ionizing photon production rate of the 780 stars shown in our HRD (for details, see Appendix\,\ref{sec:app_b}). Fig.\,\ref{fig:iotons} shows the resulting cumulative distributions. For comparison, we also calculated the ionizing photon production rate of a synthetic population of 780 stars from the single-star models of \cite{Schootemeijer19}, adopting random ages, which is displayed in Fig.\,\ref{fig:hrd_synt} (for details, see Sect.\ref{sec:rel_age_dist}).
The synthetic population of bright SMC stars emits more ionizing radiation than the observed population in Fig.\,\ref{fig:hrd_corr}, while both contain the same number of stars. The difference is largest for He\,II ionizing radiation, which depends on the presence of a few hot stars. The reason for this large difference is that we find few young stars in the observed population. We discuss this dearth of young stars in Sect.\,\ref{sec:rel_age_dist}.

%Compared to the WR stars in the SMC, non-WR stars in Fig.\,\ref{fig:hrd_corr} emit about twice as much H\,I ionizing photons. 
We find that the observed population shown in Fig.\,\ref{fig:iotons} emits about $\log Q = 2 \cdot 10^{51}$ photons per second that are energetic enough for H\,I ionization. 
In addition to this, the Wolf-Rayet (WR) stars contribute half as many,  $10^{51}$ H\,I ionizing photons per second, which means a total of $3 \cdot 10^{51}$ H\,I ionizing photons per second. 
For He\,I ionizing photons, on the other hand, WR stars emit about a factor $\text{}$four more than the observed population of Fig.\,\ref{fig:iotons}. He\,II ionizing radiation is dominated by WR stars by two to three orders of magnitude.

For H\,I ionizing photons, we can compare the numbers discussed above with observations.
%\textbf{The SMC has an} integrated H$\alpha$ luminosity of $4.8 \cdot 10^{39}$ erg\,s$^{-1}$. \textbf{Under the rough assumption that all the H\,I ionizing photons are absorbed by neutral hydrogen,} it follows that about $3.5 \cdot 10^{51}$ H\,I ionizing photons per second are produced in the SMC \citep{Kennicutt95}. To get the above number, we used the relation $Q(\mathrm{H}\,\alpha) = 7.31 \cdot 10^{11} L_\mathrm{H \, \alpha} \cdot \mathrm{erg}^{-1}$ \citep{Osterbrock06}.
The SMC has an integrated H$\alpha$ luminosity of $4.8 \cdot 10^{39}$ erg\,s$^{-1}$, which, assuming $Q(\mathrm{H}\,\alpha) = 7.1 \cdot 10^{11} L_\mathrm{H \, \alpha} \cdot \mathrm{erg}^{-1}$ , translates into a production rate of $3.5 \cdot 10^{51}$ H\,I ionizing photons per second in the SMC \citep{Kennicutt95}.
This is under the rough assumption that all the H\,I ionizing photons are absorbed by neutral hydrogen. %it follows that about $3.5 \cdot 10^{51}$ H\,I ionizing photons per second are produced in the SMC \citep{Kennicutt95}. To get the above number, we used the relation $Q(\mathrm{H}\,\alpha) = 7.31 \cdot 10^{11} L_\mathrm{H \, \alpha} \cdot \mathrm{erg}^{-1}$ \citep{Osterbrock06}.
Given the uncertainties, this number agrees well with the number of $3 \cdot 10^{51}$ that we inferred for the observed massive star population, as discussed above.
%The result is in an estimated $2 \cdot 10^{51}$ photons per second excluding WR stars, which contribute another $10^{51}$ photons per second. This total of $3 \cdot 10^{51}$ H1 ionizing photons per second is in good agreement with the aforementioned number from the H$\alpha$ luminosity.

We also estimated an upper limit to the escape fraction of H\,I ionizing radiation, $f_\mathrm{esc}$. We defined $f_\mathrm{esc}$ as the fraction of photons that reaches our galaxy unhindered after being emitted by a star in the SMC. H\,I ionizing photons can be absorbed by dust or neutral hydrogen. We emphasize that for this upper-limit estimation of $f_\mathrm{esc}$, we only took the dust component into account. \cite{Gordon03} provided extinction ratios toward $\lambda = 91$\,nm (i.e., H\,I ionization threshold). The lowest wavelength they included is 116\,nm, where $A_\mathrm{116\,\mathrm{nm}} / A_V = 7.0$. In the following estimate, we adopt this extinction value also for $\lambda < 116$\,nm. Alternatively, we could extrapolate the \cite{Gordon03} extinction law to 91\,nm to obtain $A_\mathrm{91\,\mathrm{nm}} / A_V = 9.7$. When we subtract a Milky Way foreground extinction of $A_V = 0.18$ \citep{Yanchulova17} from the values discussed in Sect.\,\ref{sec:ex_etc}, the typical $V$-band extinction of stars in the SMC itself is $A_V \approx 0.2$. In the H\,I ionizing photon regime, the expected extinction would therefore be $A_\mathrm{91\,\mathrm{nm}} = 7.0 \cdot 0.2 \approx 1.4$. %For the H\,I ionizing radiation escape fraction $f_\mathrm{esc}$, 
This translates into $f_\mathrm{esc} = 10^{-1.4/2.5} = 0.28.$ This number would decrease when the extinction continued to decrease with $\lambda$ below 116\,nm. For the extrapolated $A_\mathrm{ \lambda \lesssim 91\,\mathrm{nm}} = 9.7 \cdot 0.2 = 1.94$, $f_\mathrm{esc}$ would be 0.16. If a significant fraction of stars is deeply embedded (Sect.\,\ref{sec:implications}), this number would also decrease.
Moreover, the H\,I absorption term would still need to be added. Earlier in this section, we showed that the H\,$\alpha$ flux of the SMC matches the H\,I ionizing photon production rate well. This implies that most of the H\,I ionizing photons are already absorbed by neutral hydrogen. Combining the above, it seems most probable that $f_\mathrm{esc}$ in the SMC is far lower than our upper limit of 0.28.

\subsection{Star formation rate \label{sec:sfr}}

Using a \cite{Kroupa01} IMF to extrapolate toward lower mass stars, we calculated the SFR of the SMC. To do this, we counted the number of stars above the 18\Msol track (Appendix\,\ref{sec:app_b}), assuming constant star formation (CSF). We find a current SFR of $\sim$0.018\Msol\,yr$^{-1}$. This is below the typical literature value of $\sim$0.05\Msol\,yr$^{-1}$ \citep{Kennicutt95, Harris04, Wilke04, Bolatto11, Hagen17, Rubele15, Rubele18}. We note that previous studies (except Kennicutt et al.), which used deep photometry, were not tailored to resolve the SFR on timescales as small as the last 10\Myr.

%\textbf{In Sect.\,\ref{sec:f_complete}, we found that the completeness of the B10 data set is about 50\% (except at the bright end, but these sources do not contribute much, number-wise). This boosts the implied SFR to $\sim$0.035\Msol\,yr$^{-1}$.Thus, we conclude that the SFR implied by the number of sources in the HRD above the 18\Msol track is, within uncertainties, in agreement with the literature value of $\sim$0.05\Msol\,yr$^{-1}$. However, in Sect.\,\ref{sec:rel_age_dist} we will find that the age distribution is heavily shifted towards stars with ages of $5-10$\Myr. In that case, the present-day SFR would be lower than one would expect from the literature SFR.

In Sect.\,\ref{sec:rel_age_dist} we consider scenarios in which the SFR is not constant. In the best-fitting scenario (where the present-day SFR is relatively low), the SFR $7-10$\Myr ago was about three times higher than for the CSF scenario. This would result in an SFR of $\sim$0.05\Msol\,yr$^{-1}$ in this period, which matches the literature values mentioned above. However, then the present-day SFR would be even farther below the literature value.

Alternatively, we could assume that the IMF in the SMC is steeper. Because our method is based on counting massive stars, a steeper IMF would result in a higher inferred SFR. Then, for CSF, to obtain an SFR of 0.05\Msol\,yr$^{-1}$ instead of 0.018\Msol\,yr$^{-1}$, an IMF exponent of $\Gamma = -1.6$ is needed (Appendix\,\ref{sec:app_b}). 
%\textbf{Then, in that period the SFR would match or even exceed the literature value of $\sim$0.05\Msol\,yr$^{-1}$. However, in this best-fitting scenario the SFR in the last few Myr would be below the literature value. %We conclude that, in general, the }

%Alternatively, we could assume that the IMF in the SMC is steeper. Because our method is based on counting massive stars, a steeper IMF would result in a higher inferred SFR. Then, for CSF, \textbf{to get an} SFR of 0.05\Msol\,yr$^{-1}$ \textbf{instead of 0.018\Msol\,yr$^{-1}$,} an IMF exponent of $\Gamma = -1.6$ \textbf{is needed} (Appendix\,\ref{sec:app_b}).

%\textbf{The SFR value becomes time-dependent if we let go of our CSF assumption. For our best-fitting star formation history (Sect.\,\ref{sec:rel_age_dist}), we find that -- before it apparently started to decrease $\sim$7\Myr ago -- the SFR was about $\sim$0.045\Msol\,yr$^{-1}$. This value is in line with the literature values.}

%We notice a few features in our HRD (Fig.\,\ref{fig:hrd_corr}). These include the large number of blue supergiants with temperatures below their TAMS temperatures, an apparent lack of young sources, and an apparent lack of bright sources. In the remainder of the section, we address these features individually.

\section{Features in the Hertzsprung-Russell diagram \label{sec:hrd_features}}

\subsection{Blue supergiants \label{sec:bsgs}}
%We notice a few features in this HRD. First of all, 
Interestingly, about 200 stars in Fig.\,\ref{fig:hrd_corr} reside in the 
region between the main sequence (MS) and the RSG branch.
%\textbf{inter-MS-RSG region}, which is the area between the terminal-age main sequence (TAMS) and the region occupied by RSGs. 
%This number, however, depends on where exactly the line of the terminal-age main sequence (TAMS) is drawn. %Also, one needs to be careful not to directly compare the number of HG stars to the number of hotter stars at the same luminosity because luminosity increases as a star evolves towards cooler temperatures.
The exact number of stars in this region depends on the location of the terminal-age main sequence (TAMS). Theoretically, this is affected by the choice for the overshooting parameter \citep{Maeder76, Vink10}, and by rotation \citep[][]{vonZeipel24}.

A significant fraction of the stars roughly $0.05 - 0.1$\,dex ($2-5$\kk for $T_\mathrm{eff, \, TAMS} = 25$\kk) to the right of the TAMS shown in Fig.\,\ref{fig:hrd_corr} %(down to $T_\mathrm{eff} \approx 10^{4.35}$\,K) %, according to their B10 spectral type, relatively often stars with emission features, in particular 
are Be stars \citep[as is also the case in fig.\,13 of][]{Ramachandran19}. We highlight stars with emission features (indicated by their B10 spectral type and/or infrared excess) in Fig.\,\ref{fig:hrd_asc10}. %It shows that at $\log L \gtrsim 5.5$, these emission features disappear for stars to the right of the TAMS. 
%The presence of emission features mainly extends to $T_\mathrm{eff} \approx 19^{4.35}$\,K}. 
Emission features make it likely that these are late-MS stars that evolved toward critical velocity in isolation \citep{Ekstrom08, Hastings20} or were aided by a binary companion \citep[e.g.,][]{Gies98, Schootemeijer18b, Wang20} rather than He-burning blue supergiants. The offset of $0.05 - 0.1$\,dex is in line with a shift in $T_\mathrm{eff}$ of 0.05\,dex at critical rotation \citep{Paxton19}, uncertainties in overshooting \citep[see][]{Schootemeijer19}, and modest errors. 
For stars in Fig.\,\ref{fig:hrd_corr} at $T_\mathrm{eff} \lesssim 10^{4.3}$\,K ($\sim$20\kk) below $\log (L / L_\odot) \approx 5.5,$ it appears more likely that they are helium burning because the emission features almost completely disappear (Fig.\,\ref{fig:hrd_asc10}) and this temperature range can be covered by post-MS tracks \citep{Schootemeijer19} %shown in Fig.\,\ref{fig:hrd_corr} 
as well as by the post-binary-interaction models of \cite{Justham14}.
We argue that it is unlikely that many H-burning stars can be found in Fig.\,\ref{fig:hrd_corr} below 20\kk and $\log (L / L_\odot) < 5.5$. The reason is that very high overshooting values of $\alpha_\mathrm{ov} = 0.55$ are necessary for the TAMS to extend that far \citep{Schootemeijer19}. Even then, we do not expect many MS stars at 20\kk because these models spend 95\% of their MS lifetime at $T_\mathrm{eff} < 25$\kk.

Single-star evolution models of massive stars without efficient semiconvective mixing \citep[e.g.,][]{Brott11} cross the inter-MS-RSG region rapidly and spend only about 0.1\% of their lifetime there.
This means that if the SMC population exclusively consisted of single stars without efficient semiconvection, only about one star would reside between the MS and the RSG branch. In contrast, our HRD shows that there are least a hundred, even when the TAMS lasts 0.1\,dex longer than the TAMS of the models shown in our HRD.
%Still, given that the evolutionary models shown in Fig.\,\ref{fig:hrd_corr} spend only $\sim$0.1\% of their lifetime in the HG,
It therefore appears to be unavoidable to invoke either internal mixing \citep{Langer91, Stothers92, Schootemeijer19} or binary interaction \citep{Braun95, Justham14} to explain their presence. %Without, this part of the HRD would be virtually empty because of the short timescales on which models cross the HG \citep[e.g.,][and references therein]{Schootemeijer19}.
%Both of these phenomena can cause stars to burn helium at temperatures above the RSG temperature regime. 
%\textit{(Concluding sentence!)}

To obtain further clues, we suggest to % follow a strategy similar to \cite{Bodensteiner20}, who looked at the prevalence of MS companions of 287 Be stars (and found none, which, they argue, indicates that the binary channel is the dominant formation channel for Be stars). In this case, the aim would be to
investigate the binary properties of the 
blue stars in Fig.\,\ref{fig:hrd_corr} with $T_\mathrm{eff} \lesssim 20$\kk and $\log( L / L_\odot) \gtrsim 5$. The detection of a close MS companion would be a strong indication that such a star is following a single-star track that moves to the RSG branch on a nuclear timescale, rather than being on a blue loop or being a merger product.
There is evidence, however, that the binary fraction of massive stars decreases toward later spectral types \citep[][Sim\'{o}n-D\'{i}az et al. subm.]{Dunstall15, Mcevoy15}.
This would therefore provide a crucial test for existing evolutionary models of low-metallicity massive stars \citep[e.g.,][]{Brott11, Georgy13, Limongi18, Schootemeijer19}.  %A close MS companion would rule out the blue loop scenario because earlier interaction with the companion would have stripped the star. A merger history would be unlikely because this would require a compact triple configuration, which is likely to be unstable (ref).

   \begin{figure}
   \centering
   \includegraphics[width = \linewidth]{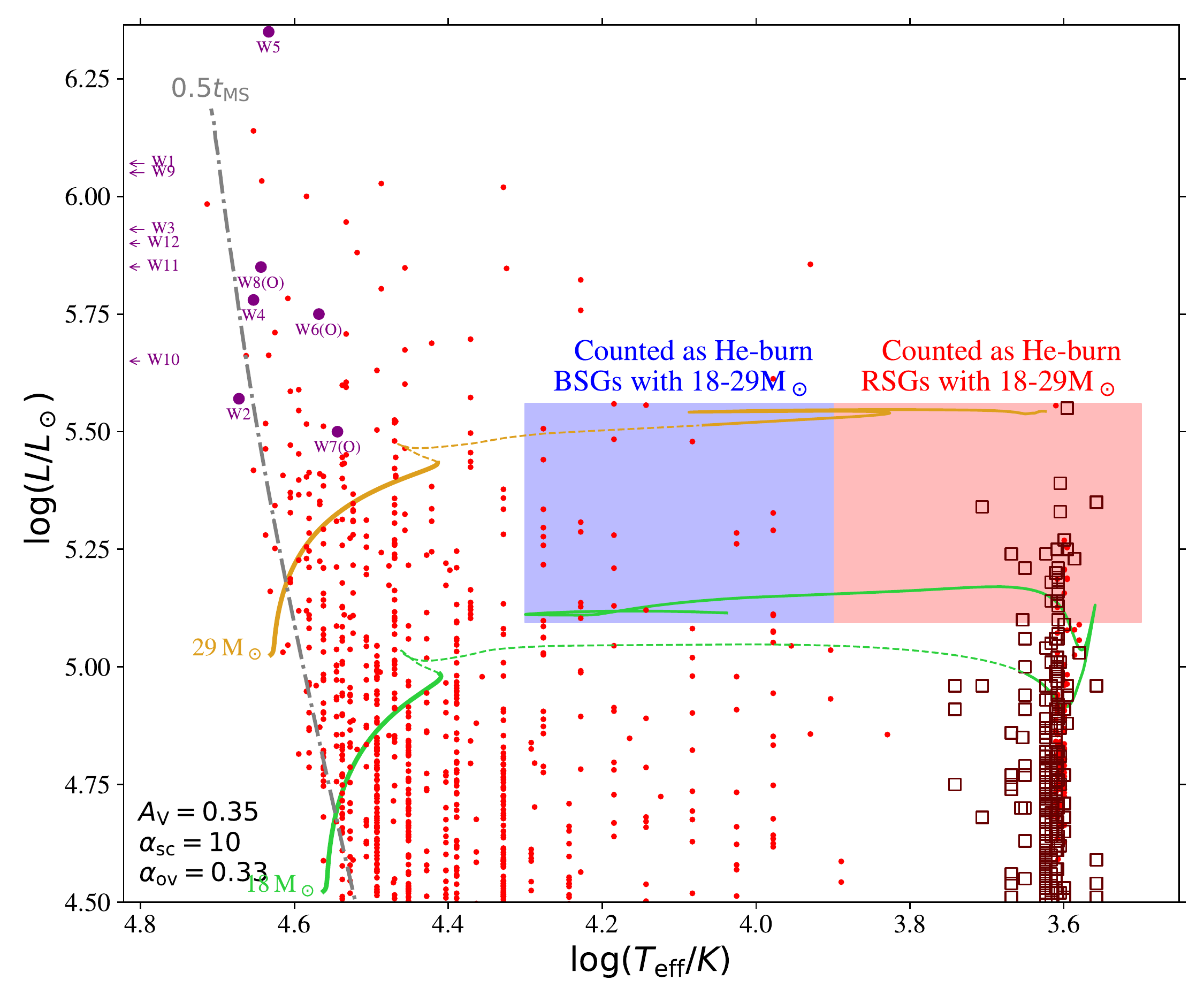}
   \caption{%Hertzsprung-Russell diagram where we indicate what sources we count as He-burning stars with initial masses between 18 and 29 \Msol. \textbf{The scatter points have the same meaning as in Fig.\,\ref{fig:hrd_corr}. The tracks are from the extended grid of \cite{Schootemeijer19}.}
   Hertzsprung-Russell diagram showing the same SMC objects as Fig.\,\ref{fig:hrd_corr}, indicating which stars are considered as He-burning stars with initial masses between 18 and 29 \Msol. The tracks are from the extended grid of \cite{Schootemeijer19}.}
   
             \label{fig:hrd_heburn_count}%
    \end{figure}
%______________________

%______________________________________________

   \begin{figure*}
   \centering
   \includegraphics[width = 0.66\linewidth]{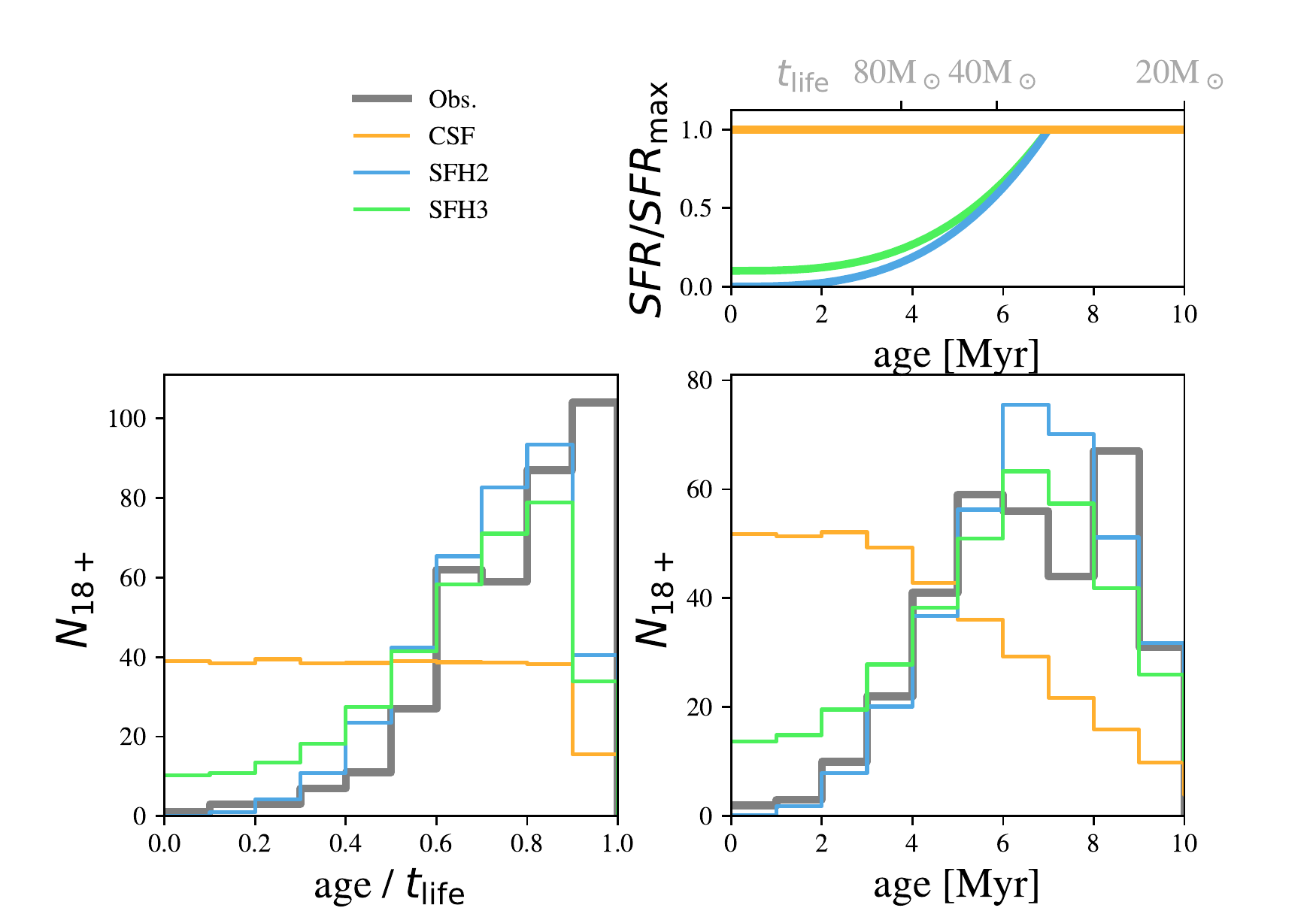}
   \caption{Fractional age (\textbf{bottom left}) and absolute age (\textbf{bottom right}) distribution of the number of stars that are above the 18\Msol track (referred to as $N_{18+}$) in Fig.\,\ref{fig:hrd_corr}. The colored distributions are theoretical predictions for CSF, and two other star formation histories (SFH2 and SFH3) where young stars have a lower probability (\textbf{top right panel}). The gray line indicates the distribution that we derived using observations.}
             \label{fig:sfh}%
    \end{figure*}
%______________________

\subsection{%Numbers of progenitors of helium-burning stars
Numbers of helium-burning stars and their progenitors \label{sec:he_burn_progenitors}}

%We count suspected helium-burning BSGs and RSGs and estimate the number of their progenitor stars. 
We counted the helium-burning stars and compared the corresponding expected number of
hydrogen-burning stars (based on stellar models) with the observed number of hydrogen-burning stars.
The lowest initial mass that we considered is 18\Msol. This is the minimum mass where the zero-age main-sequence (ZAMS) models are bright enough to be in the luminosity range of our HRD. This is shown in Fig.\,\ref{fig:hrd_heburn_count}, where we illustrate the method used in this exercise. The 18\Msol evolutionary track has a time-averaged luminosity of $\log (L/L_\odot) = 5.1$ during helium burning.
The most luminous RSG in the sample of \cite{Davies18} has $\log (L/L_\odot) = 5.55$, which corresponds to the time-averaged helium-burning luminosity of a evolutionary 29\Msol track.
The red box in Fig.\,\ref{fig:hrd_heburn_count} (with $5.1 < \log (L/L_\odot) < 5.55$) represents the temperature and luminosity range of 18 to 29\Msol RSGs, and contains observed 24 RSGs.
In Sect.\,\ref{sec:bsgs} we argued that stars in our HRD below $10^{4.3}\,\mathrm{K} \approx
20$\kk are most likely helium burning.
%In a similar way, assuming that stars with $8 < T_\mathrm{eff} / \mathrm{kK} < 20$ in this $\log L$ interval are helium-burning BSGs, we find a number of 25.
In the luminosity range described above, we count 25 blue stars below 20\kk.
%\textbf{We consider 20\kk to be a conservative value for the maximum MS temperature in this luminosity range, as we discuss in Sect.\,\ref{sec:app_a}.}
This makes a total of $24+25=49$ helium-burning stars with initial masses between 18 and 29\Msol (and a ratio of blue to red supergiants of about one).
%For helium burning stars in our model grid, this luminosity interval is covered by models of initial masses $18 < \mathrm{M}_\mathrm{ini}/\mathrm{M}_\odot < 29$ (not shown in Fig.\,\ref{fig:hrd_corr}, but existent in the denser grid).
These models have a fractional helium-burning lifetime of about $7$\% \citep[e.g.,][]{Schootemeijer19}. We therefore expect $\sim$650 progenitor stars. A total of 229 MS stars shown in Fig.\,\ref{fig:hrd_corr} lie between the 18 and 29\Msol evolutionary tracks at $T_\mathrm{eff} > 20$\kk. Corrected for a completeness of about half (Table\,\ref{tab:f_complete}), this adds up to $\sim$450 progenitor stars. This poses a modest discrepancy with the expected 650.
%; again a lower number than expected for CSF, but less tension than with the WR progenitors.
If all stars in this mass range down to 16\kk were MS stars, then only $\sim$500 progenitor stars would be required. However, we do not expect this to be the case (Sect.\,\ref{sec:bsgs}).
%Also, it is likely that we are incomplete for helium-burning blue stars, as we are . 
%Moreover, we assumed in the discussion above that B10 is 100\% complete for helium-burning blue stars.  Most likely, this is not the case. Then more progenitor stars are required, increasing the discrepancy. Thus, the presence of about two out of three (450/650) progenitor stars should be considered an upper limit.
%\textbf{Moreover, we assumed above that B10 is 100\% complete for helium-burning blue stars. Most likely, this is not the case, for example because the B10 catalog is not this complete (Sect.\,\ref{sec:f_complete}). % and because hard-to-detect stripped stars might reside in the population \citep[e.g.,][]{Gotberg18}.
On the contrary, with the already conservative value of 20\kk, we most likely exclude some number of helium-burning stars.
%If the true number is higher, 
In this case,
more helium-burning star progenitors are required, which increases the discrepancy. The presence of about two out of three (450 out of 650) helium-burning star progenitors in the $18-29$\Msol initial mass range should therefore be considered an upper limit.

We performed a similar exercise for WR stars.
These stars are thought to be complete in the SMC \citep{Neugent18}. In the 12 WR star systems, 7 are apparently single \citep{Hainich15} and 5 are binary systems \citep{Shenar16}. Nine of the 12 WR stars are so hot that only core helium-burning models can explain them \citep{Schootemeijer18}. They have inferred initial masses in excess of $\sim$40\Msol \citep{Hainich15, Shenar16, Schootemeijer18}. The fractional core helium-burning lifetime of such stars is about 7\% \citep[][]{Schootemeijer19}. This suggests the presence of at least 100 hydrogen-burning progenitors with masses in excess of $\sim$40\Msol. %Excluding these He-burning WR stars, w
We count 22 red circles above the 40\Msol track in Fig.\,\ref{fig:hrd_corr}. For an 80\% completeness (Table\,\ref{tab:f_complete}), this results in an estimated $\sim$27 progenitors, only a quarter of the expected number. %\textbf{Thus, we infer that only 27 out of 100 progenitor stars exist.} 
Even given some uncertainties in the minimum initial mass of SMC WR stars, %it seems impossible to match the expected number of WR progenitor stars. 
this discrepancy is stronger than for the blue supergiants (BSGs) and RSGs discussed above.
%of the WR stars in Fig.\,\ref{fig:hrd_corr}.

%______________________________________________

   \begin{figure*}
   \centering
   \includegraphics[width = 0.68\linewidth]{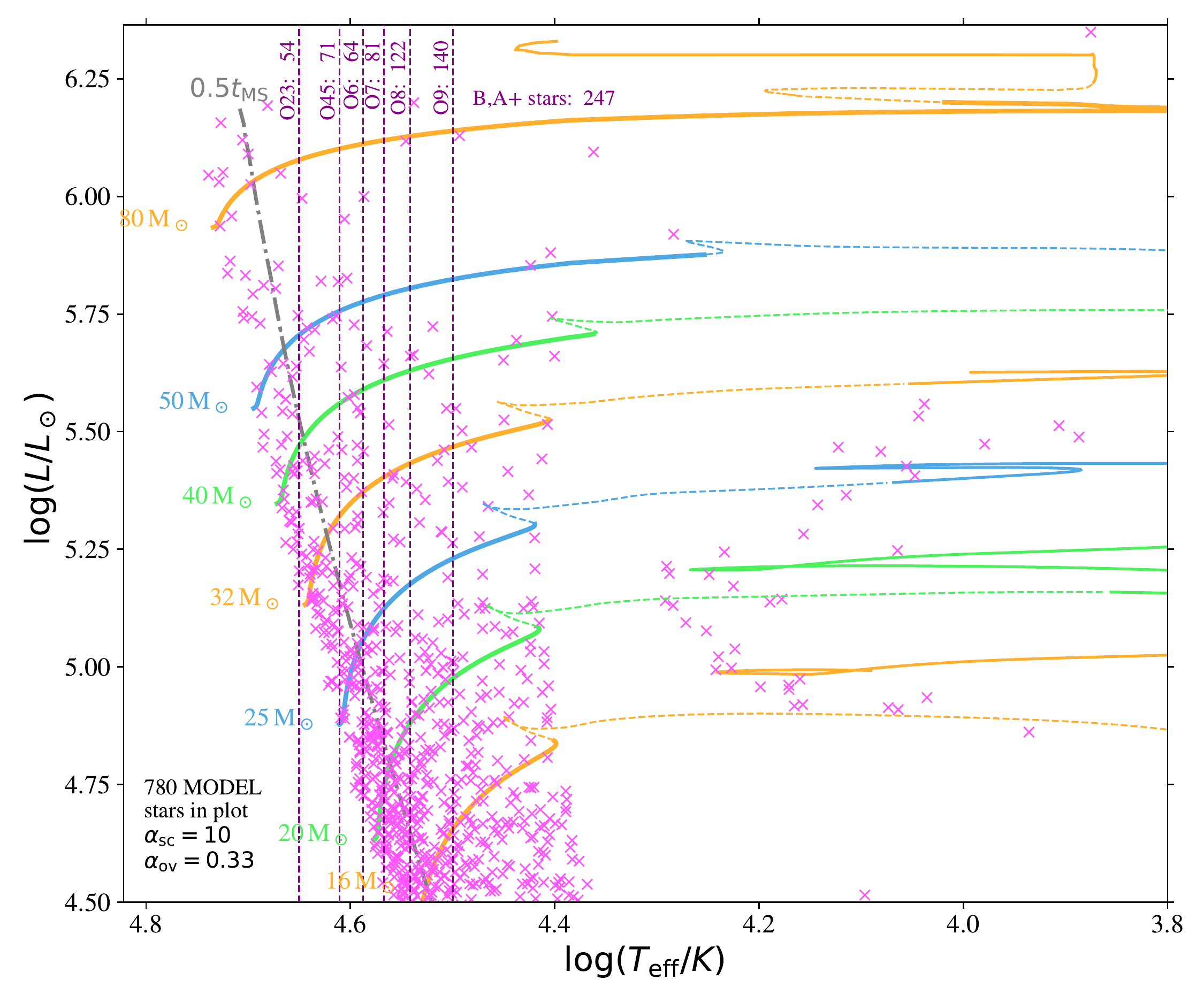}
   \caption{%Randomly drawn synthetic stars (purple crosses) in the Hertzsprung-Russell diagram. 
   Same as Fig.\,\ref{fig:hrd_corr}, except that it contains randomly drawn synthetic stars (purple crosses) and that the x-axis range does not extend to the temperature regime of red stars. The dashed lines separate the temperature ranges covered by different spectral types of luminosity class V, according to our spectral type$-$ temperature relation (Sect.\,\ref{sec:meth_lt}). 
   }
             \label{fig:hrd_synt}%
    \end{figure*}
%______________________

\subsection{Age and relative age distribution \label{sec:rel_age_dist}}
%Fig.\,\ref{fig:hrd_corr} shows only few stars that are in the first half of their MS lifetime. 
The gray dot-dashed line in Fig.\,\ref{fig:hrd_corr} denotes the location of stars that are halfway through their MS lifetime. Only a few stars lie to the left of this line.
In a scenario in which star formation is constant, we would expect roughly half of the observed sources to reside to the left of this line. To further investigate this, we obtained the age and fractional lifetime of the stars in our sample. We did this by comparing them with evolutionary models of \cite{Schootemeijer19} shown in Fig.\,\ref{fig:hrd_corr}, but using a denser grid with a spacing of 0.02\,dex in mass. %and extended to higher masses to cover the relevant luminosity range. 
The inferred age and fractional age (i.e., age divided by stellar lifetime) are those of the nearest stellar model. %We refer to these as the `observed' age and fractional age.

In Fig.\,\ref{fig:sfh} the distributions of these inferred ages (bottom right) and fractional ages (bottom left) are shown with a gray line. 
In this plot, we only show sources with $T_\mathrm{eff} > 10^{3.8}$\,K and evolutionary
masses above 18\Msol . The distribution is therefore not affected by a luminosity cutoff.
Instead of $\sim$50\% of the stars being in the first half of their MS lifetime, this number is only 7\% for the B10 sample. 
For the stars in the VSS sample, this fraction is about 20\%, which is less extreme, but still much lower than expected. 
%. Less extreme, but still a much lower number than expected. 
One explanation for this slight difference between the B10 and VSS samples is that our method could underestimate $T_\mathrm{eff}$, which would increase the inferred age. However, no such systematic trend is visible in Fig.\,\ref{fig:hrd_arr}. Another possible explanation is that there could be a stronger observational bias against hot stars (compared to cool stars) in the B10 catalog, but as we discuss in Sect.\,\ref{sec:reasons}, we do not expect this effect to be strong.
More likely, the explanation for the relatively larger fraction of young stars in the VSS sample could be that this sample is biased towards hot stars 
%are more prone to be included there 
simply because of a greater interest in them, and young massive stars tend to be hot. This is supported by Fig.\,\ref{fig:gaia_hrd}, which shows that the B10 catalog is rather complete for hot and bright stars.% are rather complete.}

We investigate in Fig.\,\ref{fig:sfh} whether a decreasing SFR can explain this heavily lopsided distribution of relative ages. From the model population, we drew $10^6$ stars with random ages between 0 and 10 Myr, and random masses between 18 and 100\Msol. Then, we weighted these model stars by the Salpeter initial mass function \citep[IMF;][]{Salpeter55} and normalized their number to the number of observed stars. Stellar models with $T_\mathrm{eff} < 10^{3.8}$\,K and models that live shorter than the drawn age were discarded. The bottom left panel of Fig.\,\ref{fig:sfh} shows that for CSF, the synthetic relative age distribution is flat% in the first 90\% of the lifetime
, as is expected, but at odds with observations. In the last bin, the number drops because RSG models are not included. 

Next, we considered a star-formation history (SFH) scenario, SFH2, that mimics the observed age distribution. There, the
%Next, we consider a scenario (SFH2) where the 
SFR is set to zero at an age of 0\Myr and then increases with age to the third power up to 7\Myr; then it stays constant (i.e., we reduce the probability to draw younger stars; Fig.\,\ref{fig:sfh}, top panel). 
%We show SFH2 because its age distribution closely resembles the inferred age distribution (bottom right panel in Fig.\,\ref{fig:sfh}). 
The resulting synthetic relative age distribution with SFH2 also fits the observed distribution well, at least up to $\mathrm{age} / t_\mathrm{life} = 0.9$.  
Finally, we considered SFH3, which is the same as SHF2 except that the SFR starts at 10\% of the final SFR. We display SFH3 to illustrate how dramatic the drop in SFR needs to be to match the observations. Even this rapid 90\% drop in SFR still strongly overpredicts the number of relatively young stars, especially at $\mathrm{age} / t_\mathrm{life} < 0.3$ and ages below $\sim$3 \Myr. 
%We conclude that a tweaked SFH can reconcile the observed relative age distribution, but only if the SFR has dropped to essentially zero in the last few Myr.
We conclude that the observed relative age distribution can be reconciled only if the SFR has dropped to essentially zero in the last few million years.
The implication of this would be that the SMC contains about 20 stars over 18\Msol in the first half of their lifetime, and more than 300 in the second half of their lifetime. In other words, assuming CSF, about 300 stars above 18\Msol in the first half of their lifetime are missing.

%__________________________________________________ One column table
\begin{table}
\centering    
    \caption[]{Number of sources of different spectral types that are shown in Fig.\,\ref{fig:hrd_corr}: $N_\mathrm{obs}$. We also show the range of the effective temperature $T_\mathrm{eff}$ covered by the spectral types; see also Table\,\ref{tab:spt_teffs}. For comparison, we display the number of stars for each spectral type in our synthetic population ($N_\mathrm{synt}$), again assuming a total of 780 stars.}
\begin{tabular}{l l | r r r }
\hline
\hline
            \noalign{\smallskip}
            Type & $T_\mathrm{eff}$ [kK] & $N_\mathrm{obs}$ & $N_\mathrm{synt}$ & $N_\mathrm{obs} / N_\mathrm{synt}$\\
            \noalign{\smallskip}
            \hline
            \noalign{\smallskip}
            O2$-$O3 & $> 44.7$ & 2 & 54 & 0.04 \\
            O4$-$O5 & $40.8 - 44.7 $ & 12 & 71 & 0.19\\ %12 in plot, low L O4
            O6 & $38.7 - 40.8 $ & 16 & 64 & 0.29\\
            O7 & $36.9 - 38.7 $ & 31 & 81 & 0.45 \\ %31 in plot
            O8 & $34.8 - 36.9 $ & 55 & 122 & 0.50 \\
            O9 & $31.6 - 34.8 $ & 116 & 140 & 0.80\\ %            %\noalign{\smallskip}
            B,A+ & $ < 31.6 $& 524 & 247 & 1.96%
            %\noalign{\smallskip}
            %\hline
            %\noalign{\smallskip}
            %Total & 723 & 1514 & 0.48%              
            %\hline
            \label{tab:type_nrs}

\end{tabular}
\end{table}
%______________________

%______________________________________________

   \begin{figure*}
   \centering
   \includegraphics[width = 0.66\linewidth]{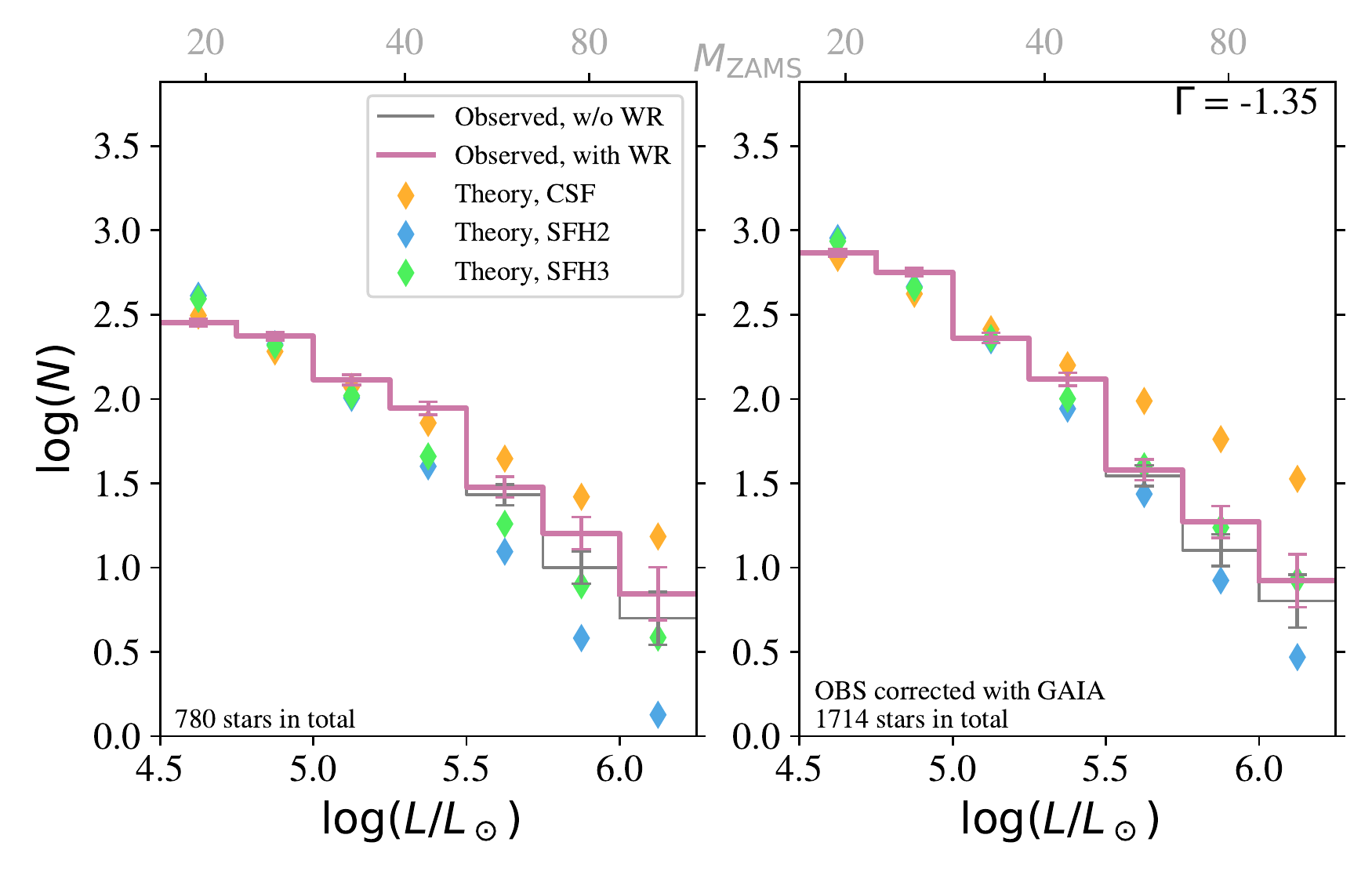}
   \caption{Diagrams showing the high end of the distribution of luminosities that we derived for sources in the \cite{Bonanos10} data set. \textbf{Left} is the uncorrected distribution. \textbf{Right} is the distribution corrected for the completeness of the B10 data set using GAIA data. For comparison, we show the theoretical values (diamonds) for different star formation history scenarios. We also show the number distributions when the brightest source in the WR star system is taken out.% \textit{Q: add RSGs? pbb easier to say: cut at logT 3.9. OR: show also dist with RSGs}
   }
             \label{fig:ldists}%
    \end{figure*}
%______________________

In the bottom right panel of Fig.\,\ref{fig:sfh} we show the corresponding absolute age distributions. 
Models that are initially more massive than $\sim$25\Msol live shorter than 10\Myr.
As a result, in the CSF scenario the shape of the age distribution is shifted toward younger ages, in stark contrast to the distribution that is observed. Again, the observed distribution matches well for SFH2, while in the SFH3 scenario the number of young stars is higher than what is inferred from observations.

Most of the time, massive stars become cooler as they evolve. Therefore our inferred lack of young stars goes hand in hand with a lack of stars with very early spectral types. This is quantified in Table\,\ref{tab:type_nrs}. The numbers that we derive are based on the amount of time spent by our evolutionary models in the different temperature ranges, weighted by the Salpeter IMF.
We show a synthetic population in Fig.\,\ref{fig:hrd_synt}. There we have drawn 780 stellar models with a probability that is based on how long-lived they are and on the IMF. The upper mass limit of the models is 100\Msol. The discrepancy is quite extreme: for example, we expect 110 stars of type O5 or earlier, but only 14 are in our sample.

\subsection{Luminosity distribution %of the brightest stars in the SMC 
\label{sec:ldist}}

%In Fig.\,\ref{fig:ldists} we show the luminosity distribution that we derived. 
The left panel in Fig.\,\ref{fig:ldists} shows the luminosity distribution of the stars shown in our HRD (referred to as `observational'; Fig.\,\ref{fig:hrd_corr}).
We consider the 780 sources from the B10 data set with $T_\mathrm{eff} < 10^{3.8}$\,K and $\log (L/L_\odot) >4.5$. %, so the visually brightest components in the WR systems are included.
%In total, this are $780 + 12 = 792$ sources.
To show that the WR sources affect the luminosity distribution to a limited extent, we also show it including the visually brightest components of the WR systems.
%We wish to avoid making a distinction between MS and post-MS stars, because the numbers would be strongly affected by both the chosen value of overshooting in the models \citep[e.g.,][]{Schootemeijer19} and small errors in temperature around the ZAMS. Thus, we also include the WR stars \citep{Hainich15, Shenar16}. In case the WR star is in a binary, we choose to show the star that is brightest in the $V$ band. 
To obtain the error bars, we randomly scattered the luminosities of all individual sources by a value drawn from a Gaussian distribution with $\sigma = 0.13$\,dex (i.e., the error on the luminosity that we derived in Sect.\,\ref{sec:meth_lt}). We did this 250 times. The shown error bars represent the standard deviation of the number of stars in each bin after scattering.

We compared this with theoretical luminosity distributions %.With orange diamonds, we show how the luminosities are distributed in our synthetic population 
\citep[models of][as in Sect.\,\ref{sec:rel_age_dist}]{Schootemeijer19} for the same total number of stars as quoted above (780).
When we computed this distribution, we assumed a Salpeter IMF while also taking the model lifetimes (where $T_\mathrm{eff} > 10^{3.8}$\,K) %, as in our HRD, Fig.\,\ref{fig:hrd_corr}) 
into account. To cover the $\log (L/L_\odot) = 6$ to 6.25 bin with ZAMS models as well, we extended the initial mass range to 126\Msol.
The observed population has fewer sources at the luminous end than the theoretical CSF populations with and without WR systems. At $\log (L/L_\odot) > 5.5$, the difference is about $0.3$\,dex, which is a factor two.
When we consider different SFHs, SFH2 (for which the SFR is essentially zero in the last few million years) on the other hand predicts fewer sources with $\log (L/L_\odot) > 5.5$ than are observed. SFR3 provides the best fit to the distribution that is not corrected for completeness, with only a slightly smaller difference to the observed distribution than for the CSF scenario. 
%We consider the cases where stars tend to burn helium as blue supergiants (with efficient semiconvection: $\alpha_\mathrm{sc} = 100$) and where they burn helium as red supergiants ($\alpha_\mathrm{sc} = 1$). The latter case is outside of the temperature range that we consider. We find that the choice for $\alpha_\mathrm{sc}$ has little effect on the theoretical $\log L$ distribution. 

%Without correction, there is a sudden steep drop in the number of very massive stars with $\log L > 5.5$, but in general the observed and theoretical $\log L$ distributions show no very significant differences.
We show the observed luminosity distributions corrected for completeness in the right panel of Fig.\,\ref{fig:ldists}. We deem these to be the most realistic. To obtain the values, the number of sources in the luminosity bins in the left panel was divided by their completeness fraction listed the right column of Table\,\ref{tab:f_complete}. The corrected distribution is steeper because the completeness correction is larger at lower luminosities. The theoretical distributions in this panel are again scaled to the total number of observed stars, which is 1714 after completeness correction. Then, the theoretical SFH3 distribution matches the observed distribution best. %, although it now slightly over-predicts the number of bright stars. 
%For the CSF distribution on the other hand, the numbers in the observed distribution\textbf{s} are lower by $\sim 0.6$\,dex (i.e., there are \textbf{about four} times less sources above $\log (L/L_\odot) = 5.5$).
The theoretical CSF distribution contains many more bright stars than the observed distributions. When observed WR stars are included, the difference is $\sim$0.6\,dex, or a factor four.

%on the other hand, the numbers in the observed distribution\textbf{s} are \textbf{much} lower. \textbf{Including WR stars,} by $\sim 0.6$\,dex (i.e., there are \textbf{about four} times less sources above $\log (L/L_\odot) = 5.5$).

%This changes when we apply the completeness correction. Given that there should then be 1320 stars with $\log L > 4.5$, the theoretical distribution predicts that 145 of them should have $\log L > 5.5$. The corrected observed number of 54, however, is significantly lower. 

%Thus, not only young stars (Sect.\,\ref{sec:rel_age_dist}) but also bright stars appear to be more rare than expected in the SMC -- at least when star formation has not come to a standstill in the last few Myr in the entire SMC.
This means that not only young stars (Sect.\,\ref{sec:rel_age_dist}), but also bright stars appear to be rarer than expected in the SMC if its star formation were constant.
We discuss possible reasons for this in Sect.\,\ref{sec:reasons}.
%Under the assumption of constant star formation, we find for $dN / dM \propto M^x$ a best fit of $x = -3.1 \pm 0.3$ (for the Salpeter IMF, $x = -2.35$) \textit{IMFs.. do we go there or not?}.
%Thus, the differences become significant as very massive stars are about three times less abundant than expected above $\log L > 5.5$
%In both cases, very massive stars with $\log L > 5.5$ in the observed population are under-abundant. Without correction, these objects are observed about 1.5 to 2 times less often than expected. However, after correction this under-abundance becomes as large as  a factor 3, because the B10 catalog is less complete for dimmer stars.

%\textit{``Small differences remain between the expected number of the most massive stars and the observed number. These differences could be explained by, for example, local star formation rate fluctuations. To illustrate: \cite{Ramachandran19} derive in a part of the SMC wing that the star formation rate five Myr ago -- the lifetime of a typical WR star progenitor of 40\Msol -- was a factor of four higher than it is now. ''}

We also varied the IMF to see where we achieve the best fit of the theoretical luminosity distribution (assuming CSF). The uncorrected luminosity distribution is fit best for $\Gamma \approx -1.9$. For the luminosity distribution that is corrected for completeness, $\Gamma \approx -2.3$ fits best. When we calculate the extinction individually per star (as we do in the HRD shown in Fig.\,\ref{fig:hrd_varAV}), these numbers become $\sim$ -1.7 and 2.1, respectively.) This means that in principle, we can resolve the lack of bright stars by assuming a steeper IMF. However, this does not resolve the problem with the dearth of young stars.

\section{Number comparisons \label{sec:numbers}}%the LORD bless you and keep you [numbers, 6:24]

%\textbf{In this section we summarize how the number of stars in our HRD (Fig.\,\ref{fig:hrd_corr}) matches the number of helium burning progenitors, ionizing photon production and SFR. We also compare with HRDs from the literature.}

%In this section we briefly summarize number comparisons from this work regarding the massive star population in the SMC. We also compare with the number of stars in HRDs from the literature.

In this section we %compare with 
discuss the number of massive SMC stars %in HRDs 
%from the literature. %regarding the massive star population in the SMC. 
obtained by previous studies. We note that we count only one star in a binary (or higher order multiple) because of the lack of completeness of the information on binarity, and the biases therein.
We compare the derived numbers with our results, which we also summarize here.

We first discuss the studies of the SMC massive star population that have been performed before \citep{Humphreys84, Blaha89, Massey95}. We note that these studies lacked stars above $V=15$, which is the ZAMS magnitude of a 40\Msol star in the SMC. %Nevertheless, the number of very massive stars that we find is comparable to what is found in these studies. 
Above the 40\Msol track, to the right of the ZAMS, we find similar numbers of stars as these studies.
When stars in WR systems are excluded and completeness is not corrected for, Fig.\,\ref{fig:hrd_corr} contains 22 of such massive stars (Sect.\,\ref{sec:he_burn_progenitors}). %(27 including WR systems). %When we calculate the extinction individually per source (Fig.\,\ref{fig:hrd_varAV}), this becomes 39 (44 with WR systems). 
The number of SMC stars above 40\Msol is 21 according to \cite{Blaha89} and 32 according to \cite{Massey95}.

%We can compare this number of 22 observed stars above 40\Msol to the number of randomly drawn stars in Fig.\,\ref{fig:hrd_synt}. For the same total number of stars in the HRD, we draw 72 stars with $M>40$\Msol, which is a factor 3.3 higher. 
Based on the Salpeter IMF, model lifetimes, random ages, and an upper mass limit of 100\Msol, we expect 75.6 model stars out of the 780 shown in Fig.\,\ref{fig:hrd_synt} to lie above the 40\Msol track. We randomly drew 72. Compared to the 22 stars in the B10 data set above the 40\Msol track, this amounts to a difference of a factor $75.6/22 = 3.4$. 
This is in line with the $0.5$\,dex difference between the uncorrected theoretical CSF luminosity distribution and the observed distribution without WR stars above $\log (L / L_\odot) = 5.5$ in the left panel of Fig.\,\ref{fig:ldists}. %\footnote{The difference in the $5.5 < \log (L / L_\odot) < 5.75$ bin is smaller, but many of the stars in this bin are located below the 40\Msol track.}. 
The difference in the $5.5 < \log (L / L_\odot) < 5.75$ bin is smaller, but we note that many of the stars in this bin are located below the 40\Msol track.
This discrepancy can be expected to be slightly larger after correcting for completeness, even after adding the WR stars (cf. the  left and right panels in Fig.\,\ref{fig:ldists}).
%Correcting for $\sim$80\% completeness (Table\.\ref{tab:f_complete}) and adding twelve stars in WR systems would boost the total number of stars above 40\Msol to $\sim$40. 

%\textbf{We compare the number of 22 stars above 40\Msol -- out of the total of 780 stars with $\log (L / L_\odot) > 4.5$ and $T_\mathrm{eff} > 10^{3.8}$\,K in Fig.\,\ref{fig:hrd_corr} -- to the 780 randomly drawn stars in Fig.\,\ref{fig:hrd_synt}. There, we have 87 stars above 40\Msol. We note that this factor four ($\sim$0.6\,dex) difference is in line with the difference between the observed luminosity distribution without WR stars and the theoretical CSF luminosity distribution in the $\log (L / L_\odot) \geq 5.75$ bins shown in the left panel of Fig.\,\ref{fig:ldists}. Adding the WR stars to the number of observed stars above 40\Msol would reduce the difference to about a factor three. On the other hand, accounting for completeness would increase the difference again, as happens in the right panel of Fig.\,\ref{fig:ldists}. }

We can also estimate how many of these very massive stars with $M > 40$ \Msol ($N_\mathrm{40+}$) are expected when a constant (typical) SFR of $\sim$0.05\Msol\,yr$^{-1}$ is assumed. %Assuming an upper mass limit of 150\Msol and integrating the IMF. 
Repeating the exercise described in Appendix \,\ref{sec:app_b} for $M > 40$\Msol yields a fraction of $f_\mathrm{40+} = 0.075$ of the stellar mass budget that is tied in stars with $M > 40$\Msol. Moreover, the IMF dictates that the average mass of stars of higher mass than 40\Msol has a value of $\overline{M}_\mathrm{40+} \approx 65$\Msol (again assuming an upper mass limit of 150\Msol). Such massive models have an average lifetime of $\overline{\tau}_\mathrm{40+} \approx 4.5$\Myr \citep[e.g.,][]{Schootemeijer19}. We have approximate values for the rate at which gas is converted into $M > 40$\Msol stars, and for their typical mass and lifetime.
We combined these for our estimate and filled in the values:
$N_\mathrm{40+} = (\mathrm{SFR} \cdot f_\mathrm{40+} \cdot \overline{\tau}_\mathrm{40+}) /  \overline{M}_\mathrm{40+} \approx 250$. This number is about a factor $6-7$ higher than the total number of $\sim$40 stars with initial masses above 40\Msol that we find in the SMC.
%is only of the order 40 (27 when corrected for completeness  -- Sect.\,\ref{sec:he_burn_progenitors} -- and then adding all twelve WR stars, which are predominantly helium burning).
%\textit{(maybe it is worthwhile to make a general relation between SFR and N stars above X Msol, taking a brief look at that now)}

%\textbf{Next, we summarize our findings from this work.} 
In Sect.\,\ref{sec:ionizing_radiation} we estimated that the H\,I ionizing photon production rate of the observed massive star population is about $3 \cdot 10^{51}$ photons per second. This agrees well with the rate of $3.5 \cdot 10^{51}$ photons per second that is inferred from the H\,$\alpha$ luminosity of the SMC.
%\textbf{In Sect.\,\ref{sec:sfr}, we derived an SFR for the SMC of $\sim$0.035\Msol\,yr$^{-1}$, which is in reasonable agreement with the typical literature value of 0.05\Msol\,yr$^{-1}$}
In Sect.\,\ref{sec:sfr} we found that the literature SFR value for the SMC ($\sim$0.05\Msol\,yr$^{-1}$) does not match the value we derived assuming CSF ($\sim$0.018\Msol\,yr$^{-1}$). However, the value did match for $7-10$\Myr ago if young stars are apparently rare. %Alternatively, a steeper IMF exponent of $\Gamma = -1.66$ would bring the SFR in agreement with the literature value.

%\textbf{As discussed in Sect.\,\ref{sec:sfr},} the typical literature value for the recent SFR in the SMC is $\sim$0.05\Msol\,yr$^{-1}$. Assuming CSF, we derived a current SFR of 0.015\Msol\,yr$^{-1}$. \textbf{However, according to our analysis} there are only few young stars in the sample. As a result, the inferred SFR $7-10$\Myr ago is higher: about 0.045\Msol\,yr$^{-1}$. This value does match well to the literature value. %\textbf{Alternatively, a steeper IMF with $\Gamma = -1.66$ would also reproduce the literature star formation rate (Appendix\,\ref{sec:app_b}).}

In Sect.\,\ref{sec:he_burn_progenitors} we showed that WR star progenitors are underabundant, even after correcting for completeness. We estimated that only 27 out of $\sim$100 of these objects are detectable in the SMC. At lower luminosity, we found the same discrepancy, although the difference was smaller. For helium-burning BSGs and RSGs with $5.1 < \log (L/L_\odot) < 5.55$, we inferred that at most two out of three progenitors are present.

%\textit{Norbert; answer to your question: 7 He burning WRs (Mini = 40\Msol) translates into $\sim$0.02\Msol/yr. This number might be higher because there might be more He-burning objects (e.g. B/RSGs) above $\log L = 5.6$. Also, for WRs this number depends more strongly on the choice for IMF slope and upper mass limit (now 150\Msol) than for the case with Mini = 18\Msol. }

%This can be compared with literature values of the current SFR that tend to be $\sim$$5\cdot 10^{-2}$\Msol\,yr$^{-1}$ \citep{Kennicutt95, Rubele15, Rubele18} or even higher \citep{Harris04}. We note that these studies, which used deep photometry, were not tailored for resolving the SFR as recent as the last 10\Myr. For the CSF scenario, we derive a lower SFR of $1.5 \cdot 10^{-2}$\Msol\,yr$^{-1}$ (Appendix\,\ref{sec:app_b}). If a steeper IMF with exponent $\Gamma = -1.6$ is assumed instead of -1.3, stars with $M>18$\Msol are more rare -- enough to boost the SFR to the literature values of $5 \cdot 10^{-2}$\Msol\,yr$^{-1}$. Instead of a steeper IMF, we can also consider non-CSF with scenario SFH2. There, the SFR would have increased from zero at present to an SFR about three times the CSF SFR at $7-10$\Myr ago. Thus, our estimate of the SFR increases to $4.5 \cdot 10^{-2}$\Msol\,yr$^{-1}$ at $7-10$\Myr ago, which is also in better agreement with literature. This explanation would also work if younger stars are hidden from view, instead of no longer being formed.

In conclusion, the numbers that we find appear to agree with literature studies on the massive star population in the SMC. 
However, the number of observed of stars with $M > 40$\Msol is much lower than what would naively be estimated from both the SFR of the SMC (by a factor $6-7$) and the total number of stars with $\log (L / L_\odot) > 4.5$ (by a factor $3-4$). The
number of helium-burning star progenitors is also significantly lower than we would expect a priori. 
%The latter
These underabundances are in line with a bias against young stars, or with them being absent due to a plummeting SFR (Sect.\,\ref{sec:rel_age_dist}).

%\subsection{Initial mass function or star formation history?}
\section{Discussion \label{sec:discussion}}
\subsection{Possible explanations for a dearth of young and bright stars \label{sec:reasons}}
Above, we have identified two types of stars that are more rare than would naively be expected: those that are in the first half of their MS lifetime (Sect.\,\ref{sec:rel_age_dist}), and those that are brighter than $\log (L/L_\odot) = 5.5$ ($M \gtrsim 40$\Msol; Sect.\,\ref{sec:ldist}). Below, we discuss the likelihood of several potential explanations. %: i) a steeper IMF in the SMC, ii) a drop in the SFR in the last few Myr, iii) embedding in birth clouds, and iv) observational biases.

\subsubsection{Steeper initial mass function}

In principle, a steeper IMF could help to explain the lack of bright stars.
%If massive, bright stars are born relatively rarely in the SMC, this could cause their low number to be low.
An exponent of $\Gamma = -2.3$ would be required to explain the completeness-corrected luminosity distribution that we show in the right panel of Fig.\,\ref{fig:ldists}, under the assumption of CSF. This is much steeper than the Salpeter exponent of $\Gamma = -1.35$.
A steeper universal IMF would have dramatic implications for the early universe because it is commonly thought that toward lower metallicity, massive stars become more common \citep{Larson71, Schneider18}. However, this is currently an unsatisfying explanation because it cannot resolve the apparent lack of young stars. %Also, such an exceptional claim should be based on detailed atmosphere analyses of a complete or at least homogeneous sample of stars.
An explanation that would have to be ruled out is a bias against young stars (Sect.\,\ref{sec:embedding}): because very massive stars (which live only for a short time) tend to be young, such a bias would make the IMF appear steeper than it really is.
%(because bright stars have short lives; we discuss this in Sect.\,\ref{sec:embedding}).
Moreover, the SFR in the SMC could be low enough for stochasticity to play a role \citep[see][]{DaSilva12}.

%\subsubsection{Homogeneous evolution above 32 Solar masses}

\subsubsection{Model uncertainties \label{sec:model_disc}}

In principle, the stellar models of \cite{Schootemeijer19} that we used could overpredict stellar temperatures and therefore the inferred stellar ages. First, an aspect that we did not consider when we derived the age distribution is rotation. Rotation can reduce the effective temperatures of stars. 
We performed a small experiment where we shifted the gray line in shown in Fig.\,\ref{fig:hrd_corr} (where stars are halfway through their MS lifetime). Our finding is that in order to have half of the sources that are shown in Fig.\,\ref{fig:sfh} (i.e., those above the 18\Msol track) on the hot side of this line, the line needs to be shifted to lower temperatures by 0.09\,dex. This is significantly larger than the shift of 0.05\,dex for stars at critical rotation \citep{Paxton19}. Furthermore, most stars in the SMC are not close to critical rotation, but typically at most at half critical \citep[e.g.,][]{Mokiem06, Ramachandran19}. In this case, the shift in effective temperatures would be about $\sim$0.01-0.02\,dex or less (see Sect.\,\ref{sec:bsgs}), which is far below the required 0.09\,dex.
In the most extreme cases, rapid rotation could induce mixing strong enough for chemically homogeneous evolution to take place \citep[e.g.,][]{Maeder87, Langer92, Brott11, Aguilera18}. Then, stars maintain high effective temperatures and stay close to the ZAMS during hydrogen burning. This means that chemically homogeneous evolution would not help either to solve the problem of too many stars being observed with relatively low temperatures. 

As a second possibility, envelope inflation could reduce stellar effective temperatures when stars approach the Eddington limit \citep{Sanyal15}. However, this is expected to occur only at luminosities of roughly $\log(L/L_\odot) \gtrsim 5.75$. Moreover, envelope inflation is less likely to occur in metal-deficient environments such as the SMC \citep{Sanyal17}. Combined, this means that at least the vast majority of the stars in our analysis should be unaffected by envelope inflation. We conclude that rotation and envelope inflation are unlikely to affect our conclusions about the dearth of young stars. Moreover, these effects would not help to resolve the issue of missing bright stars.

Alternatively, \cite{Ramachandran19} have suggested a transition at 32\Msol in the SMC, with stars below this initial mass experiencing `normal' evolution, and heavier stars evolving chemically homogeneously. %This then explains why the HRD directly above the 32\Msol track is virtually empty -- cf. their fig.\,A.1 -- since these stars have ZAMS temperatures or hotter and manifest themselves as WR stars. 
This is meant to explain why the distribution of their sample stars in the HRD avoids cool temperatures above the 32\Msol track (cf. their fig.\,A.1) because homogeneously evolving stars have ZAMS temperatures or hotter and manifest themselves as WR stars.
We note that in our Fig.\,\ref{fig:hrd_arr} the stars from \cite{Mokiem06} with an `AzV' label (that reside in the high-luminosity region between ZAMS and TAMS) are included, which is not the case in their figure. Moreover, the HRDs of \cite{Blaha89} and \cite{Massey95} contain several tens of stars to the right of the ZAMS that are above a 32\Msol SMC track. The B10 sources that we analyzed also populate this region in the HRD (Fig.\,\ref{fig:hrd_corr}).

In addition to this, it has been shown that the eight out of nine WR stars that reside to the left of the ZAMS cannot be chemically homogeneous. %\citep{Schootemeijer18}. 
The reason is that they show a significant amount of hydrogen at their surfaces, which means that if they were chemically homogeneous, they would burn hydrogen in their cores. However, they are far hotter than core hydrogen-burning models of \cite{Brott11}, for example. 
\cite{Schootemeijer18} have shown in a model-independent way that even hydrogen-poor chemically homogeneous SMC models in the WR star luminosity range burn hydrogen at $T_\mathrm{eff} \approx 50-60$\kk (their fig.\,2), whereas these eight WR stars have an average effective temperature of 93\kk \citep{Hainich15, Shenar16, Shenar18}.
%Synthetic chemically homogeneous models (i.e., models that do not depend on assumptions such rotational mixing efficiency) of \cite{Schootemeijer18} have shown that this is model independent . All their hydrogen-burning SMC models are cooler than these eight WR stars (their fig.\,2).}
Together with the absence of a void above the 32\Msol track in our (and others') HRD, this appears to rule out the scenario with homogeneous evolution for all stars above 32\Msol at SMC metallicity.

%\paragraph{SFH; a sudden halt in star formation:}
\subsubsection{Star formation history \label{sec:sfh_disc}}
At first sight, an SFH with a close-to-zero SFR in the last few million years is a tempting explanation. We have shown that it can explain both the lack of young and the lack of bright stars.
However, it cannot do this simultaneously. SFH2 fits the age distribution and SFH3 fits the luminosity distribution, but neither fits both (Sect.\,\ref{sec:hrd_features}).

We further discuss the decreasing SFR scenario in light of the SMC, LMC, and Milky Way environments.
%Also, two more problems emerge (Sect.\,\ref{sec:rel_age_dist}). First, it is questionable that star formation can have essentially stopped on the timescale of a few Myr, throughout the entire SMC. Second, a similar lack of young bright stars has been reported in the two other local, but separated, star-forming environments: the Milky Way and the LMC. This implies that a mechanism \textbf{other than a halt in star formation} is at work. %The fact that at least several hundreds of massive young stellar objects \citep[YSOs;][]{Bolatto07, Oliveira13, Sewilo13, Ward17} as well as (ultra-)compact H\,II regions (see paragraph below) have been reported in the SMC further challenges the SFR-went-to-virtually-zero explanation. \textbf{The fact that at least several hundreds of massive young stellar objects (YSOs) as well as (ultra-)compact H\,II regions (see Sect.\,\ref{sec:embedding}) are present in the SMC further challenges this explanation.}
From an analysis of the positions and proper motions of the bright sources (Appendix\,\ref{sec:app_c}) in the SMC, we conclude that they cannot be traced back to one or a few production sites on the timescale of a few million years. If the SFH is the explanation for the inferred lack of young stars, the SFR accordingly needs to have been quenched throughout the entire SMC at the same time. It might be questionable whether this scenario is realistic.
On the other hand, there is some evidence that the SFR in the SMC has changed on timescales of some tens of million years. From the high number of high-mass X-ray binaries in the SMC \citep[about 150 --][]{Haberl16}, it has been inferred that the SFR in the SMC peaked $\sim$40\Myr ago \citep{Antoniou10, Antoniou19}. \cite{Rubele15} also inferred a recent peak in SFR, $\sim$20\Myr ago. However, no such peaks are present in their more recent work \citep[][cf. their fig.\,11]{Rubele18}. Moreover, \cite{Harris04} did not show such distinct peaks.
%In the wing of the SMC, \cite{Fulmer20} report evidence for an increased SFR 25-40\Myr ago. 
However, in none of SFHs that are presented in these studies, the SFR drops by over 90\% in just a few million years, as would be required for the SFH scenario to explain the dearth of young and bright stars.

A dearth of young massive stars has been seen in different environments. 
%In the Milky Way, stars with evolutionary masses above $\sim$25\Msol close to the main sequence are non-existent in various samples \citep[][their figs.\,1 and 3, respectively]{Castro14, Holgado18}. As we discuss in Appendix\,\ref{sec:app_a} (see Fig\,\ref{fig:hrd_wvoids}), their HRDs contain a void that spans nearly the entire temperature range of $M \gtrsim 25$\Msol stars in the first half of their MS lifetime. Even if these samples of $\sim$100 stars are suffering from biases, finding zero in the first half of the MS lifetime seems highly significant. 
In the Milky Way, stars with evolutionary masses above $\sim$25\Msol close to the MS are nearly nonexistent in various samples \citep[e.g.,][]{Castro14, Holgado18}. Recently, it has been shown that this feature persists in a large and unbiased sample of 285 O-type stars \citep{Holgado20}.
As we discuss in Appendix\,\ref{sec:app_a} (see Fig\,\ref{fig:hrd_wvoids}), the HRDs shown in these studies contain a void that spans nearly the entire temperature range of $M \gtrsim 25$\Msol stars in the first half of their MS lifetime. %Even if these samples of $\sim$100 stars are suffering from biases, finding zero in the first half of the MS lifetime seems highly significant. 
In the 30 Doradus starburst region in the LMC, the reported number density of massive stars that have just been born is three times lower than at ages of $2-4$\Myr \citep[][their fig.\,3]{SchneiderOscar18}. In SMC-SGS1, a supergiant shell subregion in the wing of the SMC \citep[see, e.g.,][]{Fulmer20}, it is inferred that recently born massive stars are about five times more rare than those with an age of $\sim$8\Myr \citep[][their fig.\,17]{Ramachandran19}. %It seems highly suspicious that the same trend in SFH has to be invoked in three different environments.
%We will get back to this in Sect.\,\ref{sec:reasons}.

To summarize: it seems questionable that recently, star formation has essentially stopped on the timescale of a few million years throughout the entire SMC. 
It is even more unlikely that something similar happened in the two other local, but separated, star-forming environments: the LMC and the Milky Way.
In addition, none of the considered star formation scenarios can explain the luminosity and age distributions simultaneously.
The fact that at least several hundred massive young stellar objects (YSOs) as well as (ultra-)compact H\,II regions (see Sect.\,\ref{sec:embedding}) are present in the SMC might further challenge this decreasing SFR scenario as the explanation for a dearth of young and bright stars.

%First, it is questionable that star formation can have essentially stopped on the timescale of a few Myr, throughout the entire SMC. Second, a similar lack of young bright stars has been reported in the two other local, but separated, star-forming environments: the Milky Way and the LMC. This implies that a mechanism \textbf{other than a halt in star formation} is at work

%\paragraph{Observational biases:}
\subsubsection{Observational biases \label{sec:obs_biases}}
Related to the `embedding' scenario described below in Sect.\,\ref{sec:embedding}, observational biases might be at play.
Young massive H-burning stars are hot. As a result, they are relatively dim in the optical.
This could cause a bias in the observed sample. However, the bulk of the considered sources comes from observational studies where a magnitude cut of about 17 was used. The 2dF survey (which comprises the vast majority of B10 sources) uses $B \lesssim 17.5$ \citep{Evans04}. \cite{Ramachandran19}, who adopted $V<17$, showed that at the ZAMS their cut excludes $M < 10$\Msol stars (their fig.\,13).
This shows that such magnitude cuts are not expected to exclude any stars that would be shown in Fig.\,\ref{fig:hrd_corr}.
%(unless they would be at a temperature of at least 0.1\,dex to the left of the ZAMS). 
%A potential issue discussed in the 2dF paper is that the brightest stars are missing because they would be flagged as non-stellar based on their photometric properties \citep{Evans04}. However, the investigators were aware of this \textit{(Chris: is this ok? From your mail I understood the idea is ok, but maybe we can strengthen the point by rephrasing?)}; also, we found the bright stars to be the most complete, as one would naively expect.
Still, a potential issue is that a mixture of selection effects influenced the final 2dF sample (e.g., crowding, color cut, possible saturation of the very brightest stars, and the spatial extent of both the input astrometry and spectroscopic observations). We note that sources precluded by the 2dF survey could still be in the B10 catalog. Furthermore, we show in Sect.\,\ref{sec:f_complete} and Appendix.\,\ref{sec:fcompl_tests} that the B10 sample is complete enough for biases not to affect our conclusions about the dearth of young and bright stars.

Another potential observational bias is crowding in dense clusters. At the distance of the Magellanic Clouds, crowding can hinder an accurate analysis of the residing stars \citep{Evans11}. %Then,the cause dearth of young (and thus bright) stars could be that they are missed because they are in unresolved clusters. 
This could cause a dearth of bright young stars because they may prefer to
live in unresolved clusters.
If that is the case, these unresolved clusters should be detected as bright blue sources. 
The most luminous sources in the SMC according to GAIA photometry are shown in the top panel of Fig.\,\ref{fig:gaia_hrd}. %At least 19 from the 20 with the highest $L_\mathrm{phot}$ are consistent with being of stellar origin
The ten brightest of these have well-determined spectral types (Sect.\,\ref{sec:f_complete}, last paragraph). There is no indication that they are unresolved bright clusters. Moreover, these sources are not bright enough to contain a large number of bright stars. %\textit{Two questions: i) could bright clusters be rejected from the GAIA catalog? and ii) if so, are there other studies that could tell us if these unresolved clusters exist or not?}
From this, we conclude that the unresolved clusters required to explain the dearth of young stars are not in the GAIA catalog. 

Next, we visually inspected UV images from \cite{Cornett97}, taken with the Ultraviolet Imaging Telescope (their fig.\,1), and \cite{Hagen17}, taken with the SWIFT telescope (their figs.\,2 and 3). The dominant UV sources in these papers are the clusters NGC\,330 and NGC\,346. They have recently been resolved and do not contain many, if any, early-O stars (O5 and earlier);  see \cite{Bodensteiner20} for NGC\,330 and \cite{Dufton19} for NGC\,346. More specifically, the core of NGC\,330 contains only two O-type stars, both of type O9e. NGC\,346, which should be almost fully resolved according to \cite{Dufton19}, contains four sources of subtype O5 and earlier. All of these are in the B10 catalog. The %fact that NGC\,330 and NGC\,346 are still the prominent UV sources
lack of UV-bright sources other than NGC\,330 and NGC\,346
does not seem to leave room for unresolved clusters with a significant number of O2-O5 stars (Table\,\ref{tab:type_nrs} shows that naively, 110 of these stars are expected for CSF, but only 14 are known). %Thus, also the UV image shows no hint of unresolved clusters with many hot bright stars.
To summarize: the GAIA catalog and UV images of the SMC show no indication of unresolved clusters hiding early-type O stars, and the few stars of type O5 and earlier that are in NGC\,330 and NGC\,346 are in the B10 catalog.
Therefore we deem the unresolved cluster explanation to be highly unlikely, as long as the bright young stars are not embedded (Sect.\,\ref{sec:embedding} and \ref{sec:implications}).

%It is known that in dense clusters at the distance of the Magellanic Clouds, crowding can hinder an accurate analysis of the residing stars \citep{Evans11}. \textbf{Thus, another possible explanation for the dearth of young stars is that they are missed because they are in unresolved clusters. }\textit{ ..Move to sect 6.1? Or add to 6.1.4 observational biases? Then binaries to the same section??}
%However, there are no sources above the luminosity range of the GAIA photometry HRD (Fig.\,\ref{fig:gaia_hrd}) that would be candidates for clusters containing unresolved young stars with $\log L \gtrsim 6$. \textbf{Also, for most bright stars in clusters we derive temperatures and luminosities similar to those reported in the VSS studies (Fig.\,\ref{fig:hrd_arr}).} \textit{(Still not sure if this point is discussed optimally..if you have ideas please let me know.)}

Alternatively, %it could be the case that the real temperatures
the real temperatures of our sample stars
could be higher than those that we derived based on the B10 spectral types. This would move them to higher luminosities as well. However, given that the temperatures and luminosities that we present agree reasonably well with the temperatures derived with atmospheric analyses (Fig.\,\ref{fig:hrd_arr}), this would indicate a more severe problem than incorrect relations of spectral type and temperature.
%Moreover, aside from the biases against hot stars, observational biases against bright stars seems to be hard to explain.

%It is important to note that in a sense, it does not matter if `intrinsic biases' or `observational biases' are at play: in both cases, the missing bright and young sources do exist, but we cannot find them or do not interpret them as such.
\subsubsection{Binary companions}
About of half the sources in our HRD are expected to be binary systems \citep{Sana13} or even higher order multiples \cite[e.g., SMC WR systems AB5 and AB6:][]{Shenar16, Shenar17}. However, we treated all B10 sources as single stars (Sect.\,\ref{sec:meth_lt}).
The presence of unresolved binary companions could affect our analysis in two ways.
%\textbf{One could hypothesize that this fraction is mass dependent. If high-mass stars are more likely to reside in multiple systems, this means that we would exclude relatively more stars from the high-mass population. While this might in fact be the case to some extent \citep[e.g.,][]{Moe17}, the binary fraction would need to change dramatically between 20\Msol and 50 \Msol.
%Dedicated observations would be very helpful to learn more about this.
%Also, even if this could explain the lack of bright stars, it does not explain lack of young stars. For this it would need to affect derived Teff and logL. Below, we discuss the effects that binary companions could have on these parameters.

%In any case, for a scenario where binary  do not depend on stellar mass, we see no compelling reason for the shape of the ages and luminosity distributions to change significantly.
%In reality, more likely than not, the mass ratio and orbital do depend on the primary mass \citep{Moe17}. However, it seems unlikely that this can account for the factor-of-a-few dearth of stars with $\log (L/L_\odot) > 5.5$ (Sect.\ref{sec:ldist}). In that case, each of these sources would have to harbor a few of such bright stars. Another problem with this explanation is that the flux per star would be lower, such that in practice a lower luminosity would be derived for the individual components.}

The first possible effect on our analysis caused by the presence of unresolved binary companions could be errors in the temperature and luminosity that we derive. We discuss this below.
For a coeval equal-mass binary, we will derive a luminosity that is 0.3\,dex higher than that of its individual components. Because luminosity scales strongly with mass, the error will be significantly smaller for unequal-mass binaries. %most cases.
%We also note that in order to noticeably tilt the slope of the luminosity distribution shown in Fig.\,\ref{fig:ldists}, the binary properties (binary fraction, mass ratio distribution) might have to depend on luminosity.
A binary companion could also affect the spectral features of a source. To what extent this can change the apparent spectral types is not easy to gauge from first principles. We consider two extreme cases in a coeval binary. First, the equal-mass binaries are not affected because both components would have the same spectral type. Second, for systems where the spectral types of the components differ by more than a few subtypes, the effect should typically be minor because the spectrum is dominated by the massive component.
In between these extremes, the single-star approximation could lead to modest errors in the derived temperature. In these cases, multi-epoch spectroscopy would be highly valuable to obtain more accurate measurements.
To account for the dearth of young stars, we found that a systematic shift of 0.09\,dex in temperature is required (Sect.\,\ref{sec:model_disc}).
To illustrate: 0.09\,dex corresponds to the temperature difference between an O7\,V and a B0\,V star. 
%O7 should appear as B0
It is unlikely that unresolved binary companions could systematically shift the apparent spectral types of the population by that much.

The second possible effect of the presence of other stars within the sources that we analyzed is that it would add stars to the population that are now unaccounted for.
%We note that this adds a currently unresolved complexity to the calculation of for example an IMF.
%This would cause an underestimation in the SFR that we derive. 
This could affect the SFR and the number distributions that we derived.
For the extreme case that all sources are equal-mass binaries, a naive guess is that we underpredict the SFR by a factor two. In reality, the impact of equal-mass companions would be smaller: the flux per star would be lower, such that in practice lower luminosities (and therefore masses) would be derived for the individual components. This effect of binarity could bring the SFR of 0.018\Msol\,yr$^{-1}$ (assuming CSF; Sect.\,\ref{sec:sfr}) closer to the literature value of 0.05\Msol\,yr$^{-1}$.

For a scenario in which the binary fraction and mass ratio distribution do not depend on primary mass, we see no compelling reason for the shape of the age and luminosity distributions to change significantly. This would mainly affect the absolute numbers. A dependence of binary properties on primary mass may still exist, however \citep{Moe17}. An explanation of the factor-of-a-few underabundance of WR star progenitors (Sect.\ref{sec:he_burn_progenitors}) and of stars with $\log (L/L_\odot) > 5.5$ in the luminosity distribution (Sect.\ref{sec:ldist}) would still require 
an extreme increase in multiplicity of stars towards the high-mass end. And again, if the most luminous stars are higher order multiples, the luminosity of the individual components would decrease, which works against this explanation.
%extreme changes in these binary properties at the high-mass end. 
We conclude that the effects of binarity likely lead to typically modest individual errors, but that binarity alone cannot be expected to change the main conclusions of the paper on the dearth of young and bright stars.

\subsubsection{Embedding in birth clouds \label{sec:embedding}} 
Another possible explanation for the lack of young and bright stars could be that young stars are still embedded in the clouds from which they are born.
%The embedded phase is expected to last %relatively long for the more massive stars 
%for a longer fraction of the lifetime for more massive stars
With increasing mass, the embedded phase is expected to last for an increasingly long fraction of the total lifetime
\citep[e.g., fig.\,1 of][]{Yorke86}. 
If this causes these objects to be hidden from our view, %in this scenario 
both young stars and bright stars would become more rare. We note that this would artificially steepen the IMF because brighter stars would then have a greater likelihood to be outside the sample.

For O-type stars in the Milky Way, it has been estimated that $10-20\%$ of their MS lifetime is spent inside a molecular cloud \citep{Wood89}.
Plausibly, metallicity can affect this number. % perhaps embedding could explain the apparent lack of young and bright stars
Naively, we might expect young stars in low-metallicity environments to be less obscured because there is less dust. However, the situation might be more complex than that.
%it could be even larger in the SMC. 
For example, it might be speculated that at low metallicity, the relatively low amount of dust makes it more difficult for bright hot stars to push away the gas around them. The weaker stellar winds at low metallicity might also help to retain circumstellar material for a longer time. If these effects are significant enough, they could instead cause young stars at low metallicity to be embedded for a longer time.

Embedded young massive stars can manifest themselves as YSOs or (ultra-)compact H\,II regions \citep[see, e.g.,][for examples of the latter]{Heydari02, Testor14}. Both are known to be quite numerous in the SMC.
A few hundred intermediate- to high-mass YSOs have been reported \citep[][]{Bolatto07, Oliveira13, Sewilo13, Ward17}.
At least several dozen (ultra-)compact H\,II regions are known to exist \citep{Indebetouw04, Wong12, Lopez14}.
In Sect.\,\ref{sec:implications} below, we discuss how likely it is that they account for the dearth of young and bright stars.

\cite{Holgado20} have proposed an extended period of accretion as the reason for the lack of young massive stars in the Milky Way. For accretion timescales comparable to the MS lifetime of massive stars, \cite{Holgado20} show that the maximum effective temperatures in corresponding evolutionary tracks remain well below those of the classical ZAMS. The problem with this ansatz may be, however, that we should observe roughly half of all massive stars to currently undergo accretion, which does not seem to be the case. The only reason why we could miss these objects would be that they are still embedded in their birth cloud. In this case, the duration of the embedding may be more relevant than the duration of the accretion phase to explain the lack of young massive stars.

\subsection{Implications of the embedding scenario \label{sec:implications}}%and potential issues}

A decrease in SFR and a prolonged embedding phase appear to be the most likely to simultaneously explain the dearth of young and bright stars. In Sect.\,\ref{sec:sfh_disc} we pointed out a number of shortcomings of the SFR scenario. We discuss the plausibility and implications of the embedding scenario below.

If massive stars are embedded in their birth cloud for a relatively long part of their life, there must be a large number of such objects in the SMC. %While, as also discussed above, 
As discussed above, these could manifest themselves as compact H\,II regions or massive YSOs. Several dozen to several hundreds of both massive YSOs and compact H\,II regions are indeed observed. %The question remains if their numbers are sufficient.
Assuming CSF, we found in Sect.\,\ref{sec:rel_age_dist} that about 300 young massive stars in the first half of their MS lifetime above the 18\Msol track are missing. 
We tested this prediction by examining observational studies in different wavelength regimes.
%Here, we investigate if enough of such massive objects can be contained in the observed YSO sample.

\subsubsection{Infrared observations}

The catalog of \cite{Sewilo13} is based on photometic observations in the infrared of the Spitzer Space Telescope. It contains 984 'intermediate- to high-mass' YSO candidates. The authors present physical parameters of 452 of these, which they can fit well to YSO models. Of these, 216 are massive stars with $M \geq 8$\Msol. However, for only 12 of their YSOs do they derive masses above 18\Msol. For 3 of these, their temperatures above 32\kk fall in the O-star range (Table\,\ref{tab:spt_teffs}) while also luminosities above $\log (L/L_\odot) = 4.5$ are derived.
It is inferred that these 3 objects are embedded in circumstellar envelopes of 230, 740, and 3400\Msol. The 2 for which we find a GAIA match within 1$''$ have magnitudes of $G \approx 19$. 
The presence of this group of objects proves that  deeply embedded optically faint hot stars do in fact exist. However, their number is of the order of a few instead of a few hundred.
%\textbf{They report only 12 objects more massive that 18\Msol. However, the number of objects above 8\Msol is 216. Thus, the number of YSOs would be of the same order of magnitude only if the YSO masses would be underestimated by up to a factor of two.}
%%\textbf{In sample of \cite{Sewilo13}, there is a group of 20 objects with $T_\mathrm{eff} \approx 30$\kk and circumstellar envelope masses of the order of a hundred to a few thousand Solar masses. These objects are very faint in the optical. The presence of this group of objects proves that  deeply-embedded, optically faint hot stars do in fact exist; however, the reported number is too low. On the other hand, it is possible that not all deeply embedded hot stars in the SMC have been identified as such in their study. In the same envelope mass range, there is a sample of 189 sources with low stellar temperatures that might be unidentified hot star candidates. }
%In addition, some massive YSOs might be rejected in the \cite{Sewilo13} study because these sources illuminate a larger volume, [sect 2.1 last para] ALSO, Mottram+11 report much lower YSO lifetimes.
%To our knowledge, the current observations are not at odds with the scenario that we paint.  
%%\textbf{We conclude that the current observations are not in direct agreement with the embedding scenario. If there is uncertainty in the YSO masses from \cite{Sewilo13}, or if their sample is not complete, however, it is also not ruled out.}
From the analysis of the \cite{Sewilo13} catalog, we conclude that the current observations fail to agree directly with the embedding scenario. 
This scenario would only be plausible if many other embedded hot massive stars were not in this catalog.
%This scenario would only be plausible if many other hot massive stars would not be identified as such, or if the catalog would be incomplete.
%We encourage a dedicated future analysis.
%\textit{Question: am I missing references about YSOs and HII regions? Also, is it true that the masses are only constrained to `massive'? }

\subsubsection{Radio observations of (ultra-)compact H II regions}
With 48 sources, the catalog of \cite{Wong12} contains the largest number of (ultra-)compact H\,II regions in the SMC to our knowledge. However, it is not known how complete this catalog is, therefore the true number could be higher. Moreover,  (ultra-)compact H\,II regions could theoretically harbor more than one young bright star. Conversely, it is not evident that every (ultra-)compact H\,II region in the \cite{Wong12} catalog contains a star that matches the criteria of our missing hot and bright stars. We conclude that observations of (ultra-)compact H\,II regions neither rule out nor confirm the presence of several hundred deeply embedded young and bright stars.

%Apart from infrared wavelengths, 
%Young star-forming regions 

\subsubsection{Submillimeter observations}
Star-forming regions harboring proto- and young stellar objects
can also be identified by their submillimeter (submm) dust emission. For example, the APEX Telescope Large Area Survey of the Galaxy (ATLASGAL) has provided a map of 870\,$\mu$m emission in the Galactic disk using the Atacama Pathfinder 12\,m submm telescope \citep[APEX,][]{Schuller09}. It has identified compact submm sources %that have hot cores and/or H\,II regions, making them probable embedded young massive star candidates. 
that contain deeply embedded objects in a range of evolutionary phases, from the earliest pre-protostellar stage to ultra-compact H\,II regions.
This claim is supported by the fact that in the Milky way, deeply embedded O-type stars have been confirmed \citep[e.g.,][]{Messineo18}. %\textit{(Karl, I have a question: we are sure that this extinction arises from their local environment and not from MW disk extinction right?)} %The SMC, however, has not been (systematically) observed by APEX. 

We briefly discuss the observability of embedded stars at submm wavelengths.
\cite{Schuller09} reported that ATLASGAL would detect clumps more massive than $\sim$100\Msol at 8\,kpc, the distance to the Galactic center. At the distance of the SMC, this minimum mass would be $\sim$5600\Msol, assuming the Galactic gas-to-dust ratio and dust temperature. 
A survey of 1100 $\mu$m dust emission in the SMC with the ASTE 10-meter telescope was sensitive to condensations with molecular gas masses in excess of $10^4$\Msol \citep{Takekoshi17}, which is on the same order as the mass limit estimated from ATLASGAL.
Regardless of the exact assumptions, APEX or any other single-dish submm telescope could detect only rather large star-forming associations in the SMC. No conclusions can therefore currently be drawn from submm observations.
The Atacama Large Millimeter/submillimeter Array (ALMA) can provide much better sensitivity and angular resolution than the APEX single-dish telescope. A study of 30 Doradus in the LMC \citep{Indebetouw13} has resolved CO line and dust continuum emission regions with sub-parsec resolution. It reports that the emission arises from a sample of clumps with masses between tens and a few times $10^3$\Msol. %\textit{(Karl: is it possible to make an even stronger connection to embedded young bright stars?)} 
Similar studies of the entire SMC could use the maps obtained by \cite{Takekoshi17} as guide for ALMA observations.

\subsubsection{Theoretical expectations}
We also briefly considered the embedding scenario from the theoretical side. For this, we inspected the simulations of \cite{Geen18}, who modeled star formation in a 10\,000\Msol gas cloud with Milky Way composition. We briefly elaborate in Appendix.\,\ref{sec:xtinxion}. In this case, in the first few million years after stars start to form, at least 80\% of the massive stars are found to have an extinction toward their line of sight of $A_V> 5$ (Fig.\,\ref{fig:sam}). With such high extinction, they would most likely not pass the selection criteria for studies at optical wavelengths aimed at massive stars on the MS; if they would, our method would underestimate their luminosity by at least two orders of magnitude, so that they would not show up in our Fig.\,\ref{fig:hrd_corr}. Clearly, this needs to be explored in further detail. %We aim to do so in a forthcoming paper. %\textit{(This is still preliminary, work in progress with Sam (Geen), but we will keep this short).}
%If the massive stars in the SMC are embedded in their birth clouds for a significant part of their main-sequence lifetime, then this would have implications for their ability to ionize their surroundings.
This means that while direct observational evidence is still lacking, the theoretical studies of \cite{Yorke86} and \cite{Geen18} agree with a prolonged embedding in birth clouds. %\citep{Yorke86, Geen18}.}

For massive MS stars in the SMC, the embedding scenario would have implications for their ability to ionize their surroundings.
We estimated the effect this would have. %If the missing young stars in the best-fitting SFH2 (Fig.\,\ref{fig:sfh}) would be added in (i.e., the ones that would now be embedded), about
We adopted the approximation that the stars in the first half of their MS lifetime are deeply embedded in their birth cloud (cf. Fig.\,\ref{fig:sfh}). Then, there are two effects at play for the ionizing radiation: i) %there are twice as many stars in the total population (half of them obscured) 
half of the stars is obscured,
and ii) the ionizing photon production rate per unobscured star is lower because unobscured stars would tend to be older %(Fig.\,\ref{fig:iotons}) 
and thus cooler.

The result would be that H\,I ionizing photons are emitted by MS stars at a rate that is four times lower than if embedding does not play a role (half of the stars are obscured, the average unobscured star has half the H\,I ionizing photon emission rate; see Fig.\ref{fig:iotons}). For He\,I this number is roughly 8 times lower, and for He\,II it is about 20 times lower (but note that He ionizing radiation is dominated by WR stars in the SMC; see Sect.\,\ref{sec:ionizing_radiation}).  %In the case of the SMC, the last two would be dominated by the contribution of WR stars.
If the embedding scenario were true, then synthetic populations of massive MS stars that are typically used to calculate the number of ionizing photons per second \citep[i.e., by integrating over stellar evolution tracks, see, e.g.,][]{Leitherer08} would strongly overpredict these rates. 
This in turn would cause the SFR inferred from H\,$\alpha$ emission to be underestimated.

\section{Conclusions \label{sec:conclusions}}
We have used literature data to obtain a nearly complete overview of the brightest stars %\textbf{that can be observed} 
in the SMC.
We used our results to calculate the extinction distribution, the ionizing photon production rate, and the SFR of the massive star population in the SMC.
In general, our results appear to agree with earlier studies. The stellar population in our HRD also appears to be similar to those in \cite{Blaha89} and \cite{Massey95}. 
%and H\,I ionizing radiation production. 
However, we identify a dearth of progenitors of helium-burning stars, especially at high luminosity.

In the HRD, at $\log (L/L_\odot) > 4.5$, we identify two strong features that would not be expected a priori.
First, when we compared our data to evolutionary models, we inferred that the distributions of both the absolute age and the age as a fraction of lifetime show a lack of young stars. This confirms a feature that has been observed before for massive stars in the LMC and the Milky Way, and extends it to the SMC environment.
Second, we found that the number of very massive stars with $\log (L/L_\odot) > 5.5$ in the SMC is significantly lower than expected under the assumption of a Salpeter IMF and constant star formation.

We discussed the possible cause of these two features.
Star formation having practically come to a standstill in the last few million years might explain both features in principle. However, it seems unlikely that not only the entire SMC but also the LMC and the Milky Way (where the same feature is observed) have reduced their SFR %by a factor of at least a factor ten 
dramatically at the same time in just a few million years.
Alternatively, young and bright stars could exist in the SMC, but are either missed or not recognized as such. 
We discussed a scenario in which massive stars, especially the bright ones, are embedded in their birth cloud for a significant fraction of their MS lifetime.
%relatively long time. 
There might be support for this scenario from theoretical arguments. %This scenario is supported by theoretical studies. 
Observationally, a few deeply embedded, optically faint, hot, luminous stars have been reported in the SMC (as is also the case in the Milky Way). However, a few hundred should be present to explain the missing stars. This means that the embedding scenario is challenged as well unless more of such objects are found. Prolonged embedding would have several astrophysical implications: for example, the obscuring of hot young stars would significantly reduce feedback in the form of ionizing photons. Moreover, in this scenario, the IMF would appear steeper than it really is.
%\textbf{Therefore, the steep IMF exponent of $\Gamma 0 -2.3$ found that the luminosity distribution}
%\textbf{We have found }

We suggest to further investigate the currently unsolved issue of the dearth of young and bright stars. More detailed analyses of bright stars in the SMC and information about their binarity would be highly valuable. 
%Most likely, only then the question can be answered if a steeper (or shallower) IMF is at work in the SMC. This is, however, a crucial question to answer since 
Comprehending the dearth of young and bright stars in the SMC is crucial because
the SMC is at this point our main stepping stone toward a better understanding of massive star evolution in the early universe.

%Because of the importance of the subject and because our method does not teach anything about an influential aspect such as binarity, we advise observational follow up.

%Whether this is caused by the star formation history or an intrinsically steeper IMF at SMC metallicity is a critically important question that should be tackled with future observational campaigns.
%\textit{An important question is: can an entire dwarf galaxy dramatically change its star formation rate on timescales of a few Myr? (argument against SFH: the `very massive stars' are rather uniformly distributed. Check: could they have traveled?)}

% ABSTRACT: Essentially, we use a combination of the spectral type catalog from Bonanos et al. (2010) and GAIA magnitudes to calculate temperatures and luminosities. Then, we use literature studies to check our method and finally we use GAIA data to calculate the completeness as a function of luminosity.

\begin{acknowledgements}
We thank the referee Tomer Shenar for a very constructive referee report, which was highly valuable for improving the discussions in the paper.
A.S. would like to thank Ashley Barnes for enlightening discussions. The authors thank Nathan Grin for compiling the list of Simbad sources associated with the SMC.
\end{acknowledgements}

\bibliographystyle{aa} % style aa.4
\setlength{\bibsep}{0pt}
\bibliography{bib.bib}{}

\begin{appendix}

\section{Data sets \label{sec:data_sets}}
%In this appendix, we describe the literature data sets that we employed.
In this work we have employed different literature data sets. We describe these in this appendix.
\textbf{Bonanos et al. (2010):}
We used a spectral type catalog of the SMC \citep[their table\,1]{Bonanos10} that contains 5325 sources. Most of the sources in this catalog are also part of the 2dF survey \citep{Evans04}. To obtain information about $G$ -band magnitudes, parallaxes, and proper motions, we cross-correlated this catalog with the GAIA \cite{Gaia18} data\footnote{Using \url{http://cdsxmatch.u-strasbg.fr/}}. The maximum difference in position that we allowed was 3 arcseconds. We were able to find a matching source in the GAIA data in 5304 of 5324 cases. %We note that photometry data are more often available for GAIA magnitudes than for $UBV$ magnitudes. The B10 spectral type - photometry catalog contains $V$ magnitudes for $\sim$3000 sources.

Next, we checked whether these sources were SMC members. For this, we used two criteria. First, we removed stars from the catalog that did not match the proper motion criteria of \cite{Yang19}.
These authors found that the SMC proper motion distributions fit Gaussians with $\mu_\mathrm{ra} = 0.695 \pm 0.240$ in the right ascension direction and $\mu_\mathrm{dec} = -1.206 \pm 0.140$ in the declination direction well. Everything farther off than 5$\sigma$ was marked as foreground stars by them.
The difference is that we excluded stars 15$\sigma$ away from the SMC bulk proper motion instead of 5$\sigma$. With 5$\sigma$ we would exclude a number of apparently fast-moving O stars that we deem unlikely to be foreground stars. Second, we removed sources whose parallaxes confidently identified them as foreground sources: those that have $\pi / \sigma_\pi > 5$ \citep{Aadland18}. After this, 5269 sources remained in our sample. The 5155 sources for which we are able to convert the spectral type into an effective temperature constitute the final B10 data set as used in this work.

When multiple sources were within the 3$''$ separation (which was the case for about half the B10 sources), we took the closest source. The first exception is that when the closest source was dimmer than $G=18$, we took the brightest source within 3$''$ instead. The reason is that such dim sources are rare in the B10 catalog, where in the combined spectral type - photometry table (their table\,3) only 16 have $V>18$, making it most likely that their positions matched by chance. The second exception is that when a $V$ magnitude was listed in the B10 catalog, we took the source with the $G$ magnitude that was closest to the $V$ magnitudes from the B10 catalog.

The B10 catalog contains 12 WR sources. In the SMC, no more WR sources than that are thought to exist \citep{Neugent18}. For the WR sources we therefore simply adopted the most recently published observational parameters \citep{Hainich15, Shenar16} instead of working with their spectral types. Because we did not use their spectral types, the WR stars are not included in the final B10 data set.

\textbf{GAIA:}
The part of the GAIA data \citep{Gaia18}\footnote{Available at \url{http://gea.esac.esa.int/archive/}} that we downloaded is the following. We selected sources brighter than magnitude $G = 17$. The positional restrictions were $3^\circ < r < 25^\circ$ and $-75.5^\circ < \delta < -70^\circ$, again following \cite{Yang19}. We repeated the procedure described above for the B10 catalog to remove foreground stars. %We note that this procedure removed \textit{all} of the downloaded 11264 sources with $G<15$ that have a parallax above $\pi = 1/3$ mas (i.e., that are within 3\,kpc). Thus, we believe that foreground contamination does not significantly affect the sample we work with.
This patch on the sky encompasses all but four of the sources in table\,1 of B10; the source represented by the red dot without a black edge in Fig.\,\ref{fig:gaia_hrd} around $\log (T_\mathrm{eff / K}) = 3.85$ and $\log (L/L_\odot) = 4.95$, and three sources that are too dim to be shown in Fig.\,\ref{fig:gaia_hrd}.

\textbf{Various spectroscopic studies (VSS):}
In these studies, hereafter referred to as the VSS sample, the effective temperatures and luminosities of hot bright stars in the SMC were spectroscopically derived by comparison with model atmospheres. We included \cite{Trundle04}, \cite{Trundle05}, \cite{Hunter08b}, \cite{Bouret13}, \cite{Dufton19}, \cite{Ramachandran19}, and \cite{Mokiem06}. The last study includes eight very bright stars that are not in the Hunter et al. compilation of the VLT-FLAMES survey of massive stars \citep{Evans06}. We also applied the foreground cleaning procedure to this sample.
The sources in this sample were subjected to the most detailed analysis; therefore we deem these temperatures and luminosities to be the most trustworthy.

%__________________________________________________ One column table
\begin{table*}
\tiny
\centering    
    \caption[]{Adopted spectral type$-$temperature relation for SMC stars with different spectral types and luminosity classes. In the notes, we mention possible interpolation from neighboring spectral types (`int.'), or if the temperature is adopted from the same spectral type but other luminosity class (e.g., `from I$\_$Teff'), or if the value is taken directly from a single study such as E03 \citep{Evans03} or T18 \citep{Tabernero18}.}
\begin{tabular}{l l | r l r l r l }
\hline
\hline
            \noalign{\smallskip}
SpT&SpT$\_$nr&V$\_$IV$\_$Teff [kK]&V$\_$IV$\_$note&III$\_$II$\_$Teff [kK]&III$\_$II$\_$note&I$\_$Teff [kK]&I$\_$note\\
%& &[kK]& &[kK]& &[kK]& \\
            \noalign{\smallskip}
            \hline
            \noalign{\smallskip}

O2&2.0&51.7&from III$\_$II$\_$Teff&51.7&&51.7&from III$\_$II$\_$Teff\\
O3&3.0&46.0&&49.5&int.&49.5&from III$\_$II$\_$Teff\\
O4&4.0&43.4&&47.3&int.&47.3&from III$\_$II$\_$Teff\\
O4.5&4.5&42.8&int.&46.2&int.&46.2&from III$\_$II$\_$Teff\\
O5&5.0&42.2&&45.0&int.&45.0&from III$\_$II$\_$Teff\\
O5.5&5.5&41.2&int.&43.0&int.&43.9&from III$\_$II$\_$Teff\\
O6&6.0&40.3&&42.8&int.&42.8&from III$\_$II$\_$Teff\\
O6.5&6.5&39.3&&40.6&int.&40.6&from III$\_$II$\_$Teff\\
O7&7.0&38.1&&38.4&&34.1&\\
O7.5&7.5&37.3&int.&37.7&int.&34.1&int.\\
O8&8.0&36.5&&36.9&&33.0&int.\\
O8.5&8.5&35.1&&34.3&&31.8&int.\\
O9&9.0&34.5&&33.4&int.&30.7&\\
O9.5&9.5&33.7&&32.1&&30.0&int.\\
O9.7&9.7&32.1&&30.8&&29.4&int.\\
B0&10.0&31.1&&29.5&int.&28.6&\\
B0.2&10.2&29.6&&28.3&&27.5&int.\\
B0.5&10.5&28.3&&26.8&&26.4&\\
B0.7&10.7&26.8&&25.5&&25.0&int.\\
B1&11.0&25.3&&24.5&&23.5&\\
B1.5&11.5&23.1&&22.7&&21.3&\\
B2&12.0&21.1&&21.3&&18.9&\\
B2.5&12.5&19.4&&19.6&&16.9&\\
B3&13.0&17.7&&17.5&&15.3&\\
B4&14.0&16.9&int.&16.2&int.&14.6&int.\\
B5&15.0&16.2&&14.9&&13.9&\\
B6&16.0&15.0&int.&14.3&int.&13.3&int.\\
B7&17.0&13.8&&13.8&int.&12.7&int.\\
B8&18.0&14.0&&12.9&&12.1&\\
B9&19.0&10.6&from I$\_$Teff&10.6&from I$\_$Teff&10.6&\\
A0&20.0&9.5&from I$\_$Teff&9.5&from I$\_$Teff&9.5&E03\\
A1&21.0&9.0&from I$\_$Teff&9.0&from I$\_$Teff&9.0&int. E03\\
A2&22.0&8.5&from I$\_$Teff&8.5&from I$\_$Teff&8.5&E03\\
A3&23.0&8.0&from I$\_$Teff&8.0&from I$\_$Teff&8.0&E03\\
A4&24.0&7.875&from I$\_$Teff&7.875&from I$\_$Teff&7.875&int. E03\\
A5&25.0&7.75&from I$\_$Teff&7.75&from I$\_$Teff&7.75&E03\\
A6&26.0&7.5&from I$\_$Teff&7.5&from I$\_$Teff&7.5&int. E03\\
A7&27.0&7.25&from I$\_$Teff&7.25&from I$\_$Teff&7.25&E03\\
A8&28.0&7.08&from I$\_$Teff&7.08&from I$\_$Teff&7.08&int. E03\\
A9&29.0&6.92&from I$\_$Teff&6.92&from I$\_$Teff&6.92&int. E03\\
F0&30.0&6.75&from I$\_$Teff&6.75&from I$\_$Teff&6.75&E03\\
F1&31.0&6.575&from I$\_$Teff&6.575&from I$\_$Teff&6.575&int. E03\\
F2&32.0&6.4&from I$\_$Teff&6.4&from I$\_$Teff&6.4&int. E03\\
F3&33.0&6.225&from I$\_$Teff&6.225&from I$\_$Teff&6.225&int. E03\\
F4&34.0&6.05&from I$\_$Teff&6.05&from I$\_$Teff&6.05&int. E03\\
F5&35.0&5.875&from I$\_$Teff&5.875&from I$\_$Teff&5.875&E03\\
F6&36.0&5.803&from I$\_$Teff&5.803&from I$\_$Teff&5.803&int. E03 and T18\\
F7&37.0&5.731&from I$\_$Teff&5.731&from I$\_$Teff&5.731&int. E03 and T18\\
F8&38.0&5.650&from I$\_$Teff&5.650&from I$\_$Teff&5.650&int. E03 and T18\\
F9&39.0&5.588&from I$\_$Teff&5.588&from I$\_$Teff&5.588&int. E03 and T18\\
G0&40.0&5.516&from I$\_$Teff&5.516&from I$\_$Teff&5.516&T18\\
G1&41.0&5.081&from I$\_$Teff&5.081&from I$\_$Teff&5.081&T18\\
G2&42.0&4.804&from I$\_$Teff&4.804&from I$\_$Teff&4.804&int. T18\\
G3&43.0&4.526&from I$\_$Teff&4.526&from I$\_$Teff&4.526&T18\\
G4&44.0&4.503&from I$\_$Teff&4.503&from I$\_$Teff&4.503&T18\\
G5&45.0&4.657&from I$\_$Teff&4.657&from I$\_$Teff&4.657&T18\\
G6&46.0&4.472&from I$\_$Teff&4.472&from I$\_$Teff&4.472&T18\\
G7&47.0&4.202&from I$\_$Teff&4.202&from I$\_$Teff&4.202&T18\\
G8&48.0&4.202&from I$\_$Teff&4.202&from I$\_$Teff&4.202&T18\\
K0&49.0&4.135&from I$\_$Teff&4.135&from I$\_$Teff&4.135&T18\\
K1&50.0&4.077&from I$\_$Teff&4.077&from I$\_$Teff&4.077&T18\\
K2&51.0&4.059&from I$\_$Teff&4.059&from I$\_$Teff&4.059&T18\\
K3&52.0&4.050&from I$\_$Teff&4.050&from I$\_$Teff&4.050&T18\\
K4&53.0&4.024&from I$\_$Teff&4.024&from I$\_$Teff&4.024&T18\\
K5&54.0&3.976&from I$\_$Teff&3.976&from I$\_$Teff&3.976&T18\\
K6-9& &3.976&from I$\_$Teff&3.976&from I$\_$Teff&3.976&As K5 in T18\\
M0&55.0&3.942&from I$\_$Teff&3.942&from I$\_$Teff&3.942&T18\\
M1&56.0&3.856&from I$\_$Teff&3.856&from I$\_$Teff&3.856&T18\\
M2&57.0&3.802&from I$\_$Teff&3.802&from I$\_$Teff&3.802&T18\\
M3&58.0&3.610&from I$\_$Teff&3.610&from I$\_$Teff&3.610&T18\\
M4-6& &3.610&from I$\_$Teff&3.610&from I$\_$Teff&3.610&as M3 in T18

            \label{tab:spt_teffs}

\end{tabular}
\end{table*}
%______________________

\section{Tests. Completeness, extinction, and Hertzsprung-Russell diagrams \label{sec:app_a}}

%______________________________________________

   \begin{figure}[ht]
   \centering
   \includegraphics[width = \linewidth]{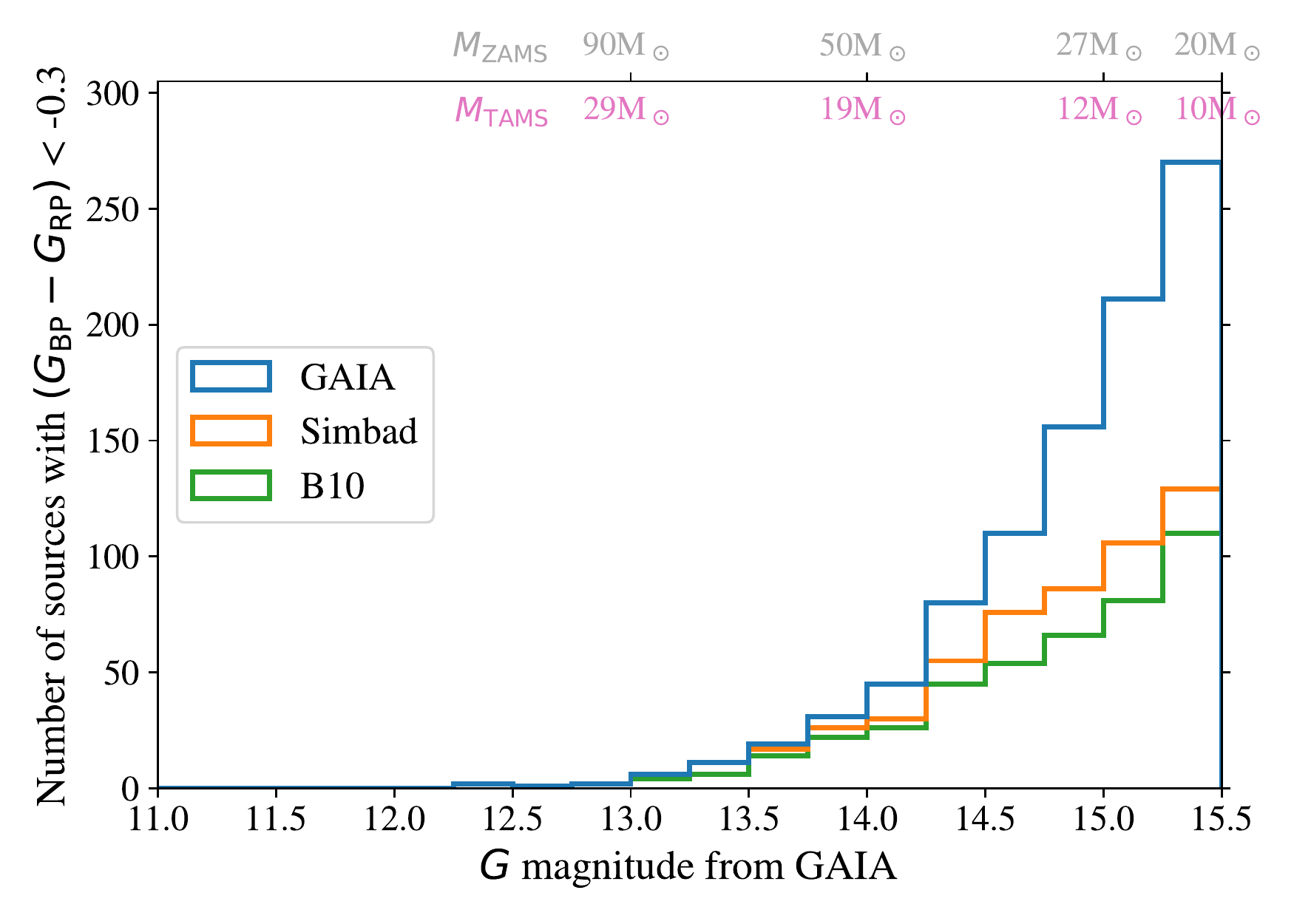}
   \caption{Number distributions of the $G$ magnitudes of sources in the GAIA, Simbad, and B10 \citep{Bonanos10} catalogs that have a GAIA color of $G_\mathrm{BP} - G_\mathrm{RP} < -0.3$. Indicated at the top axis are the apparent $G$ magnitudes from GAIA that stars of a certain mass have at the ZAMS and TAMS at the distance of the SMC, and an extinction of $A_V = 0.35$.
   }
             \label{fig:ndist_mag}%
    \end{figure}
%______________________

\subsection{Completeness \label{sec:fcompl_tests}}

%In Sect.\,\ref{sec:f_complete}, we estimated the completeness of the B10 catalog in different luminosity intervals by comparing with GAIA data, where we obtained luminosities using photometry. For this to work, the photometry-derived luminosity needs to have an at least reasonably good predictive power for the `real' luminosity.We plot in Fig.\,\ref{fig:order_of_logl} how the luminosities derived with only GAIA photometry and our method using the spectral types in the B10 catalog correlate. This figure shows that while the scatter is sizable, the luminosities obtained with GAIA photometry in general give a good indication in which luminosity segment most sources belong.

As a check, we also performed a more simple completeness test than the test described in Sect.\,\ref{sec:f_complete}. We plot the GAIA $G$ magnitude distributions of stars in the GAIA, B10, and Simbad catalogs in Fig.\,\ref{fig:ndist_mag}. We excluded sources that did not fulfill our SMC-membership criteria (Appendix\,\ref{sec:data_sets}). The shown sources are those that have $G_\mathrm{BP} - G_\mathrm{RP} < -0.3$, which for an extinction of $A_V = 0.35$ corresponds to a temperature cut of $\sim$25\kk according to MIST bolometric corrections. This roughly corresponds to the TAMS temperature of the evolutionary tracks shown in Fig.\,\ref{fig:hrd_corr}, except for the most massive stars. We note that the B10 and Simbad completeness fractions do not significantly change when a color cut of $G_\mathrm{BP} - G_\mathrm{RP} < -0.4$ (then corresponding to $T_\mathrm{eff} \gtrsim$35 \kk) or even $G_\mathrm{BP} - G_\mathrm{RP} < 0.1$ ($T_\mathrm{eff} \gtrsim$10 \kk) is adopted instead. 
At the top of Fig.\,\ref{fig:ndist_mag} we indicate for a few magnitude values to what ZAMS star mass models of \cite{Schootemeijer19} they correspond (at the distance of 60.6\,kpc and extinction of $A_V = 0.35$). For example, the models indicate that a 50 \Msol SMC star at the ZAMS has $G \approx 14$.
The shown masses should be approached with care: when the TAMS is considered instead of the ZAMS, $G = 14$, for example, is the magnitude of corresponding to a 19\Msol SMC star. For comparison, we also display these corresponding TAMS masses at the top of the plot in magenta. 
%We also note that for more evolved -- and therefore cooler -- stars, these $G$ magnitudes correspond to stars that are less massive -- where the B10 and Simbad completeness is lower. 
%Fig.\,\ref{fig:ndist_mag} shows that up to $G=14$ (the $G$ magnitude of a $\sim$55\Msol star at the ZAMS), the B10 and Simbad samples contain virtually the same number of stars as the GAIA sample. 
The resulting $G$ -magnitude distributions shown in Fig.\,\ref{fig:ndist_mag} indicate that the B10 catalog is nearly complete for the brightest hot stars. The completeness drops to about 35-40\% for dimmer sources that would still be shown in Fig.\,\ref{fig:hrd_corr}, which is comparable to the numbers we obtained in Table\,\ref{tab:f_complete}.
This confirms our earlier conclusion that the B10 catalog is close to being complete for the brightest stars above $\log ( L / L_\odot) = 5.5,$ also for the young and hot ones. 
%Also here, the B10 completeness seems to drop to about $35$\% towards $G = 15.5$, which is in line with what we derived in Sect.\,\ref{sec:f_complete}.
We conclude that this second completeness test strengthens our findings in Sect.\,\ref{sec:f_complete}. %This implies that the B10 completeness goes up slightly with ages, at least below 50\Msol. However, this is by far not enough to account for the features seen in our age distribution (Fig.\,\ref{fig:sfh}). There, we found only 7\% of the stars to be in the first half of their MS lifetime.

%______________________________________________

   \begin{figure}[ht]
   \centering
   \includegraphics[width = \linewidth]{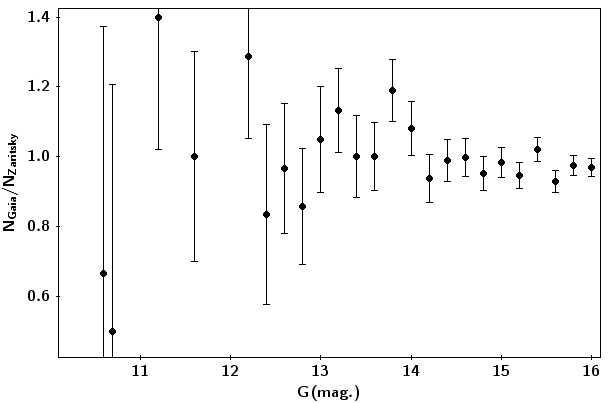}
   \caption{Ratio of sources in the GAIA and \cite{Zaritsky02} catalogs in different magnitude bins.
   }
             \label{fig:gaia_zar_rat}%
    \end{figure}
%______________________

The discussion in the paragraph above assumes that the GAIA catalog is complete for stars that would be shown in our HRDs. To test the completeness of the GAIA catalog, we compared it with the $UBVI$ catalog of \cite{Zaritsky02}. For different magnitude bins, we show the ratio of the number of stars in the GAIA catalog to the number of stars in the \cite{Zaritsky02} catalog  (Fig\,\ref{fig:gaia_zar_rat}).
For the GAIA sources included in this figure, we made a color cut of $G_\mathrm{BP} - G_\mathrm{RP} < 0.55$ to exclude red sources. Empirically, we find that sources with GAIA color $G_\mathrm{BP} - G_\mathrm{RP} < 0.55$ have a color $B-V < 0.35$ in the \cite{Zaritsky02} catalog. For the \cite{Zaritsky02} sources we therefore applied a color cut of $B-V < 0.35$ instead of 0.55.
In Fig.\,\ref{fig:gaia_zar_rat} we compare $G$ to $V$ magnitudes, but we find that they do not differ by more than $\sim$0.03 for the vast majority of the sources that we consider.
The error bars are derived from the number of sources in a bin ($N$): $\sigma = 1 / \sqrt{N}$.
Fig\,\ref{fig:gaia_zar_rat} shows that the ratio of sources in both catalogs is very close to one. For the brightest sources, there is some scatter, which is explained by the low number of sources in each bin.
This result implies that GAIA is highly complete. The same has been implied by the fact the we find a GAIA counterpart for nearly all the B10 sources (Sect.\,\ref{sec:methods}).
This good performance reflects the fact that as discussed by \cite{Arenou18}, Gaia is essentially complete except in regions of very high stellar density, such as exist near the centers of globular clusters, or for close visual binaries.

Finally, we quantitatively investigated the apparent discrepancy in the number of O stars in the SMC predicted by \cite{Massey10} and by our method based on the B10 catalog. The B10 catalog contains 259 O stars. For a typical completeness of 40\% (Fig.\,\ref{fig:ndist_mag}), we would expect about 600 O stars in total. In their table\,1, \cite{Massey10} list a much larger number of 2800 O stars above 20\Msol alone. 
To look for clues, we cross-correlated table\,8b from \cite{Massey02} with the B10 catalog. We again used {\tt cds-xmatch} 
and cross-correlated within 3$''$. This table\,8b from \cite{Massey02} contains 2671 blue stars (of which 1901 rather than 2800 have a listed $T_\mathrm{eff} > 30$\kk, required to be considered O stars). For the 2671 blue stars from \cite{Massey02} (with median $V$ = 14.84), we find 841 matches. This is slightly worse than but comparable to the $\sim$40\% completeness we have around that $G$ magnitude in Fig.\,\ref{fig:ndist_mag}.
Of the stars with a listed $T_\mathrm{eff} > 30$\kk that are also in B10, 191 of 513 (37\%) have the expected O type in B10. The rest have a later type, predominantly B. A simple estimate then gives a number of $0.37 \cdot 1901 = 703$ of the Massey O stars to be O stars in reality. This number agrees well with our expected number of about 600. We note that this comparison likely suffers from biases that could go in both directions (e.g., O stars possibly being more likely to end up in the B10 catalog than later-type stars, or the adopted cut at $M_\mathrm{bol} = -7$ for the sources included in \cite{Massey02}'s table\,8b).

%Also, we test how the spectral type - temperature relations affect the shape of our B10 HRD (Fig.\,\ref{fig:hrd_corr}). For this, instead of the relations described in Sect.\,\ref{sec:methods} (see Fig.\,\ref{fig:spt_teff}) we adopt those from \cite{Pecaut13}. The result is shown in Fig.\,\ref{fig:hrd_p13}. Modest differences exist for individual sources, but the general shape of the HRD is not affected.

\subsection{Extinction \label{sec:xtinxion}}

We start this section by providing a few in-depth remarks on how we calculate the extinction.
We did find unphysical negative extinction values, but only for less than 1\% of the sources, with $A_V = -0.13$ in the worst case. We consider this result to be comforting. Conversely, we also find very high extinction values of $1.95 \leq A_V \leq 5.5$ for 22 cases. Inspection shows that 14 of these make up the entire upper end of the 14 highest $G$ magnitudes (i.e., they are not bright). By far the most of these (apparently) very high extinction sources have uncertain spectral types, for example, B1-5\,III. The earliest spectral type is O9. Therefore they appear to be unable to make up part of the missing early O stars discussed in Sect.\,\ref{sec:rel_age_dist}. We deem it most likely that these sources are chance alignments of cool (i.e., relatively red) dim sources. Our motivation for this is as follows. About half of the sources from the GAIA - B10 cross-match have more that one match, that is, a chance alignment, with a typically cool, dim source. Also, we find no GAIA match for 20 B10 sources. Because the probabilities to have or lack a chance alignment are about equal, we would also expect 20 B10 sources without their true match but only chance alignments. The number of 22 likely chance alignments fits this expected number of 20 surprisingly well. 
In Fig.\,\ref{fig:AVdist} and Fig.\ref{fig:hrd_varAV}, if we encountered a value of $A_V > 1.95$, we therefore assume our standard value of $A_V = 0.35$ instead.

%Finally, we check what happens if we use a different spectral type catalog. Instead of using B10, we consider Simbad. The resulting HRD (Fig.\,\ref{fig:hrd_sim}) contains 1023 sources instead of the 723 in the B10 HRD. Again, the shape of this HRD is similar and our conclusions about the lack of bright stars and young stars persist.

%______________________________________________

   \begin{figure}[ht]
   \centering
   \includegraphics[width = \linewidth]{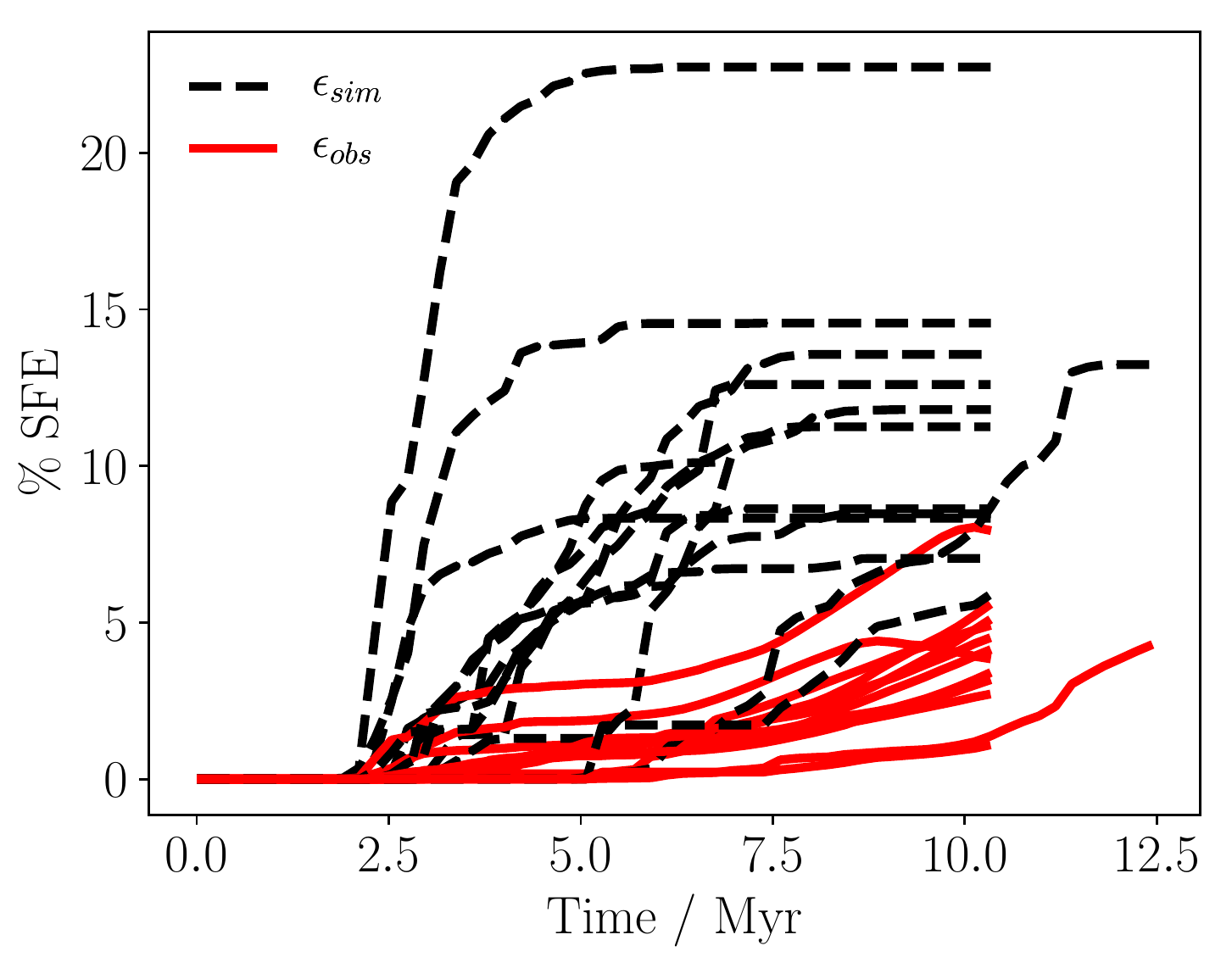}
   \caption{Star formation efficiency, defined as $\mathrm{SFE} = M_\mathrm{stars} / (M_\mathrm{stars} + M_\mathrm{gas})$, as a function of time in a simulation of \cite{Geen18}. Dashed lines indicate the total fraction of mass that is converted into stars, red is the same fraction, but with the requirement that the star has an extinction lower than $A_V = 5$. The different dashed and red lines represent simulations with different initial conditions.}
             \label{fig:sam}%
    \end{figure}
%______________________

We also briefly considered extinction toward our sources from a theoretical side. As discussed in Sect.\,\ref{sec:implications}, we took a closer look at the simulations of \cite{Geen18}. In this work, a 10\,000\Msol gas cloud was simulated, including the formation of stars and their feedback. In Fig.\,\ref{fig:sam} we show the fraction of gas that has been converted into stars. We compared the whole ensemble of stars to the subset of stars that would most likely be visible. The latter was defined as those with an extinction of $A_V \leq 5$ toward their line of sight.
The extinction was obtained from the hydrogen column density in the simulation using the relation $A_V = N_H / (1.8\cdot10^{21} \mathrm{cm}^{-2})$\,mag \citep{Lombardi14}.
It shows that immediately after formation, most of the model stars have an extinction of $A_V > 5$, most likely preventing them from being included in observational studies in the optical. % towards their line of sight. %, which would mean that they would not be included in our HRD shown in Fig.\,\ref{fig:hrd_corr}: most likely, this would result in them no longer meeting the selection criteria for studies aimed at bright massive stars. Also, such high extinction, if not accounted for, would mean that their inferred luminosity would go down at least two orders of magnitude.

%\textbf{In Fig.\,\ref{fig:hrd_synt} } we show a synthetic population with stellar models from \cite{Schootemeijer19}. The stellar models

%______________________________________________

   \begin{figure}
   \centering
   \includegraphics[width = \linewidth]{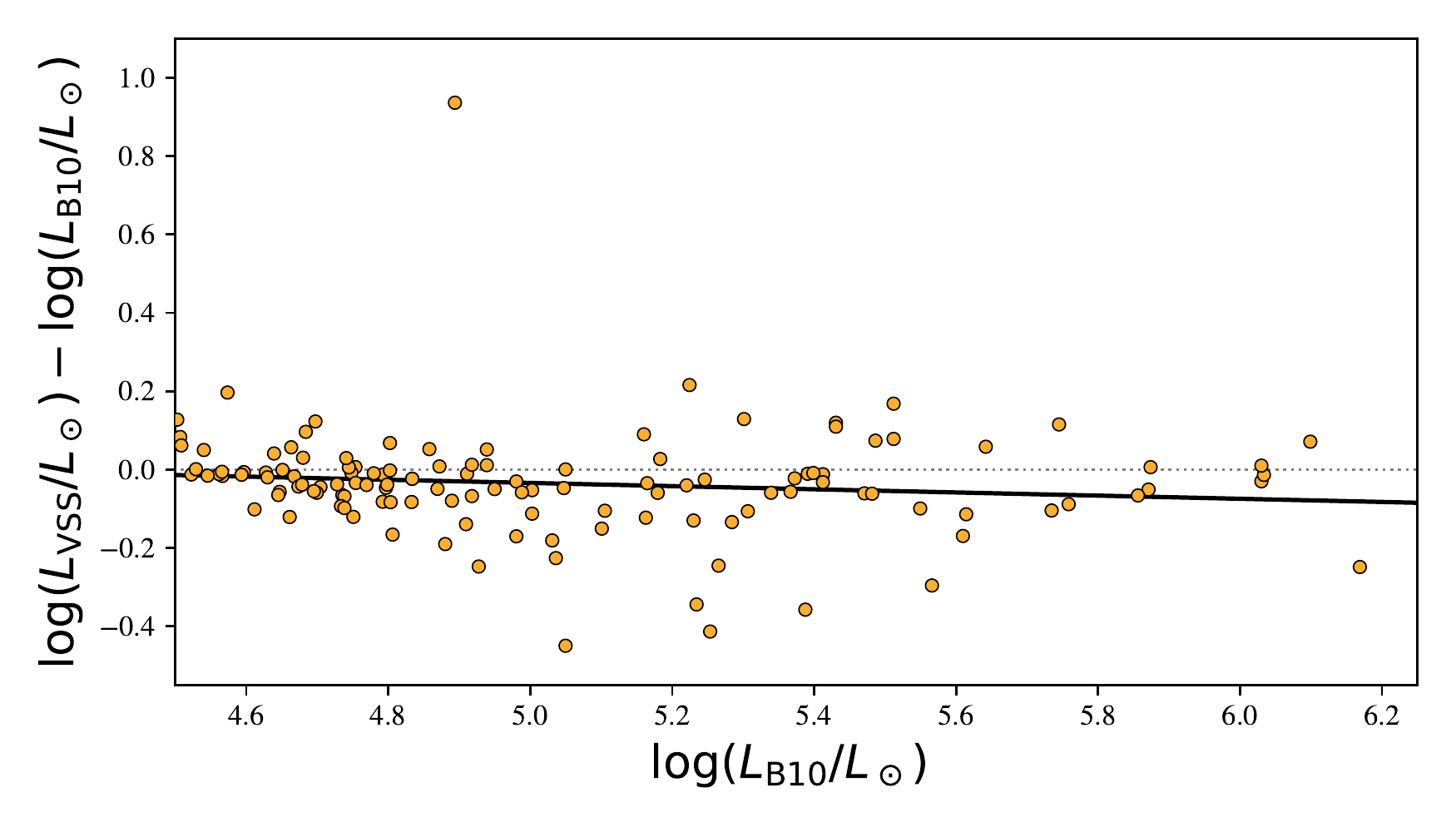}
   \caption{Same as Fig.\,\ref{fig:logl_dlogl}, but here the B10 luminisoties are calculated differently. Instead of assuming a constant visual extinction of $A_V = 0.35$, $A_V$ is calculated for each star individually.}
             \label{fig:logl_dlogl_varAV}%
    \end{figure}
%______________________

\subsection{Hertzsprung-Russell diagrams \label{sec:test_hrds}}

\begin{figure*}
   \centering
   \includegraphics[width = 0.68\linewidth]{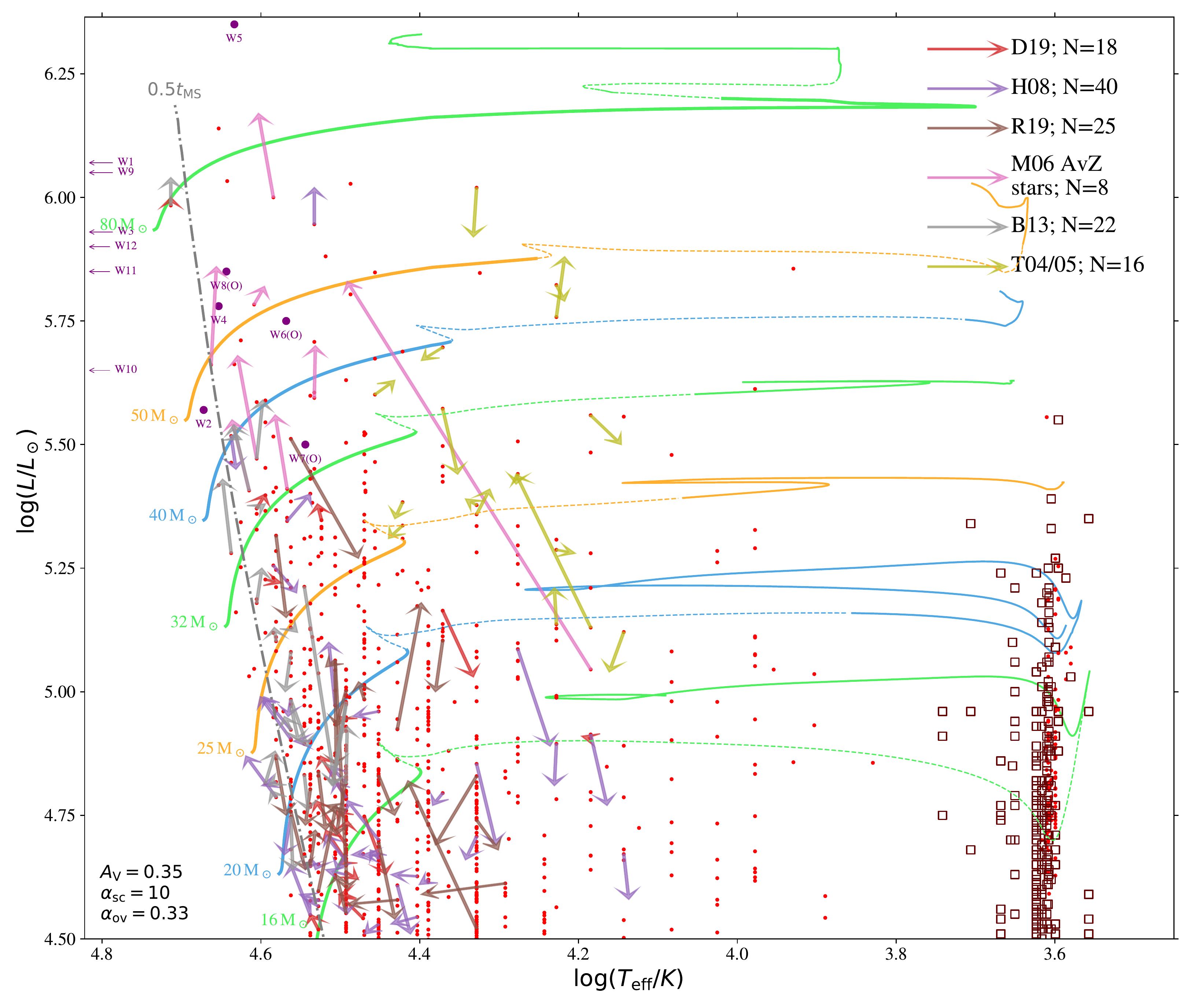}
   \caption{Same as Fig.\,\ref{fig:hrd_corr}, but the arrows point at temperatures and luminosities of objects in various spectroscopic studies. D19 stands for \cite{Dufton19}, H08 for \cite{Hunter08}, R19 for \cite{Ramachandran19}, M06 for \cite{Mokiem06}, B13 for \cite{Bouret13}, and T04/05 for \cite{Trundle04} and \cite{Trundle04}.
   }
             \label{fig:hrd_arr}%
    \end{figure*}

The first test HRD that we provide (as discussed in Sect.\,\ref{sec:results}) is Fig.\,\ref{fig:hrd_arr}. It compares our result to the results of studies in the VSS sample.
Wenn a source within 1$''$ is in the VSS sample, we point an arrow to its literature values.
%We note that the large outlier, classified as a B3 star in \cite{Bonanos10} is listed as an O9 star in \cite{Mokiem06}. With this classification, the offset to the VSS temperature and luminosity would be about 0.04\,dex and 0.12\,dex instead of 0.3\,dex and 0.8\,dex, respectively.
We note that the outlier around $\log (L_\mathrm{VSS} / L_\odot)- \log (L_\mathrm{B10}/L_\odot) = 0.8$ with $\log (L_\mathrm{B10} / L_\odot) \approx 5.05$ is classified as a B3 star in the B10 catalog, but as an O9 star in \cite{Mokiem06}. If the latter type is assumed, the luminosity difference is only 0.1\,dex. 
In general, we find that temperatures and luminosities we derived for the B10 sources agree reasonably well with the HRD positions predicted by the studies in the VSS sample.
In this HRD, we do not include the yellow supergiants from \cite{Neugent10} because those temperatures are based on photometry and not atmosphere analysis. The stars from \cite{Castro18} are also omitted because of the low quality of their input data (this is the RIOTS4 survey of \citealt{Lamb16}, aimed at SMC field stars). It did not allow the analysis of metal lines. Therefore they had to resort to the ionization balance of hydrogen and helium, which could not reliably constrain $T_\mathrm{eff}$ in all corners of the HRD. Moreover, they stated that the surface gravities of the brightest stars are unreliable, causing stars in the spectroscopic HRD to artificially move up in luminosity. We note that these authors reported that there were only few massive stars close to the ZAMS in this sample from the RIOTS4 survey, as they also noted in their earlier study in the Milky Way \citep{Castro14}.
%; as a result, telling apart 20\kk and 30\kk stars was difficult.

\begin{figure*}
   \centering
   \includegraphics[width = 0.72\linewidth]{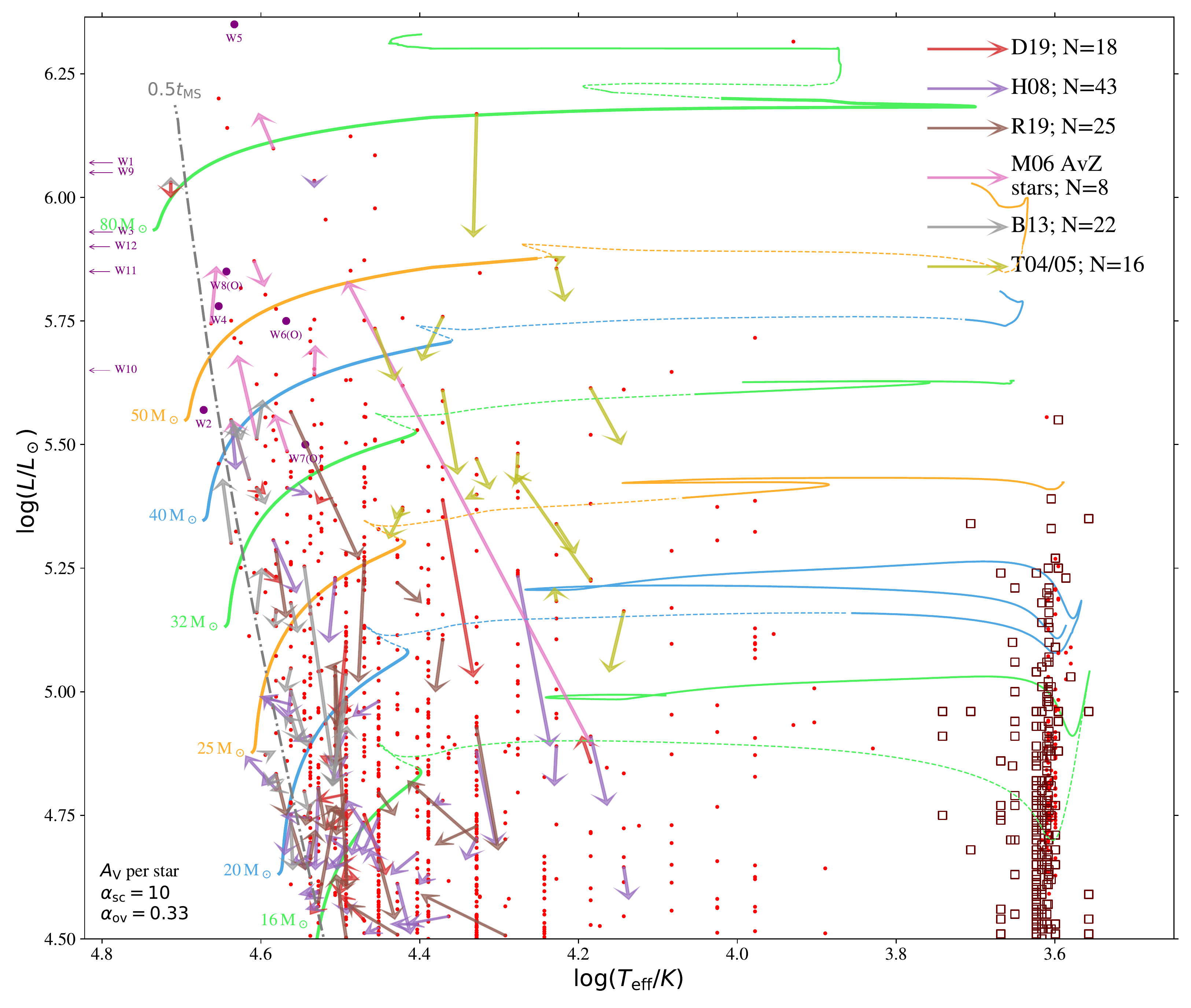}
   \caption{Same as Fig.\,\ref{fig:hrd_arr}, but here the extinction is calculated per star, rather than assuming a constant value of $A_V = 0.35$.
   }
             \label{fig:hrd_varAV}%
    \end{figure*}

Throughout the paper, we assume a constant extinction value %\footnote{In this work we use GAIA colors and magnitudes to estimate completeness -- then we have no information about extinction for individual sources.} 
of $A_V = 0.35$. As a test, we also calculated the extinction for each source individually (as explained in Sect.\,\ref{sec:ex_etc}%the second paragraph of Appendix\,\ref{sec:app_a}
; see also the $A_V$ distribution in Fig.\,\ref{fig:AVdist}). This results in slightly more luminous sources in the HRD (Fig.\,\ref{fig:hrd_varAV}), but the difference is not large enough to affect our conclusions.

\begin{figure*}
   \centering
   \includegraphics[width = 0.72\linewidth]{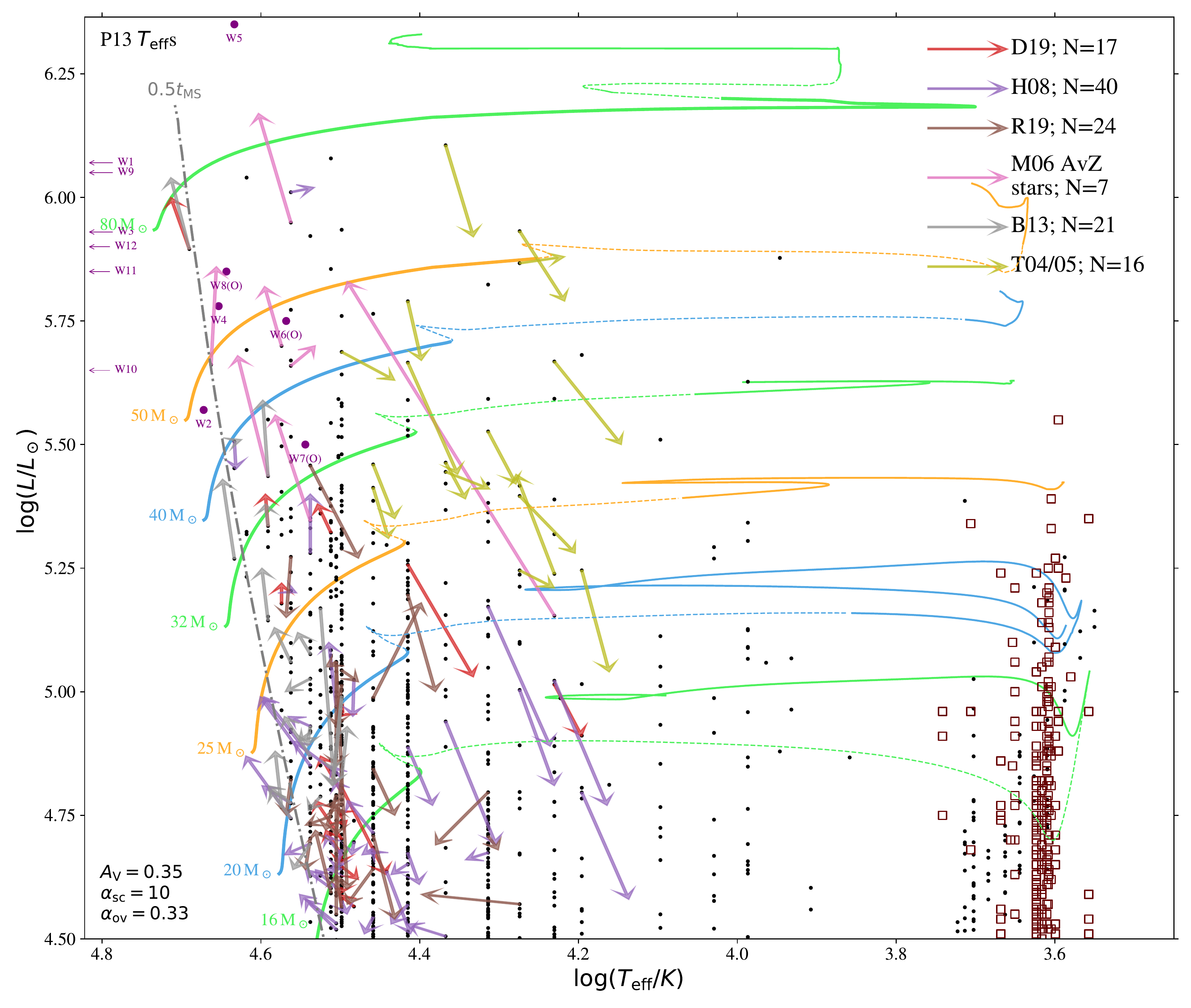}
   \caption{Same as Fig.\,\ref{fig:hrd_arr}, but with spectral type - $T_\mathrm{eff}$ relations of \cite{Pecaut13} instead of those from Table \,\ref{tab:spt_teffs}.
   }
             \label{fig:hrd_p13}%
    \end{figure*}

\begin{figure*}
   \centering
   \includegraphics[width = 0.72\linewidth]{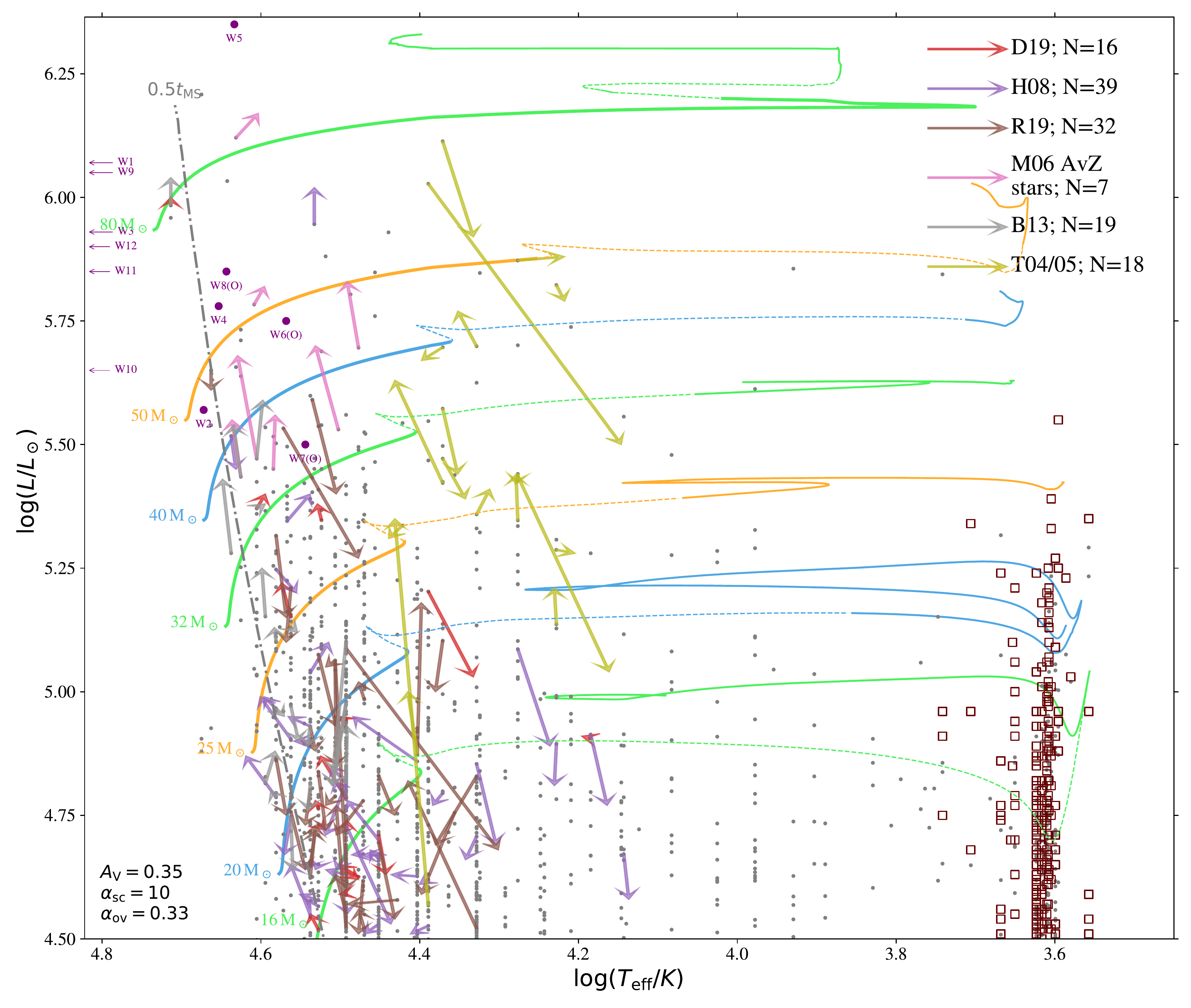}
   \caption{Same as Fig.\,\ref{fig:hrd_arr}, but with with spectral types taken fron Simbad instead of \cite{Bonanos10}. It contains 1095 sources instead of the 780 that are in our B10 sample HRD. 
   }
             \label{fig:hrd_sim}%
    \end{figure*}

We provide two more HRDs to test the robustness of our results (discussed in Sect.\,\ref{sec:results}). %In one, we extracted spectral types of SMC sources from Simbad instead of the B10 catalog. 
%In the first of these, instead of the B10 catalog we cross-correlated Simbad sources instead of B10 sources with the GAIA data. 
In one (Fig.\,\ref{fig:hrd_p13}), we have determined the temperatures and luminosities using the spectral type - temperature relations of \cite{Pecaut13} instead of those presented in Table\,\ref{tab:spt_teffs}. This results in slightly different individual values, but a population that again has almost the same shape as the one shown in Fig.\,\ref{fig:hrd_corr}.
In the other test, Fig.\,\ref{fig:hrd_sim}, the source catalog of the spectral types is Simbad instead of B10. The main motivation for making this Simbad HRD is to test whether 
%an apparent lack of young stars that we discuss in Sect.\,\ref{sec:rel_age_dist} 
the apparent lack of young and bright massive stars
might result from a biased input catalog. The resulting HRD (Fig.\,\ref{fig:hrd_sim}) contains more sources (1023 instead of 780), but it has the same features as the HRD shown in Fig.\ref{fig:hrd_corr}. 
%This implies that our results are robust against selection biases. 
This suggests that our results are robust from drawing on a (slightly) different compendium of observational types.

\begin{figure*}
   \centering
   \includegraphics[width = 0.665\linewidth]{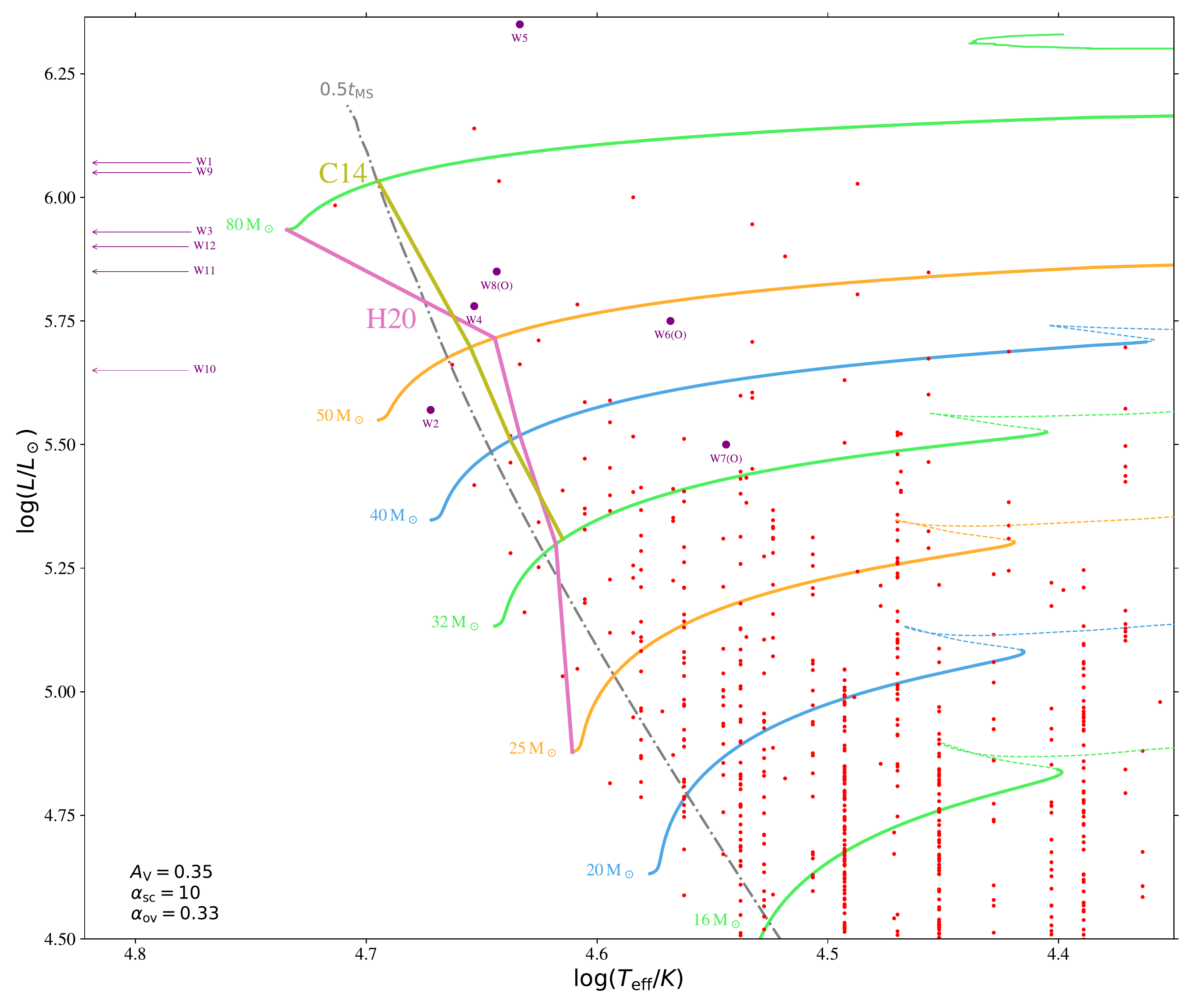}
   \caption{Same as Fig.\,\ref{fig:hrd_corr}, but here we indicate the regions that were found to be devoid of stars in earlier studies. To be more precise, there are no stars at the hot side of these lines in those studies. C14 means \cite{Castro14}, and H20 means \cite{Holgado20}. We note that we only show the hot stars in this diagram.
   }
             \label{fig:hrd_wvoids}%
    \end{figure*}

\begin{figure*}
   \centering
   \includegraphics[width = 0.665\linewidth]{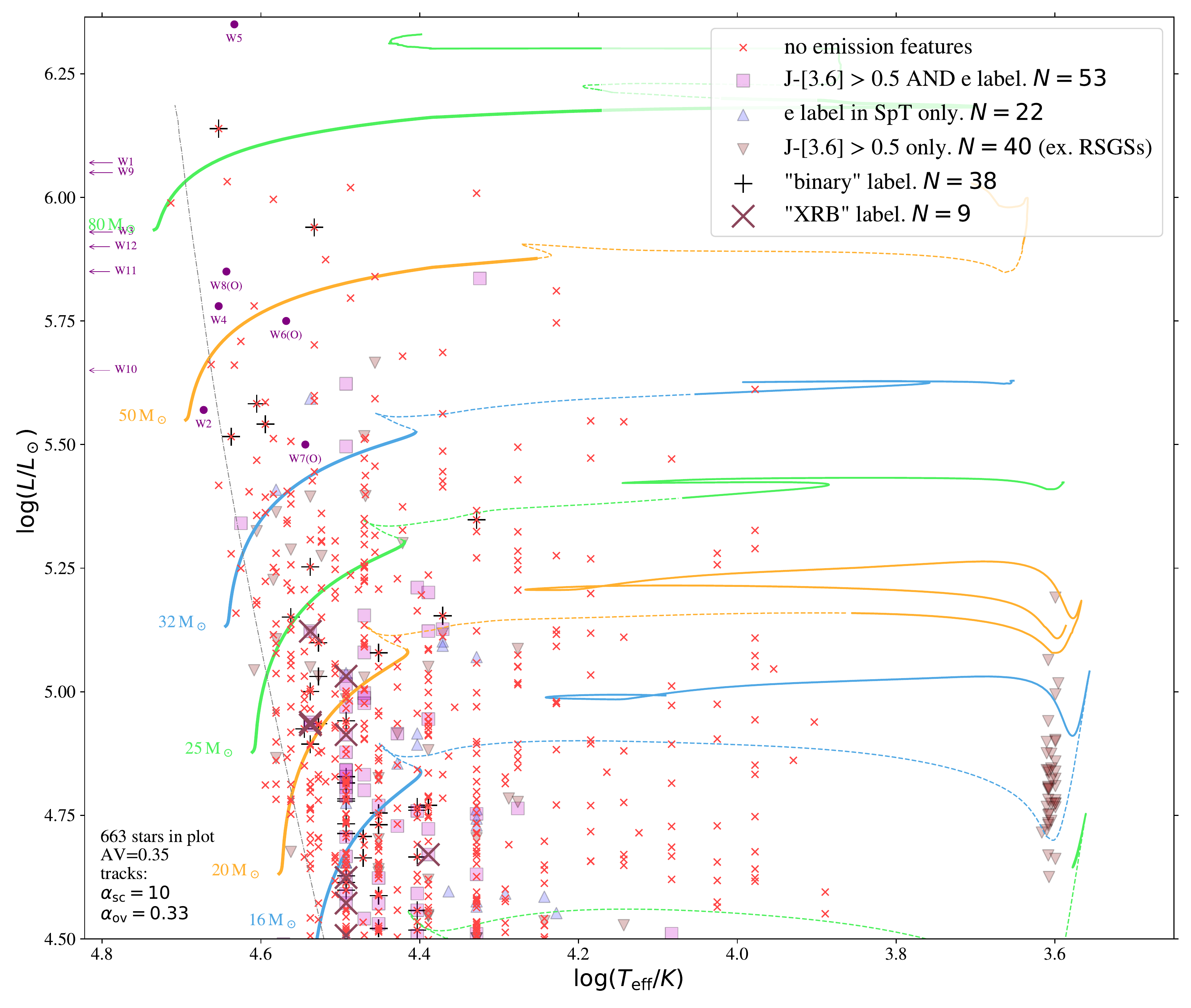}
   \caption{Hertzsprung-Russell diagram where we highlight sources with the label "binary" or "XRB" (X-ray binary) in \citealt{Bonanos10} (B10), and sources that show emission features. Emission features can be indicated by spectral type information in the B10 catalog, or by an infrared color larger than $J - [3.6] = 0.5$\, mag in B10. This figure contains slightly fewer sources than for example Fig.\,\ref{fig:hrd_corr} because it only shows those that have both a known $J$ and $[3.6]$ magnitude.
   }
             \label{fig:hrd_asc10}%
    \end{figure*}

The next HRD in this appendix is Fig.\,\ref{fig:hrd_wvoids}, which highlights the regions that were found to be completely devoid of stars in other studies. This is discussed in Sect.\,\ref{sec:sfh_disc}. These studies are \cite{Castro14} and \cite{Holgado20}, in which luminous Milky Way stars were spectroscopically analyzed. 
We have drawn the lines in our figure such that the distance in $\log ( T_\mathrm{eff} / K)$ from the ZAMS to the edge of the void is the same as in these studies.
The sample of \cite{Holgado20} contains 285 stars (O-types only) with evolutionary masses in excess of 25\Msol. The exact number is not clear for the HRD of \cite{Castro14}, where shading indicates the number density of stars, but it can be expected to be $\sim$100.
Fig.\,\ref{fig:hrd_wvoids} shows that these voids extend to temperatures comparable to the half-MS temperature. Inspection of the \cite{Brott11} models shows that this holds for Milky Way metallicity. This shows that in these subsamples with about (a few) 100 stars, fewer than a handful of stars are in the first half of their MS lifetime according to the tracks. In the evolutionary mass interval of $32-50$\Msol, this number reaches zero.
Finally, we show Fig.\,\ref{fig:hrd_asc10}, which highlights observed stars with emission features (discussed in Sect.\,\ref{sec:bsgs}) and (X-ray) binaries (discussed in Sect.\,\ref{sec:meth_lt}).

%Finally, we show two more HRDs. % for various purposes.
%Fig.\,\ref{fig:hrd_asc10} highlights observed stars with emission features (discussed in Sect.\,\ref{sec:bsgs}) and (X-ray) binaries (discussed in Sect.\,\ref{sec:meth_lt}). %Fig.\,\ref{fig:hrd_heburn_count} illustrates how we count helium burning stars between 18 and 29\Msol and their progenitors (Sect.\,\ref{sec:he_burn_progenitors}). %\textbf{In this interval, we have drawn the line between H-burning objects and He-burning objects at 20\kk. We argue that it is unlikely that many H-burning stars can be found at cooler temperatures. The reason is that very high overshooting values of $\alpha_\mathrm{ov} = 0.55$ are necessary for the TAMS to lie at 20\kk \citep{Schootemeijer19} in this case. Even then, we do not expect many MS stars around 20\kk because these models spend 95\% of their MS lifetime at $T_\mathrm{eff} < 25$\kk.}
%Fig.\,\ref{fig:hrd_synt} is the HRD with synthetic stars that we describe in Sect.\,\ref{sec:rel_age_dist}.

%______________________________________________

%______________________

%______________________________________________

%______________________

%______________________________________________

%______________________

%______________________________________________

%______________________

%______________________________________________

%______________________

%______________________________________________

%______________________

\clearpage 

\section{Calculation of star formation rate and ionizing radiation emission\label{sec:app_b}}

%______________________________________________

   \begin{figure}[h]
   \centering
   \includegraphics[width = \linewidth]{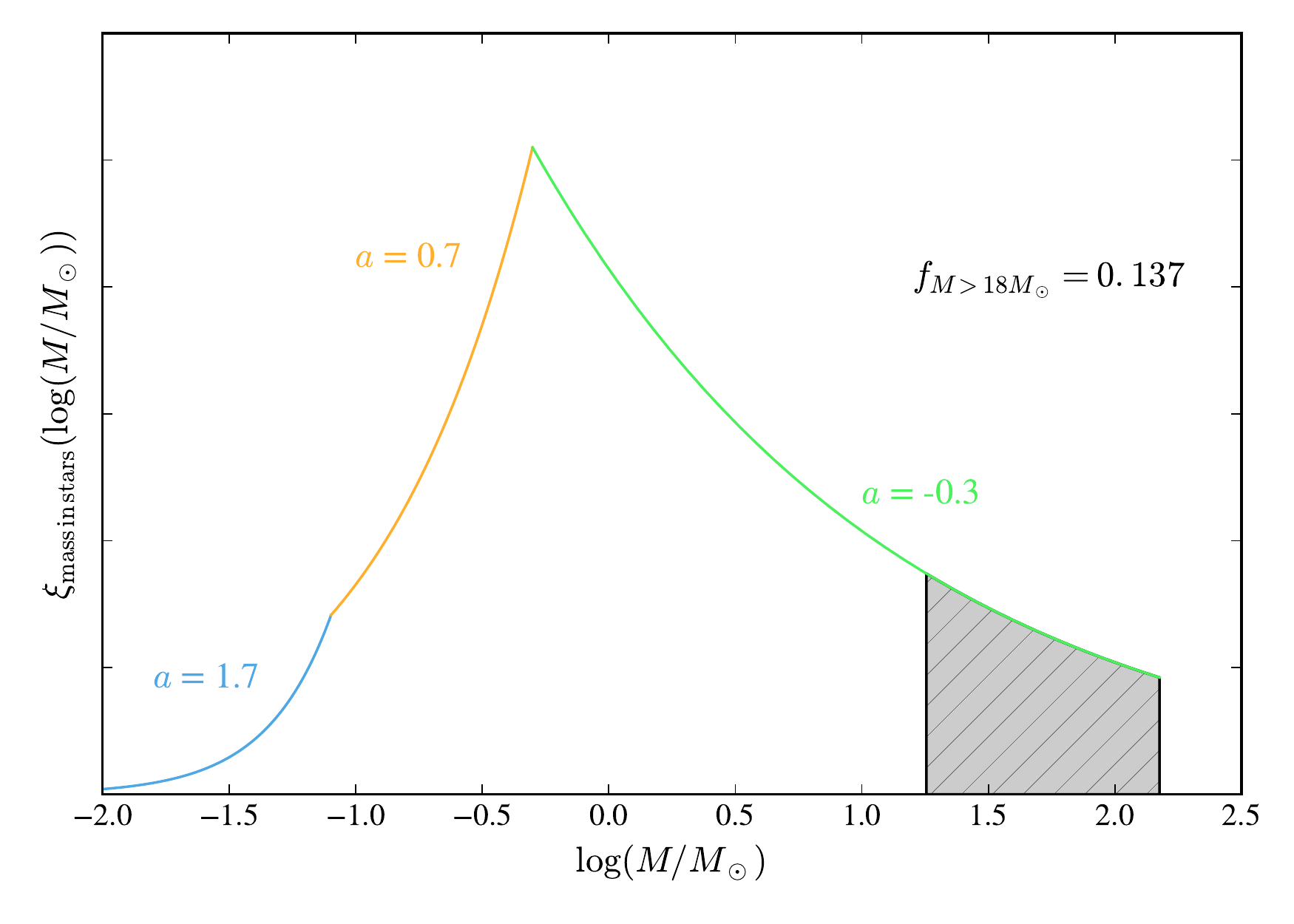}
   \caption{Distribution of stellar mass as a function of birth mass, upon star formation. Above the curve, we show the numbers that indicate the slope of the mass distribution. The hatched region is the region where stars have a mass between 18\Msol and the assumed upper mass limit of 150\Msol.
   }
             \label{fig:sfh_b}%
    \end{figure}
%______________________

\subsection{Star formation rate}

Here we describe how we have estimated the SFR discussed in Sect.\,\ref{sec:ex_etc}. For this we used the IMF of \cite{Kroupa01}. There, the exponent $\Gamma$ of the IMF, $\xi(\log M) = dN / d\log M \propto M^\Gamma$, with $M$ in solar units, has the following values in different mass intervals:
\begin{itemize}
    \item[] $\Gamma = +0.7$, \qquad for $ 0.01 \leq M < 0.08$
    \item[] $\Gamma = -0.3$, \qquad for $ 0.08 \leq M < 0.50$
    \item[] $\Gamma = -1.3$, \qquad for $ 0.50 \leq M$.
\end{itemize}

However, we are not interested in the number of stars, but in the yearly amount of mass that is converted into stars. 
Therefore the distribution has to be multiplied by a factor $M$ because more massive stars contain more mass. We consider $\xi_\mathrm{mass\,in\,stars}(\log M)  = dN / d\log M \propto M^a$. Thus,
\begin{itemize}
    \item[] $a = +1.7$, \qquad for $ 0.01 \leq M < 0.08$
    \item[] $a = +0.7$, \qquad for $ 0.08 \leq M < 0.50$
    \item[] $a = -0.3$, \qquad for $ 0.50 \leq M$.
\end{itemize}
Fig.\,\ref{fig:sfh_b} shows the resulting stellar mass distribution as a function of birth mass. Assuming an upper mass limit (this limit is more important now because the most massive stars contribute more in mass than in numbers) of 150\Msol, 13.7\% of the mass converted into stars ends up in stars born with 18\Msol or more. Therefore $f_{M > 18M_\odot} = 0.137$. This corresponds to the area of the hatched region divided by the total area below the curve in Fig.\,\ref{fig:sfh_b}.
This number changes slightly if, for example, 100\Msol (11.9\%) or 250\Msol (15.5\%) is chosen as upper mass limit.

Then, we proceed to estimate the SFR for CSF in the following steps. Clearly, the CSF scenario does not match the inferred age distribution, but we use it as a starting point in this estimation.
\begin{enumerate}
    \item We consider the orange line in the bottom right panel of Fig.\,\ref{fig:sfh}. It shows that to derive the amount of stars with $M > 18$\Msol that is in the HRD with the B10 sources (Fig.\ref{fig:hrd_corr}), they need to be produced at a rate of about 50\,Myr$^{-1}$ (after some million years, the curve drops because stars start to die). 
    \item We find that these stars have an average mass of $\sim$25\Msol. This results in approximately $1250$\Msol\,Myr$^{-1}$ or 0.00125\Msol\,yr$^{-1}$.
    \item The completeness of the stars above the 18\Msol track should be about $40-50$\% (Sect.\,\ref{sec:f_complete}). This doubles the inferred SFR to $\sim$0.0025\Msol\,yr$^{-1}$ in the considered mass range.
    \item Above, we found that $f_{M > 18M_\odot} = 0.137$. Integrated over the entire mass spectrum, the SFR therefore equals 0.0025\Msol\,yr$^{-1} / 0.137$ = 0.018\Msol\,yr$^{-1}$.
\end{enumerate}

A typical literature value for the SFR in the SMC at recent times is $\sim$0.05\Msol\,yr$^{-1}$ (Sect.\,\ref{sec:ex_etc}). This value is significantly higher than the value we just derived for a CSF scenario. If the SFR is not constant, as is the case for SFH2 and SFH3, it is necessary (to explain the observed number of $M > 18$\Msol stars) that $7-10$\Myr ago, the SFR was about a factor three higher than the SFR in the CSF scenario (bottom right panel in Fig.\,\ref{fig:sfh}). At that point in time, we therefore estimate that the SFR was about 0.045\Msol\,yr$^{-1}$. We note that this argument works the same way if, instead of a drop in the SFR, young stars are not observed due to biases.

We also investigated how much steeper the IMF should be to derive the typical literature value of $\sim$0.05\Msol\,yr$^{-1}$ %\textbf{(instead of $\sim$0.018\Msol\,yr$^{-1}$) under the assumption of CSF}. 
For this, we changed the exponent $a$ in the highest mass interval (green line in Fig.\,\ref{fig:sfh_b}). For $a=-0.6$ (i.e., a slightly steeper IMF of $\Gamma -1.6$) we find $f_{M > 18M_\odot} = 0.049$, which then results in the literature SFR of 0.05\Msol\,yr$^{-1}$. While tuning the IMF can fix the lack of bright stars, however, it does not fix the lack of young stars.

\subsection{Ionizing radiation}
We considered POWR models%\footnote{\url{{\tt www.astro.physik.uni-potsdam.de/~wrh/PoWR}}
\footnote{\url{www.astro.physik.uni-potsdam.de/~wrh/PoWR}}
\citep[SMC grid OB\,I of][see also \citealt{Todt15}]{Hainich19} and fit their $\log Q$, which is the ionizing photon production rate. We did this for H\,I, He\,I, and He\,II using the following fit formula:
%$\log Q_\mathrm{H1} = 105.008589 \log T_\mathrm{eff} - 11.1406044(\log T_\mathrm{eff})^2 + 0.235248626\log L + 0.0827383002(\log L)^2 - 201.969690.$

%\begin{equation}
%    \log Q = C_\mathrm{T, \,1} \log T_\mathrm{eff \, m4} + C_\mathrm{T, \,2} (\log T_\mathrm{eff\, m4})^2 + C_\mathrm{T, \,3} (\log T_\mathrm{eff\, m4})^3   + C_\mathrm{L, \,1} \log L + C_\mathrm{L, \,2} (\log L)^2 + D
%\end{equation}

\begin{eqnarray}\nonumber
 \log Q&=& C_\mathrm{T, \,1} \log T_\mathrm{eff \, m4}
+ C_\mathrm{T, \,2} (\log T_\mathrm{eff\, m4})^2 + C_\mathrm{T, \,3} (\log T_\mathrm{eff\, m4})^3\\
&& + C_\mathrm{L, \,1} \log (L / L_\odot) + C_\mathrm{L, \,2} (\log (L / L_\odot))^2 + D.
\end{eqnarray}

Here, $\log T_\mathrm{eff \, m4}$ is defined as $\log(T_\mathrm{eff} / K) -4$, which allowed a better fit than $\log ( T_\mathrm{eff} / K )$. The fitting constants that we find for H\,I, He\,I, and He\,II are tabulated in Table\,\ref{tab:iotons}. The POWR grid extends to 50\kk. Because we are hesitant to extrapolate outside this range, we treated hotter stars as if their $T_\mathrm{eff}$ were 50\kk. The observed population has only one star above $T_\mathrm{eff} = 50$\kk, at 51.7\kk.
However, the synthetic population shown in Fig.\,\ref{fig:hrd_synt} contains more stars above $T_\mathrm{eff} = 50$\kk, and even hotter stars. It is therefore possible that we underestimate $Q_\mathrm{He\, II}$ in the synthetic population. This would strengthen our conclusions that $Q_\mathrm{He\, II}$ is much lower in the observed population than in the synthetic population.
This limit of 50\kk is expected to play less of a role for the total of $Q_\mathrm{H\, I}$ and $Q_\mathrm{He\, I}$, because these are not dominated by the $\sim$10 hottest stars.

In Table\,\ref{tab:iotons} we show the ionizing photon production rate of the 780 stars in the observed population (Fig.\,\ref{fig:hrd_corr}) and the synthetic population (Fig.\,\ref{fig:hrd_synt}).
Because the brightest, hottest stars are expected to be fairly complete (about 80\%), we do not expect the real values to be much higher in practice.

%This fit formula is accurate up to 0.1\,dex, at least at the higher end. We find $Q_\mathrm{HI} = 1.4 \cdot 10^{51}$ to be the sum of all B10 sources shown in our main HRD (Fig.\,\ref{fig:hrd_corr}). Most of this is contributed by the first 100 sources (Fig.\,\ref{fig:iotons}). Bright sources should be relatively complete: about 75\% (Table\,\ref{tab:f_complete}). Thus, we estimate a corrected number of $Q_\mathrm{HI} = 2 \cdot 10^{51}$.

POWR models %\citep{Todt15, Hainich19} 
of WR stars (all have $\log (L/L_\odot) = 5.3$) hotter than 50\kk emit $\log Q_\mathrm{HI} = 10^{49.15}$. Above this temperature value, this number changes only by about $0.02$\,dex for different temperatures, transformed radii, and hydrogen surface abundances. We can accordingly assume for the SMC WR stars analyzed by \cite{Hainich15} and \cite{Shenar16} that they emit $Q_\mathrm{HI} = 10^{49.15 - 5.3} = 10^{43.85}$ photons per second per unit of solar luminosity. Summing the luminosities of all these SMC WR stars yields a total luminosity of $\log (L/L_\odot) = 7.17$. Their total rate of HI ionizing photon production is then $Q_\mathrm{H\,I} = 10^{43.85 + 7.17} = 10^{51.02}$. %(\textit{meh add units}). %photons\,s$^{-1}$.
For He\,I we adopt $Q_\mathrm{H\,I} = 10^{48.95 - 5.3} = 10^{43.85}$ photons per second per unit of solar luminosity as a representative value after looking up POWR models. This leads to $Q_\mathrm{He\,I} = 10^{50.82}$. Similarly, for He\,II, we obtain $Q_\mathrm{He\,II} = 10^{46.55 - 5.3} = 10^{41.25}$ photons per second per unit of solar luminosity, making a total of $Q_\mathrm{He\,II} = 10^{48.42}$.

We summed the H\,I ionizing emission of WR and non-WR stars. This results in $Q_\mathrm{H\,I} = 3 \cdot 10^{51}$. %photons per second.
$Q_\mathrm{He\,I,\, tot}$ is dominated by WR stars, although in the synthetic population the contribution of MS stars is comparable. On the other hand, $Q_\mathrm{He\,I}$ is completely dominated by the WR stars in any case by some orders of magnitude. 

%% COMMENTED PROMISE
%\textit{Instead of assuming an average value, I will just add up individual values for $Q_\mathrm{He\,I}$ and $Q_\mathrm{He\,II}$, but the conclusions won't change.}

%__________________________________________________ One column table
\begin{table}
\centering    
    \caption[]{Values used for the $\log Q$ fit. The resulting values of ionizing photon production rate $Q$ are also shown in units of photons per second for the observed population (Fig.\,\ref{fig:hrd_corr}), the synthetic population (Fig.\,\ref{fig:hrd_synt}, for reference), and the SMC WR stars. }
\begin{tabular}{l | r r r }
\hline
\hline
            \noalign{\smallskip}
             & H\,I & He\,I & He\,II\\
            \noalign{\smallskip}
            \hline
            \noalign{\smallskip}
            $C_\mathrm{T, \,1}$ & 8.11803352896 & 153.255573939 & 2394.03074992\\
            $C_\mathrm{T, \,2}$ & 10.2526683624 & -203.063138384 & -3731.75720934\\ 
            $C_\mathrm{T, \,3}$ & -17.1132439362 & 89.5572474026 & 1946.01231826\\
            $C_\mathrm{L, \,1}$ & 3.68149521625 & 3.22654049538 & 0.717552405798\\
            $C_\mathrm{L, \,2}$ & -0.226236345452 & -0.186185975219 & 0.043861662701\\
            $D$ & 31.0059348278 & -1.84481792354 & -475.168132327\\
            Fit range & $20-50$\kk & $28-50$\kk & $36-50$\kk\\
            \noalign{\smallskip}
            %\hline
            %\noalign{\smallskip}
            %$\sigma_\mathrm{log \, Q \, fit}$ & 0.066 & 0.079 & 0.066\\
            %\noalign{\smallskip}
            \hline
            \noalign{\smallskip}
            $\Sigma Q_\mathrm{obs}$ & $1.89 \cdot 10^{51}$ & $2.22 \cdot 10^{50}$ & $4.94 \cdot 10^{45}$\\
            $\Sigma Q_\mathrm{synt}$ & $4.27 \cdot 10^{51}$ & $8.79 \cdot 10^{50}$ & $4.82 \cdot 10^{46}$\\
            $\Sigma Q_\mathrm{WR}$ & $1.05 \cdot 10^{51}$ & $6.61 \cdot 10^{50}$ & $2.63 \cdot 10^{48}$%         
            %\hline
            \label{tab:iotons}

\end{tabular}
\end{table}
%______________________

\section{Positions and motions of sources \label{sec:app_c}}

%______________________________________________

   \begin{figure*}
   \centering
   \includegraphics[width = 0.8\linewidth]{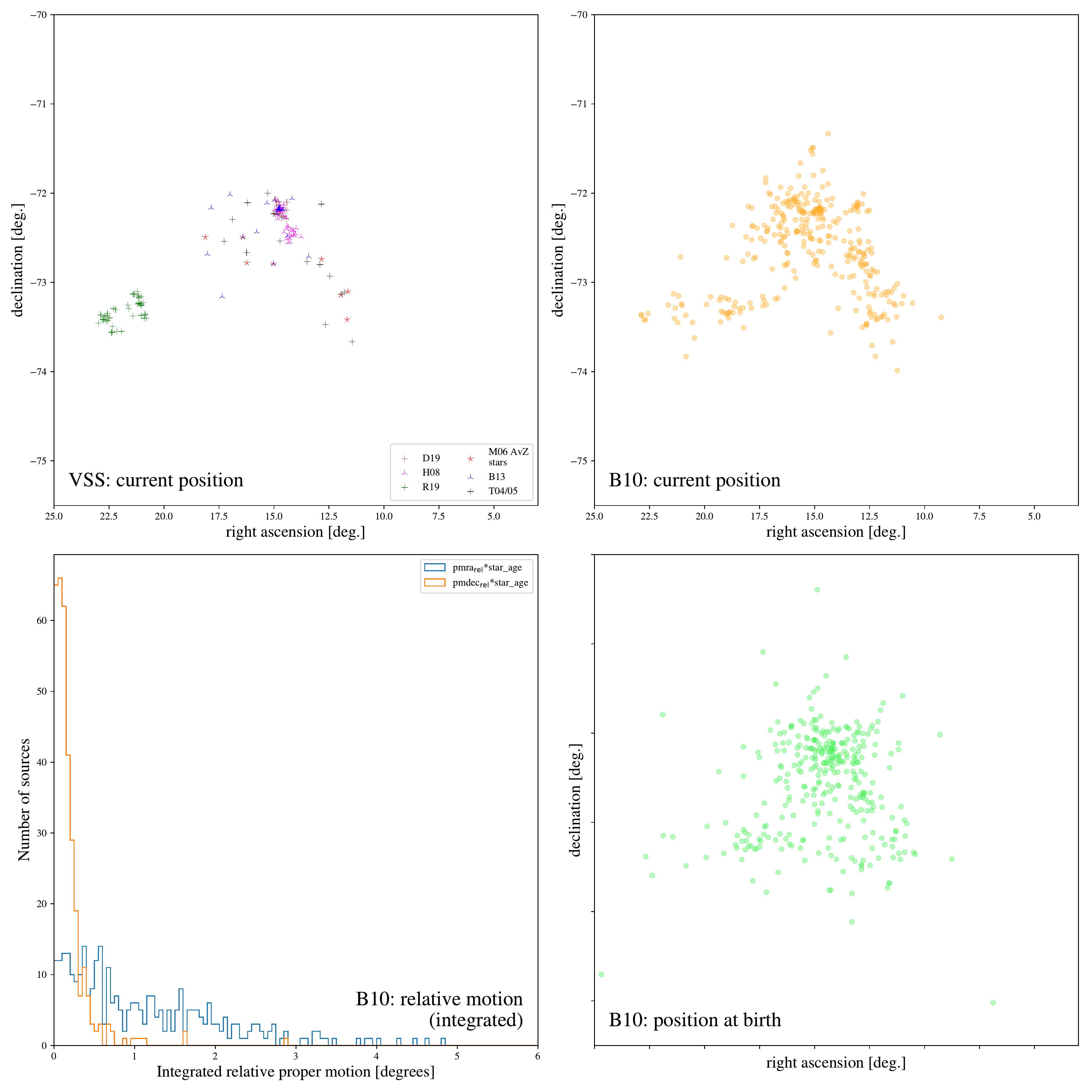}
   \caption{Diagrams showing positions and motions of sources in the Small Magellanic Cloud. The top left panel shows the positions of the sources that are in the VSS. %and also in Fig.\,\ref{fig:hrd_corr}. %The bottom left panel shows proper motions relative to the SMC proper motions of \cite{Yang19}.
   {The bottom left panel shows the distribution of the integrated motion relative to the SMC of B10 sources (that are above the 18\Msol track in Fig.\,\ref{fig:hrd_corr}) since their birth (see text).}
   The top and bottom right panel show the positions of these B10 sources now and integrated to their their moment of birth in the SMC, respectively.% $N^\mathrm{th}$ brightest source according to the 
   }
             \label{fig:rel_motion}%
    \end{figure*}
%______________________

Below, we briefly investigate whether the inferred drop of the SFR (Sect.\,\ref{sec:rel_age_dist}) would have needed to take place over the entire SMC, or if the stars younger than 10\Myr can come from only a few birth places.
For this, we considered the GAIA proper motions of the sources in the B10 data set that are shown in Fig.\,\ref{fig:sfh}, that is, those with evolutionary masses in excess of 18\Msol and ages younger than 10\Myr.
%The proper motions of the sources in the B10 sample (bottom left panel in Fig.\,\ref{fig:rel_motion}) indicate that they typically move at most a few arcminutes per Myr within the SMC \citep[compared to the proper motions of][]{Yang19}. This means that in the timescale of a few Myr, they do not cross significant parts of the SMC, which has an angular size of the order of a few (in declination) to ten (in right ascension) degrees.
For each of these sources, we calculated their proper motion relative to the SMC by subtracting an SMC bulk motion of $\mu_\mathrm{ra} = 0.695$\,mas/yr and $\mu_\mathrm{ra} = -1.206$\,mas/yr \citep{Yang19} from their GAIA proper motions. Then, we used their inferred age (Sect.\,\ref{sec:rel_age_dist}) to estimate how far they have traveled during their lifetime so far, in  terms of both right ascension and declination. The bottom left panel of Fig.\,\ref{fig:rel_motion} shows the distribution of the absolute values of how far the sources traveled in both directions. In right ascension, the median is 0.82 degrees. In declination, the median is 0.13 degrees. The main reason for this difference is that the SMC, with its declination of about -72.5 degrees, is close to the south pole.

We can compare this with motions in the line-of-sight direction. \cite{evans08} provided radial velocities of 2045 2dF sources (about half of the sample). The value that they find for the radial velocity dispersion is $\sim$30\kms. For source with this velocity and a typical age of 7\Myr (Fig.\,\ref{fig:sfh}), this indicates a typical lifetime travel distance of about 200\,pc. This distance would correspond to a projected angular distance of 0.17 degrees in the declination direction. This shows that we obtain similar results from the proper motions and the radial velocity dispersion.
For both methods, the lifetime travel distances are likely amplified by the errors (observational errors, and for the radial velocities, reflex motion due to binary companions). The real travel distances within the SMC would therefore most likely be smaller than the values mentioned above.

The top right panel of Fig.\,\ref{fig:rel_motion} shows the current position of the sources that are displayed in Fig.\,\ref{fig:sfh}. %(i.e., those that are above the 18\Msol track in Fig.\,\ref{fig:hrd_corr}). 
The SMC has an angular size on the order of a few (in declination) by ten (in right ascension) degrees. This is more than a factor ten larger than the typical angular distance that these sources have traveled with respect to the SMC during their lifetime. During their lifetimes, these sources therefore cannot have traveled through a significant part of the SMC.
%. See Fig.\,\ref{fig:rel_motion}. 
Combining this with the fact that the B10 sources %with $\log (L/L_\odot) > 4.5$ 
above 18\Msol
are currently spread out all over the SMC (top right panel in Fig.\,\ref{fig:rel_motion}), it is implied that the hypothetical halt in star formation %(as discussed above) 
would have had to take place over the entire SMC at roughly the same time.
%We deem this unlikely because the SFR is typically thought to change on much longer timescales \citep[e.g.,][]{Weisz13} -- see also our discussion in Sect.\,\ref{sec:sfh_disc}. %(to discussion).

In the bottom right panel of Fig.\,\ref{fig:rel_motion}, we give an impression of the birth locations within the SMC of the sources that are displayed in Fig.\,\ref{fig:sfh}. Similar to what is described above, we used relative proper motions and inferred ages of these sources to trace back where they resided in the SMC when they were born.
This figure implies that at the moment of birth, these sources were also spread out over the SMC.
The angular scale of the plot is the same as in the top right panel of the figure. This strengthens our original assumption that the stars above 18\Msol originate from a multitude of birth places in the SMC, rather than being born in only a few sites.
For reference, we also show the positions of the sources in the VSS sample in the top left panel of Fig.\,\ref{fig:rel_motion}.

\end{appendix}
\end{document}